\documentclass[%
 reprint,
superscriptaddress,
 amsmath,amssymb,
 aps,
rmp,
bm,
]{revtex4-2}

\usepackage[english]{babel}
\usepackage{xpatch}
\makeatletter
\xpatchcmd{\@ssect@ltx}{\@xsect}{\protected@edef\@currentlabelname{#8}\@xsect}{}{}
\xpatchcmd{\@sect@ltx}{\@xsect}{\protected@edef\@currentlabelname{#8}\@xsect}{}{}
\makeatother

\usepackage{nameref} 
\usepackage{graphicx}
\usepackage{dcolumn}
\usepackage{bm}
\usepackage{upgreek} 
\usepackage{tabularx} 
\usepackage{booktabs} 
\usepackage{diagbox} 
\usepackage{multirow} 
\usepackage{cellspace} 
\usepackage{xcolor} 
\usepackage{makecell} 
\usepackage{ulem} 
\usepackage[rightcaption]{sidecap} 

\usepackage[breaklinks]{hyperref}

\begin{document}


\title{
Charge, heat, and spin transport phenomena in metallic conductors}
\graphicspath{{./Figures/}}

\newcommand{\nynke}[1]{\textcolor{red}{#1}} 
\newcommand{\hbl}[1]{\textcolor{blue}{#1}} 
\newcommand{\hblc}[1]{\textcolor{red}{\textbf{#1}}} 

\newcommand{\stbg}[1]{\textcolor{brown}{#1}} 
\newcommand{\NV}[1]{\textcolor{teal}{#1}} 
\newcommand{\NVL}[1]{\textcolor{violet}{#1}} 

\newcommand{\rholong}{\rho_{\mathrm{long}}}
\newcommand{\Vlong}{V_{\mathrm{long}}}
\newcommand{\rhotrans}{\rho_{\mathrm{trans}}}
\newcommand{\Vtrans}{V_{\mathrm{trans}}}
\newcommand{\Jvec}{\mathbf{J}_\mathrm{c}}
\newcommand{\Jc}{J_\mathrm{c}}
\newcommand{\Jvecheat}{\mathbf{J}_\mathrm{h}}
\newcommand{\Jvecspin}{\mathbf{J}_\mathrm{s}}
\newcommand{\Jvecspintilde}{\tilde{\mathbf{J}}_\mathrm{s}}
\newcommand{\Jvecspintildesigma}{{\Jvecspintilde}{_{,\bm{\hat{\sigma}}}}}
\newcommand{\Jvecspinx}{\mathbf{J}_{\mathrm{s},x}}
\newcommand{\Jvecspiny}{\mathbf{J}_{\mathrm{s},y}}
\newcommand{\Jvecspinz}{\mathbf{J}_{\mathrm{s},z}}
\newcommand{\Jvecspini}{\mathbf{J}_{\mathrm{s},i}}
\newcommand{\sigmavec}{\bm{\sigma}_\mathrm{c}}
\newcommand{\sigmavecheat}{\bm{\sigma}_\mathrm{h}}
\newcommand{\sigmavecspin}{\bm{\sigma}_\mathrm{s}}
\newcommand{\muspin}{\mu_\mathrm{s}}
\newcommand{\muspinx}{\mu_{\mathrm{s},x}}
\newcommand{\muspiny}{\mu_{\mathrm{s},y}}
\newcommand{\muspinz}{\mu_{\mathrm{s},z}}
\newcommand{\muspini}{\mu_{\mathrm{s},i}}
\newcommand{\mucharge}{\mu_\mathrm{c}}
\newcommand{\muheat}{\mu_\mathrm{h}}
\newcommand{\rhovec}{\bm{\rho}_\mathrm{c}}
\newcommand{\kappavec}{\bm{\kappa}}
\newcommand{\Ec}{E_\mathrm{c}}
\newcommand{\Efield}{\mathbf{E}}
\newcommand{\Efieldc}{\mathbf{E}_{\mathrm{c}}}
\newcommand{\tauc}{\tau_\mathrm{c}}
\newcommand{\tauphonon}{\tau_\mathrm{p}}
\newcommand{\taumagnon}{\tau_\mathrm{m}}
\newcommand{\taucharge}{\tau_\mathrm{c}}
\newcommand{\sigmaspin}{\sigma_\mathrm{s}}
\newcommand{\sigmac}{\sigma_\mathrm{c}}
\newcommand{\rhoc}{\rho_\mathrm{c}}
\newcommand{\thetaSH}{\theta_\mathrm{SH}}
\newcommand{\thetaSN}{\theta_\mathrm{SN}}
\newcommand{\charge}{q}

\newcommand{\Jvectrans}{\Jvec^\mathrm{trans}}
\newcommand{\Jhvectrans}{\Jvecheat^\mathrm{trans}}
\newcommand{\Jsvectrans}{\Jvecspin^\mathrm{trans}}
\newcommand{\Jsvectildetrans}{\Jvecspintilde^\mathrm{trans}}
\newcommand{\Jveclong}{\Jvec^\mathrm{long}}
\newcommand{\Jhveclong}{\Jvecheat^\mathrm{long}}
\newcommand{\Jsveclong}{\Jvecspin^\mathrm{long}}
\newcommand{\Jsvectildelong}{\Jvecspintilde^\mathrm{long}}
\newcommand{\potc}{\phi_{\mathrm{c}}}
\newcommand{\poth}{\phi_{\mathrm{h}}}
\newcommand{\pots}{\phi_{\mathrm{s}}}
\newcommand{\mingradmuc}{-\boldsymbol{\nabla} \mu_{\mathrm{c}}}
\newcommand{\gradmuc}{-\boldsymbol{\nabla} \mu_{\mathrm{c}}}
\newcommand{\mingradT}{-\boldsymbol{\nabla} T}
\newcommand{\gradT}{-\boldsymbol{\nabla} T}
\newcommand{\mingradmus}{-\boldsymbol{\nabla} \mu_{\mathrm{s}}}
\newcommand{\gradmus}{-\boldsymbol{\nabla} \mu_{\mathrm{s}}}
\newcommand{\textleft}{\leftskip=4pt\makecell[l]}
\newcommand{\Hvec}{\mathbf{H}}
\newcommand{\Mvec}{\mathbf{M}}
\newcommand{\Bvec}{\mathbf{B}}
\newcommand{\Bhat}{\mathbf{\hat{b}}}
\newcommand{\Hhat}{\mathbf{\hat{h}}}
\newcommand{\Mhat}{\mathbf{\hat{m}}}
\newcommand{\sigmapol}{\hat{\bm{\sigma}}}

\newcommand{\captionclose}{-23pt}
\newcommand{\captionhalfclose}{-10pt} 

\newcommand{\figuretop}{-20pt}
\newcommand{\textfigures}{0.65\hsize} 
\newcommand{\Axes}{0.26\hsize} 
\newcommand{\textfiguresPerp}{0.47\hsize} 
\newcommand{\planarFigs}{3cm} 
\newcommand{\textfigureThomson}{0.6\hsize} 

\newcommand{\firstcolumn}{0.3\columnwidth} 
\newcommand{\firstcolumnsingle}{0.2\columnwidth} 
\newcommand{\centerfirstt}{-1.4ex} 
\newcolumntype{Y}{>{\centering\arraybackslash}X} 
\newcolumntype{Z}{>{\hsize=1.1\hsize}Y} 
\newcolumntype{?}{!{\vrule width 1.5pt}} 
\newcolumntype{M}{>{\raggedright\arraybackslash}X} 
\definecolor{drive}{HTML}{333399}
\definecolor{response}{HTML}{CC3333}
\definecolor{grey-text}{HTML}{C0C0C0}
\setlength{\cellspacetoplimit}{6pt}
\setlength{\cellspacebottomlimit}{6pt}

\newcommand{\figurewidth}{0.95\hsize} 

\newcommand{\kboltz}{k_\mathrm{B}}

\author{Nynke Vlietstra}
\affiliation{Walther-Mei{\ss}ner-Institut, Bayerische Akademie der Wissenschaften, 85748 Garching, Germany}
\affiliation{School of Natural Sciences,  Technische Universit\"{a}t M\"{u}nchen, 85748 Garching, Germany}

\author{Sebastian T. B. Goennenwein}
\affiliation{Department of Physics, University of Konstanz, D-78457 Konstanz, Germany}

\author{Rudolf Gross}
\affiliation{Walther-Mei{\ss}ner-Institut, Bayerische Akademie der Wissenschaften, 85748 Garching, Germany}
\affiliation{School of Natural Sciences,  Technische Universit\"{a}t M\"{u}nchen, 85748 Garching, Germany}
\affiliation{Munich Center for Quantum Science and Technology (MCQST), 80799 M\"{u}nchen, Germany}

\author{Hans Huebl}
\email{huebl@wmi.badw.de}
\affiliation{Walther-Mei{\ss}ner-Institut, Bayerische Akademie der Wissenschaften, 85748 Garching, Germany}
\affiliation{School of Natural Sciences,  Technische Universit\"{a}t M\"{u}nchen, 85748 Garching, Germany}
\affiliation{Munich Center for Quantum Science and Technology (MCQST), 80799 M\"{u}nchen, Germany}

\date{\today}

\begin{abstract}
In solid state materials, gradients of the electro-chemical potential, the temperature, or the spin-chemical potential drive the flow of charge, heat, and spin angular momentum, resulting in a net transport of energy. 
Beyond the primary transport processes—such as the flow of charge, heat, and spin angular momentum driven by gradients in their respective potentials—a wide range of coupled or cross-linked transport responses can occur, giving rise to a rich variety of transport phenomena. These transport phenomena are commonly categorized under (anomalous) thermoelectric, thermomagnetic, and galvanomagnetic effects, along with their spin-dependent counterparts. However, establishing a systematic classification and comparison among them remains a complex and nontrivial task. This paper attempts a didactic overview of the different transport phenomena, by categorizing and briefly discussing each of them based on charge, heat, and spin transport in conducting solids. The phenomena are structured in three categories: collinear, transverse, and so-called `planar' transport effects. The resulting overview attempts to categorize all effects in a consistent manner.

\end{abstract}


\maketitle
\tableofcontents
\section{Introduction}
\label{sec:Intro}
Transport phenomena play a key role in condensed matter physics and have been studied intensively in the past two centuries, resulting in the discovery of various electrical, thermal, thermoelectric, thermomagnetic, galvanomagnetic, and related spin-dependent effects.\citep{Gross,OHandley,BookTakahashi,Maekawa:2017gn, BauerCaloritronics,BauerCaloritronicsReview} Transport effects describe the flow of quantized entities such as charge carriers, phonons, magnons, flux quanta etc.\, under the action of generalized forces that can be expressed as gradients of potentials. The flow of the quantized entities is expressed by the respective current densities.  Typically, the treatment of transport phenomena  is restricted to the linear response regime, where the current density depends linearly on the corresponding potential gradient. The associated proportionality constant is called the linear transport coefficient. Examples of such linear response transport effects are  Ohm's law, the Seebeck effect, the Hall effect, or the Nernst effect.\cite{BookOhm, Seebeck, Hall, Nernst}  Although this concept appears simple at first glance, there are many facets contributing to the richness of the mentioned phenomena. First, the transport coefficients usually are not just numbers but second rank tensors for crystalline materials. \cite{Gross, birss1966symmetry} Second, the different entities can support different types of current densities at the same time, leading to a pronounced coupling of multiple transport phenomena. 
Given the multitude of contributing factors and the different transported entities at play, it is important to recognize that these multi-faceted transport phenomena  can, in fact, be consistently categorized. It is the aim of this paper to summarize and categorize the associated effects.

It has been recognized early on that charge carriers in solids transport both charge and heat and hence a thermal gradient causes the transport of not only heat, but also charge. The resulting coupling of charge and heat transport  is called  {\it\bfseries thermoelectric coupling} and results in the corresponding {\it\bfseries thermoelectric effects} (Seebeck, Peltier, and Thomson effect), which have been studied already in the early 1800's \citep{Seebeck,Seebeck2,Peltier,Thomson1,Thomson2}. Since potential differences (voltages) often are easier to detect experimentally, some of the established transport effects describe voltage or temperature differences induced by a driving gradient. We address this point in more detail below. 

Later on, it was found that the application of a magnetic field perpendicular to the applied potential gradients results in transverse charge and heat currents under the action of longitudinal thermal or electrochemical potential gradients. These transport phenomena were discovered at the end of the  19$^\mathrm{th}$ century and are often referred to as {\it\bfseries thermomagnetic effects} (transverse charge and heat current caused by a longitudinal temperature gradient, Nernst and Righi-Leduc/thermal Hall effects) and  {\it\bfseries galvanomagnetic effects} (transverse charge and heat current caused by a longitudinal electric field, Hall and Ettingshausen effects).\citep{Thomson,PHE,Hall,HallAHE,AHE2,Nernst,Nernst2,Ettingshausen,Righi,Leduc,Butler,Zahn,SmithLeduc,PNE,PNE2}. In materials with finite magnetization a (transverse) Hall and Nernst effect have been found even without any applied magnetic field and called the anomalous Hall and Nernst effect. \cite{HallAHE, Nernst2}

In the general case, the magnetic field can also have a finite projection oriented along the direction of the applied potential gradients. If the transport is additionally restricted to a plane, which is common for many experiments, this \textit{in-plane} magnetic field component gives rise to the so-called \textit{planar} effects (e.g.  planar Hall and planar Nernst effect).\citep{PHE,PNE,PNE2} This behavior is typically observed in materials with a finite magnetic moment, where the modification of the transport is associated with the orientation of the magnetization.\cite{OHandley, Gross, PHE,PNE,PNE2} While the nomenclature ''planar effects'' is established in literature, this choice of naming unfortunate, as the effects of this category are of magnetoresistive nature in magnetic metals.\citep{Ky:1968,PHE,PNE,PNE2}

With the discovery of the spin degree of freedom of electrons, the next level of complexity was introduced into transport phenomena. Since electrons can transport angular momentum in addition to charge and heat, it is immediately obvious that this can result in a {\it\bfseries spin-thermo-electric coupling} and, in turn, in a large variety of additional {\it\bfseries spin-thermo-electric effects}. 
Moreover, the application of a magnetic field perpendicular to the applied potential gradients results in additional effects, which in analogy to the thermoelectric phenomena may be classified into spin-thermomagnetic and spin-galvanomagnetic effects as well as their planar versions. 

Interestingly, it took till the end of the 20$^\mathrm{th}$ century until the first transport phenomena related to the spin degree of freedom were reported\citep{DyakonovPerel1,DyakonovPerel2,Hirsch}. In the meantime, many of them have been identified and studied in detail under the umbrella of the field of spin- and spincaloritronics, with prominent examples such as the spin Hall Effect, the spin Seebeck effect, spin Peltier effect, and the spin Nernst effect.\citep{SHE1,SHE2,Uchida1,SdPE_Gravier,SdPE_Flipse,SPE_Flipse,Katsura,Onose,SNEtheory1,SNEtheory2,NernstWMI}  

The fact that the transport of angular momentum related to  the spin and orbital degrees of freedom can be driven by a thermal gradient leads to a coupled transport of heat and spin (spin-thermal coupling). This led to the very active research field of spin-caloritronics (study of the coupled spin and heat transport) \cite{BauerCaloritronics}. Whereas the coupling of charge and heat transport already results in a variety of thermoelectric effects, the coupling of charge, heat, and spin transport further increases complexity, thus making transport phenomena surprisingly rich and complex \citep{Gross,Overview1,Grimmer:2017,HeremansReview,HarmanHonig,BookZiman,OHandley,BehniaBook,Goldsmid}. Moreover, additional features such as spin-orbit coupling or the presence of topological spin textures make the situation even more interesting. However, this can also lead to confusion. This renders the  classification of the different transport phenomena and their complex interplay a formidable task, resulting also in some inconsistencies in nomenclature in the past.
   
In this article, we provide a basic overview of the most common charge, heat and spin related transport phenomena, which have been discovered and studied in the past centuries, and set them in context to each other. Doing so we take an experimentalists perspective in the sense that we do not rigorously derive expressions for transport coefficients from microscopic theory, but rather classify the various transport phenomena using a consistent classification scheme. We will see that there are some phenomena which have been named inconsistently in the literature (like the spin Hall effect or the magnon Hall effect) and that there are some phenomena which even have not yet been experimentally demonstrated (to the best of our knowledge). We will restrict our discussion to the linear response regime, where the generated currents depend linearly on the generalized forces, which can be expressed as gradients of suitable potentials (see also Refs.\citep{VlietstraThesis,Gross,Overview1,HeremansReview,BookZiman,OHandley,HarmanHonig,Schmid,BehniaBook,Goldsmid,BauerCaloritronicsReview}). Moreover, we focus on the diffusive transport phenomena in context with the transport of charge carriers, phonons and spins in metallic conductors. Since charge carriers can also carry heat and angular momentum, coupled transport properties arise, which we discuss in the specific sections. Note that there are a large variety of (quasi-) particles that  can transport charge, heat, and spin in solids. For example, charge can be transported also by ions, excitons, or polarons,\cite{Gross,Ashcroft:1976ud,BookZiman} heat can be carried by phonons,\cite{Gross,BookGrosso,BookZiman}, magnons\cite{Boona:2014fh,Pan:2013gga,Sato:1955hb,Akhiezer:1958tb} and many other quantized excitations, while angular momentum can be transported by magnons or the nuclear spin. The plethora of transport phenomena associated with all these excitations are beyond the scope of this article. Moreover, for the sake of simplicity, for systems with a finite magnetization, we restrict the discussion to the ferromagnetically ordered case and assume the absence of complex magnetic textures and phenomena originating from a non-trivial topology. We also will not discuss transport phenomena in the mixed state of superconductors, where the motion of quasiparticles and flux quanta under the action of an applied temperature gradient results in longitudinal and transverse voltages. In analogy to normal metals these effects have been named (flux) Seebeck and Nernst effect\citep{Huebener:1969,Zeh:1990,Palstra:1990,Galffy:1990,Ri:1991,Ri:1993,Viewpoint,Behnia2,Behnia}. 

Moreover, we do not specifically include the categorization of effects linked to the field of orbitronics in this article, however, we note that they could conceptually be included in the discussed categorizaiton scheme \citep{Go:2018,Go:2020, Jo:2024, Kontani:2009, Bernevig:2005, Zhang:2005, Wang:2025, Jo:2018, Ding:2020, Choi.2023}.

This article is organized as follows: In section\,\ref{sec:Theory} we start with introducing the general linear transport equation and the transport coefficients, from which all phenomena described in the continuation of this article can be derived. Section\,\ref{sec:Figs} introduces the definitions used in the figures corresponding to each described transport phenomenon. Thereafter, the individual transport phenomena are discussed, subdivided in three categories: Collinear transport phenomena (Section\,\ref{sec:Col}), transverse transport phenomena (Section\,\ref{sec:TransNew}), and planar transport phenomena (Section\,\ref{sec:Planar}). Each section starts with a brief introduction of the phenomena belonging to the described category, followed by an overview of all effects in the form of a table. We also allude to boundary conditions, which are typically chosen in the context of experiments. Afterwards the transport phenomena are individually detailed in separate subsections. 

\section{General transport equation}\label{sec:Theory}
In this section, we rationalize the concept of a generalized transport equation which eventually can be used to  describe and categorize all charge, heat and spin related transport phenomena (in electrical conductors), including the (anomalous) thermoelectric, thermomagnetic, and galvanomagnetic effects as well as their spin-related counterparts. Here, it is noteworthy that charge and heat current densities $\Jvec$ and $\Jvecheat$ characterize the transport direction and amplitude of the scalar transport quantities charge $q$ and heat $k_\mathrm{B}T$. In contrast, the spin-current density $\Jvecspin$ describes the transport of spin angular momentum and thus represents a tensor. 
Specifically, the spin angular momentum is described by the vector of the Pauli spin matrices $\boldsymbol{\sigma} =(\sigma_x , \sigma_y , \sigma_z)$. Therefore, we have to express $\mathbf{J}_\mathrm{s}$ via the dyadic (or tensor) product $\boldsymbol{\sigma} \otimes \mathbf{v}$ of the vector of the Pauli spin matrices and the drift velocity $\mathbf{v}$ of the entity carrying the spin.  For charge carriers with density $n$, charge $q$, drift velocity $\mathbf{v}$, and spin $\mathbf{\hat{s}}=\tfrac{\hbar}{2}\boldsymbol{\sigma}$ we obtain \citep{BauerCaloritronicsReview}
  \begin{equation}
  \mathbf{J}_\mathrm{s} = n \, \frac{\hbar}{2} \, \langle \boldsymbol{\sigma} \otimes \mathbf{v} \rangle \: ,
  \label{eq:SS_01}
  \end{equation}
where $\langle \ldots \rangle$ is the thermodynamic expectation value. With the electrical current density $\Jvec=n q \mathbf{v}$ this can be rewritten as
  \begin{equation}
  \mathbf{J}_\mathrm{s} = \frac{\hbar}{2q} \, \langle \boldsymbol{\sigma} \otimes \Jvec \rangle \: .
  \label{eq:SS_02}
  \end{equation}
We see that the ratio of the magnitudes of $\Jvec$ and $\mathbf{J}_\mathrm{s}$ is given by the ratio of the transported charge $q$ and angular momentum $\hbar/2$.

In order to simplify the treatment of spin currents, usually a two-spin-fluid model is considered.\citep{OHandley, BauerCaloritronicsReview, BauerCaloritronics, Maekawa:2017gn} In this approach one makes use of the fact that the spin of the charge carriers (e.g. electrons) can only assume two values -- spin-up ($\uparrow$) and spin-down ($\downarrow$) -- relative to a quantization axis $\sigmapol$. If the two spin species form two independent transport channels (if spin is a good quantum number), we obtain the illustrative description of the spin current by
  \begin{equation}
  \mathbf{J}_\mathrm{s} =  \frac{\hbar}{2q} \, \left( \Jvec^\uparrow - \Jvec^\downarrow  \right)  \: .
  \label{eq:SS_03}
  \end{equation}
In many publications, $\mathbf{J}_\mathrm{s}$ (with units angular momentum per area and time, J/m$^2$) is divided by the spin $\hbar/2$ and multiplied by the charge $q$ transported by each charge carrier to obtain a spin current density $\Jvecspintilde$, sharing the same units with the charge current density $\Jvec$ (charge per area and time, A/m$^2$):\citep{Gross,BauerCaloritronics}
  \begin{equation}
  \Jvecspintilde = \mathbf{J}_\mathrm{s} \frac{2q}{\hbar}  = \left( \Jvec^\uparrow - \Jvec^\downarrow  \right) = \sigmac \left( -\boldsymbol{\nabla}\mu_\mathrm{s}/2q \right) 
  \: .
  \label{eq:SS_04a}
  \end{equation}
This definition  is appealing from the perspective of a charge transport picture, where a spin current is directly connected to the charge current and originates from an imbalance or counterflow of $\Jvec^\uparrow $ and $\Jvec^\downarrow$.  We also note that this allows one to redefine the gradient $\boldsymbol{\nabla}\pots$ in terms of a  gradient of the spin chemical potential $\muspin$ via  $\tilde{\mathbf{J}}_\mathrm{s}= \mathbf{J}_\mathrm{s}\cdot (2q/\hbar) = - \sigma_\mathrm{s} \boldsymbol{\nabla}\mu_\mathrm{s}\cdot (2q/\hbar^2)$ and the fact that the ratio $\sigma_\mathrm{s}/\sigmac$ of the spin and charge conductivity is given by $(\hbar/2q)^2$. We see that in Eq.\,(\ref{eq:SS_04a}) the gradient $\mathbf{\tilde{E}}_\mathrm{s}=  -\boldsymbol{\nabla}\mu_\mathrm{s}/2q$ then has the units of an electric field. 

The coupled transport of charge, heat, and spin can  be expressed in the linear response regime in  vector form as \cite{BauerCaloritronics, NernstWMI,BauerCaloritronicsReview, uchida-review}
\begin{equation}\label{eq:GeneralGeneralTransportEquation}
\left(\begin{array}{c}
\Jvec \\
\Jvecheat \\ 
\Jvecspin
\end{array}\right)=\mathbf{L}\left(\begin{array}{c}
-\boldsymbol{\nabla} \potc \\
-\boldsymbol{\nabla} \poth \\
-\boldsymbol{\nabla} \pots
\end{array}\right)+
\mathbf{L^{\perp}}\left(\begin{array}{c}
-\boldsymbol{\nabla} \potc \times \Bhat \\
-\boldsymbol{\nabla} \poth \times \Bhat\\
-\boldsymbol{\nabla} \pots \times \Bhat
\end{array}\right).
\end{equation}

In linear approximation, the charge, heat and spin current densities (responses) depend linearly on the acting forces expressed by the gradients of the corresponding charge $\potc$, thermal $\poth$, and spin $\pots$ potentials. The proportionality constants are called the general transport coefficients. The first term of the right hand side of Eq.\,(\ref{eq:GeneralGeneralTransportEquation}) summarizes the collinear transport phenomena, where the current densities are collinear with the driving potential gradients. The second term describes the transverse transport phenomena. Here, the current densities are perpendicular to the driving gradients and a generalized magnetic field $\Bvec$. Here, $\Bhat$ is the unit vector parallel to $\Bvec$, which e.g. can be caused by an applied magnetic field $\mu_0 \Hvec$, a finite magnetization of the material $\mu_0 \Mvec$, or an effective field proportional to the spin quantization direction $\sigmapol$, which describes, for example, the spin-selective deflection due to spin-orbit interaction in non-magnetic metals. The quantities $\mathbf{L}$ and $\mathbf{L^\perp}$ are $3 \times 3$ matrices that represent the linear transport coefficients. They describe the transport of the individual entities as well as their interplay. Each element of $\mathbf{L}$ and $\mathbf{L^\perp}$ itself is a  second rank tensor in the most general case.\footnote{The description of the transport coefficients by second rank tensors is required by the anisotropy of crystalline solids, with the number of independent tensorial components determined by crystal symmetry.}  For simplicity, we will treat these coefficients as scalars in the following. Due to Onsager's reciprocity relation\citep{Onsager,Onsager2} the coefficients of matrix $\mathbf{L}$ are symmetric, $L_{ij}=L_{ji}$, while the entries of $\mathbf{L^\perp}$ are antisymmetric $L_{ij}^\perp(\Bvec)=L_{ji}^\perp(-\Bvec)$,  i.e. both $\mathbf{L}$ and $\mathbf{L^\perp}$  can each be described by six independent transport coefficients. 

All transport phenomena discussed in Sec.\,\ref{sec:Col} and \ref{sec:TransNew} of  this article are contained in  Eq.\,(\ref{eq:GeneralGeneralTransportEquation}). Within linear response, the current densities (response) are proportional to the generalized forces, originating from gradients of the corresponding potentials. The corresponding proportionality constants are given by the transport coefficients, each associated with a particular transport phenomenon. Unfortunately, historically some transport phenomena have been introduced in a different manner not following the scheme of Eq.\,(\ref{eq:GeneralGeneralTransportEquation}), resulting in a  sometimes confusing nomenclature. Typically, the origin of those definitions can be found in the way the original experiments have been performed. For example,  it is often easier to drive currents (e.g. charge or heat currents) and detect the resulting potential gradients (e.g. electric field and temperature gradient).  Therefore, many transport coefficients such as electrical and thermal conductivity as well as the Seebeck and Nernst coefficients have not been introduced  as linear transport coefficients in the spirit of Eq.(\ref{eq:GeneralGeneralTransportEquation}). However, the conversion of the $\mathbf{L}$-matrix into the matrix containing the historically introduced transport coefficients typically is straightforward.\citep{Gross}   In this article, the transport phenomena will be discussed from the experimentalist's perspective, using the historically introduced transport coefficients. Therefore, some drive and response terms may look different from the theoretical description conveyed by Eq.\,(\ref{eq:GeneralGeneralTransportEquation}), but their origin can in all cases be related back to this general transport equation.

\section{Scheme for the graphical representation}\label{sec:Figs}
To assist the reader in understanding and comparing the various charge, heat, and spin transport phenomena, we illustrate each effect with a sketch, showing its simplistic response based on electrons as carriers, which transport the quantities charge, heat, and spin. The sign of the resulting response is determined by the corresponding material dependent transport coefficient of each effect. For each figure, the assumed sign of the corresponding coefficient is indicated in the figure caption. 

All figures share the same color scheme: Blue vectors represent the generalized driving force originating from the gradient in the corresponding potential, black vectors are external control parameters like the applied magnetic field $\Hvec$, the magnetization of the material $\Mvec$, or the spin polarization $\sigmapol$, while the red vectors denote the response direction. 

To be specific regarding the directions of the depicted vectors, we remind the reader that, by definition, the electric field vector $\Efieldc$ points from positive charge ($+$) towards negative charge ($-$), and the temperature gradient $\mathbf{\nabla T}$ is directed from cold to hot (note that all figures show $-\mathbf{\nabla T}$). The gradient of the spin accumulation $\nabla \muspin$ points from the region with smaller spin-up concentration to the region with larger spin-up concentration. By definition, $\Jvec$ is parallel to $\Efieldc=-\boldsymbol{\nabla} \potc = - \boldsymbol{\nabla} \mucharge/q$. Note, however, that $\Jvec=nq\mathbf{v}$ is defined as the technical current density, i.e., $\Jvec$ is parallel to the drift velocity $\mathbf{v}$ of charge carriers with positive charge $q$. This means that the drift velocity of electrons ($q=-e$) is anti-parallel to $\Jvec$. For the spin current density $\Jvecspin$, the direction marks the flow direction of the angular momentum. If we are dealing with spin $\pm 1/2$ carriers, we can define a "technical spin current direction" as the flow direction of spin-up ($s=+\hbar/2$) carriers. This is analogous to charge transport of carriers with $q=\pm e$ and the definition of the "technical charge current direction" as the flow direction of $q=+e$ carriers. By this definition the drawn $\Jvecspin$ is opposite to $\Jvecspintilde$, as we defined $\Jvecspintilde=\frac{2\charge}{\hbar}\Jvecspin$ and $q=-e$ for electrons carrying the spin.  

In the figures the spin polarization  direction (quantization axis) $\sigmapol$ is taken along
the $z$-axis (spin-up points in the $+z$-direction). 
Furthermore, the gradient of the potential/the driving force (blue vector) is always aligned with the $x$-axis. For the definition of the coordinate system, we refer to Fig.\,\ref{fig:Ohm}. 

In addition, the applied temperature gradients are visualized by a red (=hot) to white (=cold) color gradient. Grey-colored boxes indicate that the effect is characteristic for a magnetically ordered material.

\section{Collinear transport phenomena}\label{sec:Col}

In this section, we focus on various collinear transport phenomena. Here, the resulting current densities are a consequence of the generalized driving forces or applied potential gradients and they are oriented along their directions. These phenomena are contained in Eq.~(\ref{eq:GeneralGeneralTransportEquation}) for $\Bhat=0$, and by neglecting transverse deflection of charge carriers due to spin-orbit interaction. In detail, the collinear transport phenomena are described by\footnote{Note that in general each $L_{ij}$ is a second rank tensor by itself reflecting e.g. anisotropy of crystalline solids. To keep the discussion simple, we assume isotropic materials for which the second rank tensor can be replaced by a simple number. }
\begin{equation}\label{eq:GeneralCollinearTransportEquation}
\left(\begin{array}{c}
\Jvec \\
\Jvecheat \\ 
\Jvecspin
\end{array}\right)=\left(\begin{array}{ccc}
L_{11} & L_{12} & L_{13} \\
L_{12} & L_{22} & L_{23} \\
L_{13} & L_{23} & L_{33}
\end{array}\right)\left(\begin{array}{c}
-\boldsymbol{\nabla} \potc \\
-\boldsymbol{\nabla} \poth \\
-\boldsymbol{\nabla} \pots
\end{array}\right).
\end{equation}
Some of the $L_{ij}$ are straightforward to interpret: For example the coefficients $L_{11}=\sigmac$, $L_{22}=\sigma_\mathrm{h}$, and $L_{33}=\sigmaspin$ characterize Ohm's law (Sec.\,\ref{subsec:ohmslaw}), thermal transport (Sec.\,\ref{subsec:themalconductivity}) and spin transport (Sec.\,\ref{subsec:spinconductivity}), while the off-diagonal elements describe the Seebeck and Peltier effect (Sec.\,\ref{sec:Seebeck}\&\ref{sec:peltier}) as well as their spin-related counterparts (Sec.\,\ref{subsec:SpindependentSeebeckEffect}\&\ref{subsec:spinpeltier}). Here, $\sigma_c$ is the electrical conductivity, $\sigma_h=\kappa T$ the electronic heat conductivity $\kappa$ multiplied by $T$ and $\sigma_s= \sigma_c (\hbar/2q)^2$ the spin conductivity.\footnote{Note that $\kappa$ accounts exclusively for the heat conductivity of the charge carriers and does not include e.g. phononic or magnonic contributions.} For the phenomena described by the off-diagonal elements, the parameters are mostly defined in a historical context, as we will discuss in detail in the corresponding subsections. Using these common definitions, the transport coefficients $L_{ij}$ take the form \cite{BauerCaloritronics,BauerCaloritronicsReview}
\begin{widetext}
\begin{equation}\label{eq:GeneralCollinearTransportEquation2}
\left(\begin{array}{c}
\Jvec \\
\Jvecheat \\ 
{\Jvecspintilde}
\end{array}\right)=\sigmac
\left(\begin{array}{ccc}
1 & S T & P\\
S T & \kappa T /\sigmac & P' S T\\ 
P& P'S T & 1
\end{array}\right)
\left(\begin{array}{c}
-\boldsymbol{\nabla} \mucharge / \charge \\
-\boldsymbol{\nabla} T /T \\ 
-\boldsymbol{\nabla} \muspin /(2\charge)
\end{array}\right),
\end{equation}
\end{widetext}
where $S$ is the Seebeck coefficient, $P$ is the spin polarization of the conductivity, and $P^\prime$ is its energy derivative. 
By comparing Eqs.\,(\ref{eq:GeneralCollinearTransportEquation}) and (\ref{eq:GeneralCollinearTransportEquation2}), we can relate  the historically introduced transport coefficients in (\ref{eq:GeneralCollinearTransportEquation2}) with the generalized linear transport coefficients $L_{ij}$ in (\ref{eq:GeneralCollinearTransportEquation}). Coming back to the two-spin-channel model, we can express the electrical conductivity as $\sigmac=\sigmac^\uparrow+\sigmac^\downarrow$, the electronic heat conductivity $\kappa=\kappa^\uparrow+\kappa^\downarrow$, and the  Seebeck coefficient as  $S=(\sigmac^\uparrow S^\uparrow +\sigmac^\downarrow S^\downarrow )/(\sigmac^\uparrow + \sigmac^\downarrow)$ [cf. Eq.\,(\ref{eq:Seebeck-twofluid3})], the electrochemical potential as $\mucharge=(\mucharge^\uparrow + \mucharge^\downarrow)/2$ and the spin chemical potential as $\muspin=\mucharge^\uparrow - \mucharge^\downarrow$ (corresponding to the spin accumulation), as well as the charge, spin and heat current densities as $\Jvec= \Jvec^\uparrow + \Jvec^\downarrow$, $\Jvecspintilde=\Jvec^\uparrow - \Jvec^\downarrow$ and $\mathbf{J}_\mathrm{h}=\mathbf{J}_\mathrm{h}^\uparrow + \mathbf{J}_\mathrm{h}^\downarrow$. In the Sommerfeld approximation, the spin-dependent Seebeck coefficients are given by $S^{\uparrow,\downarrow}=qL_0T (\sigmac^{\uparrow,\downarrow})^\prime /\sigmac^{\uparrow,\downarrow}(\epsilon_\mathrm{F})$ (Mott's relation), where $(\sigmac^{\uparrow,\downarrow})^\prime$ is the energy derivative of the spin-dependent conductivity at the Fermi energy $\epsilon_\mathrm{F}$ and the Lorenz number $L_0=(\pi^2/3) (k_\mathrm{B}/e)^2$ .
\cite{BauerCaloritronics,Gross} 
Moreover, we introduce the spin  polarization of the conductivity and its energy derivative. (cf. Secs.\,\ref{subsec:spinconductivity} and \ref{subsec:SpindependentSeebeckEffect})
\begin{equation}
    P = \frac{\sigmac^\uparrow - \sigmac^\downarrow}{\sigmac^\uparrow + \sigmac^\downarrow} \quad \text{and}\quad
    P^\prime = \frac{(\sigmac^\uparrow)^\prime - (\sigmac^\downarrow)^\prime}{(\sigmac^\uparrow)^\prime + (\sigmac^\downarrow)^\prime}. \label{eq:SS_04}
\end{equation}
The Seebeck coefficient $S$ can be replaced by the Peltier coefficient $\Pi$ via the Thomson relation $\Pi = TS$.

Coming back to the collinear transport phenomena mediated by charge carriers, we summarize in Table\,\ref{tab:Lin} the collinear transport equations\,(\ref{eq:GeneralCollinearTransportEquation}) and (\ref{eq:GeneralCollinearTransportEquation2}), also indicating the names of the effects typically used in literature. The transport coefficients can be derived from Boltzmann transport theory or more complex theories \cite{Gross,BookGrosso}. We also note that Eq.\, (\ref{eq:GeneralCollinearTransportEquation}) holds as long as the transport is diffusive.

For completeness, the Thomson effect as well as the spin(-dependent) Thomson effect will also be discussed here. Together, the Thomson, Peltier and Seebeck effect define the so-called {\it\bfseries thermoelectric effects} originating from thermo-electric coupling. Their spin-related counterparts are named {\it\bfseries spin-electric} and {\it\bfseries spin-thermal effects}, originating from coupled spin/charge and spin/heat transport, respectively. 

To disentangle the different  transport phenomena  in the following subsections, we always assume that only a single  characteristic potential gradient is applied. For example, to discuss the Seebeck effect, defined as the longitudinal electric response  caused by an applied temperature gradient, we assume vanishing electrical and spin chemical potential gradients ($\boldsymbol{\nabla} \potc =\boldsymbol{\nabla} \pots=0$) and focus exclusively on the longitudinal electrical response. Actually, the assumption $\boldsymbol{\nabla}\potc=\boldsymbol{\nabla}\pots=0$ represents a specific boundary condition describing a particular experimental situation.  Using this step by step approach, we discuss each  element of the $3\times 3$ matrix in Eqs.\,(\ref{eq:GeneralCollinearTransportEquation}) and (\ref{eq:GeneralCollinearTransportEquation2}) as well as the corresponding effect (see Table \ref{tab:Lin}) in an isolated fashion. However, the reader should keep in mind that, a single potential gradient can cause multiple current densities at the same time, since the transported entities may carry different properties simultaneously (e.g. charge, heat and spin by electrons). In this case one has to analyze in detail what kind of boundary conditions are realized in a specific  experimental situation in order to correctly describe the transport by Eqs.\,(\ref{eq:GeneralCollinearTransportEquation}) and (\ref{eq:GeneralCollinearTransportEquation2}).

\begin{table*}
\textsf{
    \begin{tabularx}{\textwidth}{ r?Y|Y|SY } 
    	\diagbox[width=\firstcolumn]{\color{response}\textbf{Response}}{\color{drive}\textbf{Drive} \hspace{2pt}} &  \textcolor{drive}{\textbf{Electrical}} & \textcolor{drive}{\textbf{Thermal}} & \textcolor{drive}{\textbf{Spin}} \\
   		\midrule[1.5pt]
   		\multirow{3}{*}[\centerfirstt]{\textcolor{response}{\textbf{Electrical}} \hspace{10pt}}
        & ${\color{response}\Jvec} \propto {\color{drive}\mingradmuc}$ & ${\color{response}\Jvec} \propto {\color{drive}\mingradT}$ & ${\color{response}\Jvec} \propto {\color{drive}\mingradmus}$ \\
        & Electrical conductivity & Seebeck effect & Spin-polarized charge transport \\
        &  &  & (Inverse spin-electric effect) \\
        \hline
    	\multirow{2}{*}[\centerfirstt]{\textcolor{response}{\textbf{Thermal}} \hspace{10pt}} 
        & ${\color{response}\Jvecheat} \propto {\color{drive}\mingradmuc}$ & ${\color{response}\Jvecheat} \propto {\color{drive}\mingradT}$ & ${\color{response}\Jvecheat} \propto {\color{drive}\mingradmus}$ \\
        & Peltier effect & Thermal conductivity & Spin(-dependent) Peltier effect \\ 
        \hline
    	\multirow{3}{*}[\centerfirstt]{\textcolor{response}{\textbf{Spin}} \hspace{10pt}}
        & ${\color{response}\Jvecspin} \propto {\color{drive}\mingradmuc}$ & ${\color{response}\Jvecspin} \propto {\color{drive}\mingradT}$ & ${\color{response}\Jvecspin} \propto {\color{drive}\mingradmus}$ \\
        & Charge based spin transport & Spin(-dependent) Seebeck effect & Spin conductivity  \\
        & (Spin-electric effect) &  &  \\
    \end{tabularx} 
    }    
    \caption{\label{tab:Lin}
	Summary of the so-called collinear (or longitudinal) transport phenomena, as discussed in Sec.\,\ref{sec:Col}. Here, the direction of the response is oriented collinear with the gradient of the respective chemical potential ($\mingradmuc$, $\mingradT$, and $\mingradmus$). The diagonal elements in the table are the direct transport effects, while the off-diagonal elements result from the coupling of charge, heat and spin transport. 
	}
\end{table*}


\subsection{Charge, heat, and spin transport}
We first discuss the diagonal elements of the transport coefficient matrix $\mathbf{L}$ (see Eqs.\,(\ref{eq:GeneralCollinearTransportEquation}) and (\ref{eq:GeneralCollinearTransportEquation2})), which represent the conductivities of the charge, heat and spin transport. 

\subsubsection{Electrical conductivity (Ohm's law)}\label{subsec:ohmslaw}
\textit{The electrical conductivity relates the charge current density to the gradient of the electrochemical potential.}\footnote{We assume the boundary conditions $\poth=\pots=0$.}

\begin{figure}[h]
	\includegraphics[width=\Axes]{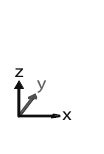}
    \includegraphics[width=\textfigures]{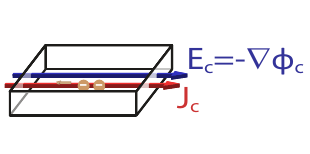} 
    \vspace{\captionhalfclose}
    \caption{\label{fig:Ohm}  Schematic of  charge transport caused by an applied electric field or gradient of an electrochemical potential (Ohm's law). The coordinate system defined on the left is used for all depictions in the entire article.}
\end{figure}

The concept of charge transport, as sketched in Fig.\,\ref{fig:Ohm}, has been studied by many scientists in the past several centuries \citep{HistoryElectricity}. In particular, Georg Ohm derived the relation between current density and electric field  in 1827 \citep{BookOhm}, based on the work of Fourier, who described the thermal analogue in 1822 \citep{Fourier,Fourier2}. Typically, Ohm's law is written as
\begin{equation}\label{eq:ohmslaw}
    \Jvec=-\sigmac \boldsymbol{\nabla}\potc =-\sigmac \boldsymbol{\nabla} \mucharge / \charge = \sigmac \Efieldc.
\end{equation}
Here, the charge current density $\Jvec$ is the linear response to an applied electric field $\Efieldc$, a relation which holds down to a length scale given by the mean free path \cite{Weber2012}. 

The  conductivity $\sigmac$ (assumed here as a scalar quantity for simplicity) quantifies the response of the material and can be understood e.g. using the Drude ansatz or Boltzmann transport theory \cite{Gross,BookGrosso}.

\vspace{\figuretop} 

\subsubsection{Thermal conductivity (Fourier's law)}\label{subsec:themalconductivity}
\textit{The thermal conductivity relates the heat current density to the  thermal gradient and thus is closely analogous to the electrical conductivity.}\footnote{We assume the boundary conditions $\potc=\pots=0$.}

\begin{figure}[h]
	\includegraphics[width=\textfigures]{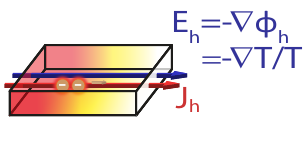} 
    \vspace{\captionhalfclose}
    \caption{\label{fig:ThermCond} Schematic for the thermal transport based on electronic charge carriers (Fourier's law).}
\end{figure}

In fact, the concept of thermal conductivity was proposed by Jean-Baptiste Joseph Fourier before Ohm's law was coined \citep{Fourier,Fourier2}.\footnote{Fourier described his law already in 1807, but his work was accepted for publication only in 1822.} It reads 
\begin{equation}
    \Jvecheat = -\sigma_\mathrm{h} \boldsymbol{\nabla}\poth= -\kappa \nabla T.
\end{equation}
The minus sign indicates that heat flows from the hot to the cold part, while the thermal gradient $\nabla T$ points from cold to hot (see Fig.\,\ref{fig:ThermCond}). 

Besides by charge carriers, heat can also  be transported in solids by a variety of particles, quasi-particles or elementary excitations. In this article, we focus on heat transport mediated by charge carriers \cite{Gross,BookGrosso,Ashcroft:1976ud,BookZiman,WiedFranz}.

\subsubsection{Spin conductivity}\label{subsec:spinconductivity}
\textit{The spin conductivity relates the spin current density to the gradient of the spin accumulation (or the spin chemical potential).}\footnote{We assume the boundary conditions $\potc=\poth=0$}

The spin conductivity $\sigmaspin$ is defined in analogy to the thermal and charge conductivity as the proportionality constant (in linear response) between the gradient of the spin potential and the spin current density via \citep{BauerCaloritronics,Chen:2013kf} \begin{equation}
\label{eq:ohmslawspin0}
    \Jvecspin=-\sigmaspin\boldsymbol{\nabla} \pots.
\end{equation}
As discussed in Sec.\,\ref{sec:Theory},  it can be convenient to rewrite the spin current density as 
\begin{equation}
\label{eq:ohmslawspin}
    \Jvecspintilde=-\sigmac\boldsymbol{\nabla} \muspin /(2\charge)
\end{equation}
in cases where the spin current is mediated by charge carriers. Note that $\Jvecspintilde$ has the units of A/m$^2$, the units of a charge current density. Moreover, the gradient of the spin chemical potential, $\boldsymbol{\nabla} \muspin /(2\charge)=\boldsymbol{\nabla}(\mucharge^\uparrow-\mucharge^\downarrow)/(2\charge)$  can be viewed as a spin field  sharing the units of the electric field. Hence, to drive a spin current, we have to establish a gradient in the spin chemical potential or an imbalance in the charge chemical potentials for the spin-up and the spin-down channel.\cite{BauerCaloritronics,Chen:2013kf}

\begin{figure}[h]
	\includegraphics[width=\textfigures]{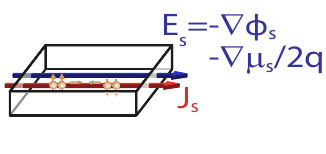} 
    \vspace{\captionhalfclose}
    \caption{\label{fig:SpinCond} Schematic for pure spin current transport based on electronic charge transport. A gradient in the spin chemical potential generates a spin current density. For the case of a pure spin current with net-zero associated charge current, the situation can be intuitively sketched as the compensated counterflow of charge carriers with the same density but opposite spin orientation.}
\end{figure}

In this context, a pure spin current describes the situation of having a finite $\Jvecspin$, while $\Jvec=0$. This requires $P=0$ (see Eq.\,\ref{eq:GeneralCollinearTransportEquation2})), that is, materials with a vanishing spin polarization at the Fermi energy such as non-magnetic metals. The intuitive picture for having a net-spin current in the presence of vanishing charge current is the  compensated counterflow of charge carriers with opposing spin orientation
(see Fig.\,\ref{fig:SpinCond}). However, this spin current cannot be driven directly by an electric field, as the charge $q$ is identical for spin-up and spin-down carriers. This is overcome by more elaborate schemes such as spin pumping employing magnetic heterostructures 
\cite{Urban:2001hr,Heinrich:2003en,MosendzPRL,CzeschkaPRL, Weiler:2013ko,Tserkovnyak:2002ju}, which are used to generate a gradient in the spin chemical potential and study  $\sigma_\mathrm{s}$ \citep{Zutic:2004fo, Takahashi:2008ir,Maekawa:2017gn}.

\vspace{\figuretop} 
\begin{figure}[h]
	\includegraphics[width=\textfigures]{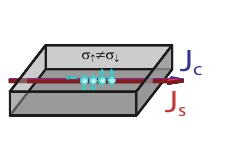} 
    \vspace{\captionhalfclose}
    \caption{\label{fig:SpinPol} Schematic for spin-polarized charge transport occurring in materials with different densities of states at the Fermi level for the two spin species (as typically present in magnetically ordered materials). In this case, an electric field drives a charge current, which is associated with a finite spin current (spin-polarized charge current). Note that at the same time an applied spin field drives a spin current, which is associated with a finite charge current (charge-polarized spin current). }
\end{figure}


\subsection{Spin-polarized transport}
\label{sec:Spin-Pol}
Up to now, spin transport was discussed in the context of a pure spin current. However, spin and charge transport are intimately linked in metallic systems, as both entities are transported by the same carriers. Hence, spin and charge currents always coexist. It is evident that a finite spin current density $\Jvecspin$ is always associated with a charge current density driven by an electric field,  if there is an imbalance of the charge current transport in the spin-up and spin-down channels. This imbalance can be expressed by different conductivities $\sigmac^\uparrow$ and $\sigmac^\downarrow$ (see Fig.\,\ref{fig:SpinPol}), which may be caused e.g. by different densities of state at the Fermi level for the two spin species. At the same time, a finite charge current density is always associated with a spin current density driven by a spin field. That is, in this situation
$\Jvec$ and $\Jvecspin$ are both finite. This phenomenon is caused by spin-electric coupling and is called spin-polarized charge transport or, equivalently, charge-polarized spin transport. It is represented by $L_{13}$ (=$L_{31}$) in Eq.\,(\ref{eq:GeneralCollinearTransportEquation}), as  discussed in this section. 

\textit{Spin-polarized charge transport (charge-polarized spin transport) describes the phenomenon that a charge (spin) current density is accompanied by a spin (charge) current density.}

Historically, Mott discussed this transport phenomenon within a two-fluid model based on two co-propagating transport channels with opposing spin orientation \cite{Mott1936,OHandley}. In this two-current model, it is further assumed that  the fluids carrying the two spin species do not intermix, allowing to describe the situation by two  conductivities, $\sigmac^\uparrow$ and $\sigmac^\downarrow$, for the spin-up and spin-down channel. In this simple model, the total charge current density is given by the sum of the individual transport channels 
\begin{eqnarray}\label{eq:spinpol5}
\Jvec &=& {\Jvec}^\uparrow + {\Jvec}^\downarrow \nonumber\\
\end{eqnarray}
Correspondingly, the spin current density is given by
\begin{eqnarray}\label{eq:spinpol6}
\Jvecspintilde &=& {\Jvec}^\uparrow - {\Jvec}^\downarrow \nonumber\\
&=& \left( \sigmac^\uparrow - \sigmac^\downarrow \right) (- \boldsymbol{\nabla} \mucharge / \charge) \nonumber\\ &=& - \sigmac P \boldsymbol{\nabla} \mucharge / \charge,
\end{eqnarray}
using Ohm's law (Eq.\,(\ref{eq:ohmslaw})) and the definition of $P$ given by Eq.\,(\ref{eq:SS_04}). Therefore, we expect that in systems with finite $P$, charge current transport is associated with a finite, net spin current transport and vice versa.  Typically, systems with   $P\neq 0$  are related to  a finite spin polarization at the Fermi energy (e.g. as in ferromagnetic metals). However, also different scattering times for the two spin species may result in $P\neq 0$. Note that, in systems with $P=0$ there is net spin (charge) current associated with a charge (spin) current flow as  expected for non-magnetic materials.

\subsection{Seebeck, Peltier, and Thomson effect}\label{subsec:Spin_Seebeck_Peltier}
We next turn to the coupled transport of charge and heat, originating from the thermo-electric coupling. This phenomenon is quantified by the off-diagonal elements $L_{12}$ ($=L_{21}$) of the collinear transport matrix (see Eq.\,({\ref{eq:GeneralCollinearTransportEquation}})). The corresponding effects are known as the Seebeck and the Peltier effect.
In Sec.\,\ref{subsec:Spin_Seebeck_Peltier2}, we extend this discussion to spin-thermal coupling leading to the coupled transport of spin and heat. This phenomenon is represented by the  off-diagonal elements $L_{23}$. Due to spin-thermal coupling, a spin (heat) current driven by a gradient of the spin-chemical potential (temperature gradient) is associated with a heat current (spin current).  

\subsubsection{Seebeck effect}\label{sec:Seebeck}
\textit{The Seebeck effect describes the generation of a charge current density (an electrical field for open circuit conditions) parallel to an applied temperature gradient.} 

In 1821, Thomas Johann Seebeck discovered the Seebeck effect via the deflection of a compass needle when the latter was brought in proximity to a heated junction made of two different electrical conductors.\citep{Seebeck,Seebeck2} 

\begin{figure}[h]
	\includegraphics[width=\textfigures]{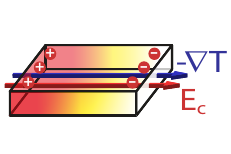} 
    \vspace{\captionhalfclose}
    \caption{\label{fig:Seebeck} Schematic of the Seebeck effect in open circuit conditions, as discussed in Eq.~(\ref{eq:Seebeck-std}), assuming a negative Seebeck coefficient. Note that in steady state, the thermally induced charge motion is exactly balanced by the electric field $\Efieldc$ (no net charge flow).}
\end{figure}

In its simplest form (boundary conditions $\Jvecheat=\Jvecspin=0$ and $\boldsymbol{\nabla}\potc = \boldsymbol{\nabla}\pots =0$), the Seebeck effect describes the presence of a charge current as the linear response to an applied gradient $\boldsymbol{\nabla}\poth=\boldsymbol{\nabla} T/T$ 
\begin{equation} \label{eq:Seebeck-current}
\Jvec= - L_{12} \boldsymbol{\nabla}\poth =-\sigmac S \boldsymbol{\nabla} T.
\end{equation}
Assuming a free electron gas model, the transport direction of the charge carriers can be rationalized as follows: The Fermi-Dirac distribution of the charge carriers on the hot side has a larger width compared to the cold side, causing a charge particle flow from the hot to the cold side. Thus, for electrons as charge carriers the Seebeck coefficient $S$ should always be negative (see also Fig.\,\ref{fig:Seebeck}). The fact that a positive $S$ is observed for some metals clearly demonstrates that the simple free electron gas model is too simple for their description and that details of the band structure have to be taken into consideration.\citep{BehniaBook, Gross, FrankThesis} 

At elevated temperatures, where the Sommerfeld approximation of a linear relation between electrical conductivity and temperature holds, the Seebeck coefficient $S$ is described by the Mott relation, which connects $S$ to $\kappa$ and $\sigmac$.\cite{Mott,BauerCaloritronics,Gross,FrankThesis} In the Sommerfeld approximation, the spin-dependent Seebeck coefficients are given by $S^{\uparrow,\downarrow}=qL_0T (\sigmac^{\uparrow,\downarrow})^\prime /\sigmac^{\uparrow,\downarrow}(\epsilon_\mathrm{F})$ (Mott's relation), where $(\sigmac^{\uparrow,\downarrow})^\prime$ is the energy derivative of the spin-dependent conductivity at the Fermi energy $\epsilon_\mathrm{F}$ and $L_0=(\pi^2/3) (k_\mathrm{B}/e)^2$ the Lorenz number.
\cite{BauerCaloritronics,Gross}

Within the two-spin current model and accounting for finite electrical fields in the system (cf. Eq.\,(\ref{eq:GeneralCollinearTransportEquation2})), we can write the charge current density $\Jvec$ as
\begin{equation}
\Jvec^{\uparrow,\downarrow} = \sigmac^{\uparrow,\downarrow} \Efieldc - \sigmac^{\uparrow,\downarrow} S^{\uparrow,\downarrow}  \boldsymbol{\nabla}T \; .
\label{eq:Seebeck-twofluid1}
\end{equation}
Here, we assign  its own ``spin-dependent'' Seebeck coefficient to each spin species.
The total charge current density is then given by 
\begin{eqnarray}
\Jvec &=& \Jvec^\uparrow + \Jvec^\downarrow
\nonumber\\
&=& \left( \sigmac^{\uparrow} + \sigmac^{\downarrow} \right) \Efieldc  - \left( \sigmac^{\uparrow} S^{\uparrow} + \sigmac^{\downarrow} S^{\downarrow} \right) \; \boldsymbol{\nabla}T  \;.
\label{eq:Seebeck-twofluid2}
\end{eqnarray}
For open boundary condition ($\Jvec=0$), which corresponds to the original design of the experiment by Seebeck,\citep{Seebeck,Seebeck2} the charge accumulation at the open ends of the conductor leads to an electric field causing a drift current which just compensates the charge current due to the temperature gradient in a steady state situation. This results in 
\begin{equation}
	\Efieldc = S \; \boldsymbol{\nabla}T \; 
\label{eq:Seebeck-std}
\end{equation}
with
\begin{equation}
S =  \frac{ \sigmac^{\uparrow} S^{\uparrow} + \sigmac^{\downarrow} S^{\downarrow} }{ \sigmac^{\uparrow} + \sigmac^{\downarrow} }  \;.
\label{eq:Seebeck-twofluid3}
\end{equation}

\begin{figure}[h]
		\includegraphics[width=0.8\columnwidth]{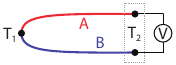}
		\vspace{\captionhalfclose}
        \caption{\label{fig:Seebeck-Schematic}
		Schematic drawing of the circuit used to observe the Seebeck effect and determine the Seebeck coefficients of Material A and B with respect to each other. Material A and B have different Seebeck coefficients $S_\textrm{A}$ and $S_\textrm{B}$. This circuit is the basis of standard thermocouples.}
\end{figure}

It is experimentally challenging to directly measure the Seebeck electrical field $\Ec$ in a homogeneous material.\footnote{From an experimental standpoint the observable of Eq.~(\ref{eq:Seebeck-std}) is hard to be accessed directly. This owes to the fact that the voltage or potential probes attached to the two ports of a voltmeter need to be on the same temperature in order to avoid the occurrence of the Seebeck voltage inside this voltmeter. This requirement is in conflict with a temperature difference between the two ends of the sample. This problem can be overcome by using two different materials, as is schematically shown in Fig.\,\ref{fig:Seebeck-Schematic}. Here, the Seebeck voltage is independent of the temperature profile along the material between the junctions, but just depends on the temperature difference between the contact points \citep{Ashcroft:1976ud}.}\citep{Ashcroft:1976ud, Gross} Therefore, the Seebeck coefficient of a material is commonly determined and referenced against a material with a known Seebeck coefficient (alternatively, the relative Seebeck coefficient is determined) via \citep{Gross,BookGrosso,HarmanHonig}
\begin{equation}
	V = \int_{T_1}^{T_2} (S_\textrm{A} - S_\textrm{B}) \mathrm{d}T,
\end{equation}
where $T_1$, $T_2$, $S_\textrm{A}$ and $S_\textrm{B}$ are defined in Fig.\,\ref{fig:Seebeck-Schematic}. The circuit drawn in Fig.\,\ref{fig:Seebeck-Schematic} is also called a thermocouple and can be used even at the nanoscale.\citep{Nanoscale}

\subsubsection{Peltier effect} \label{sec:peltier}
\textit{The Peltier effect describes the generation of a heat current density by an applied electric field or, equivalently,  a gradient in the charge chemical potential. From an experimental point of view, this effect is usually described as the generation (heating) or removal (cooling) of a heat current at the contact of two dissimilar materials ($\Pi_A \neq \Pi_B$) when a charge current is fed accross their shared interface.}

In 1834, Jean-Charles Peltier discovered the  Onsager reciprocal\citep{Onsager,Onsager2}) of the Seebeck effect, named the Peltier effect.\citep{Peltier} The Peltier effect describes the generation of a heat current density along a gradient of the charge chemical potential. Historically, the Peltier coefficient $\Pi$ has been introduced as the proportionality constant between the charge and the heat current density via $\Jvecheat=\Pi\Jvec$. 

Within the two-spin current model, using Eq.\,(\ref{eq:GeneralCollinearTransportEquation2}) and assuming $\boldsymbol{\nabla}\pots=0$, we can write the heat current density $\Jvecheat$ as
\begin{equation}
\Jvecheat^{\uparrow,\downarrow} = - \sigmac^{\uparrow,\downarrow} S^{\uparrow,\downarrow} T \boldsymbol{\nabla} \mucharge / \charge - \kappa^{\uparrow,\downarrow} \boldsymbol{\nabla}T  \; ,
\label{eq:Peltier-twofluid1}
\end{equation}
where we again assign each spin species to its own ``spin-dependent'' Seebeck coefficient. Assuming a vanishing or short-circuited thermal gradient $\boldsymbol{\nabla}T=0$, the total heat current density can be written as
\begin{equation}\label{eq:peltier1}
   \Jvecheat = \Jvecheat^\uparrow+  \Jvecheat^\downarrow = - \left( \sigmac^\uparrow S^\uparrow+ \sigmac^\downarrow S^\downarrow\right) T \, \boldsymbol{\nabla} \mucharge / \charge.
\end{equation}

Combining Eqs.\,(\ref{eq:peltier1}) and (\ref{eq:Seebeck-twofluid3}) with $\Jvec=\Jvec^\uparrow+\Jvec^\downarrow = -(\sigmac^\uparrow + \sigmac^\downarrow) \boldsymbol{\nabla} \mucharge / \charge$, we find
\begin{equation}
	\label{eq:Peltier}
	\Jvecheat  = ST \Jvec = \Pi \Jvec.
\end{equation}
 This result implies that the charge current direction controls the heat current direction. Additionally, Eq.\,(\ref{eq:Peltier}) reveals the second Thomson relation\footnote{The first Thomson relation is discussed in the section of the Thomson effect (see Sec.\,\ref{sec:thomson}) } 
\begin{equation}
\label{eq:thomsonrelation}
\Pi = ST.
\end{equation}
This relation was originally found by William Thomson (later Lord Kelvin)\citep{Thomson1,Thomson2} and later could be understood in terms of the Onsanger's reciprocity relation.\citep{Onsager,Onsager2} 

\begin{figure}[h]
	\includegraphics[width=\textfigures]{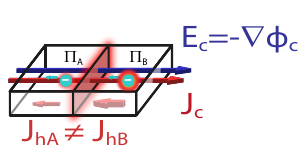} 
    \vspace{\captionhalfclose}
    \caption{\label{fig:Peltier}  Schematic representation of the Peltier effect observed using two dissimilar materials, as is typical for experiments and applications. Here we assume $\Pi_A > \Pi_B$ and that both Peltier coefficients are negative. This results in heating of the junction between the two materials. Inversion of the charge current direction will result in cooling of the junction.} 
\end{figure}

In a simplistic approach, where the heat is solely carried by electrons, the Peltier effect can be depicted as in Fig.\,\ref{fig:Peltier}. A charge current flowing in the two materials A and B is accompanied by different  heat currents when the Peltier coefficient of  A and B differ ($\Pi_\mathrm{A} \neq \Pi_\mathrm{B}$). If the charge current across the contact area is conserved, this results in a discontinuity of the associated heat current. This discontinuity in $\Jvecheat$ generates (removes) heat at (from) the contact area resulting in a local heating (cooling) effect via
\begin{equation}\label{eq:peltier2}
   \frac{d Q_\mathrm{int}}{d t} = {\Jvecheat}_\mathrm{A}-{\Jvecheat}_\mathrm{B} = \left(\Pi_\mathrm{A} - \Pi_\mathrm{B}\right)  \Jvec. 
\end{equation}
Here, $Q_\mathrm{int}$ is the heat density created (removed) at the contact per unit time and area. This principle is commercially used in heating and cooling applications (Peltier coolers). Notably, these devices can switch from heating to cooling by inversion of $\Jvec$. Note that Joule heating, which originates from power dissipation due to resistive losses in the materials, scales markedly different with $\Jvec^2$. As most Peltier coolers are based on resistive conductors, Joule heating eventually limits the minimum accessible temperatures.

\vspace{\figuretop} 

\subsubsection{Thomson effect}\label{sec:thomson}
\textit{The Thomson effect describes the generation of extra heating (or cooling) of a material when a charge current is sent through the material in presence of a temperature gradient.}

The Seebeck and Peltier effects are obtained under the boundary conditions that either the applied electric or thermal field vanishes, respectively. William Thomson considered the more general case that both fields are present at the same time. The related thermoelectric effect, which can be considered a superposition of Seebeck and Peltier effect, is named after him -- the Thomson effect.\citep{Thomson1,Thomson2}

The Thomson effect is caused by the temperature dependence of the Seebeck and thus the Peltier coefficient, $S$ and $\Pi$, respectively. Thus, when a temperature gradient is present, the magnitude of the Seebeck and Peltier coefficients changes along the temperature profile in the material.  
This situation is locally equivalent to an interface between two materials with different Peltier coefficients and hence gives rise to a local Peltier effect. For the entire material we therefore expect a `continuous' Peltier effect along the thermal gradient.\citep{BookGrosso,Ashcroft:1976ud} 
Several contributions add to the heat production per unit volume $\dot{Q}$: (i) the thermal continuity equation $\dot{Q}+\boldsymbol{\nabla} \Jvecheat=0$ in combination with the heat flow due to a thermal gradient $\Jvecheat = -\kappa \boldsymbol{\nabla} T$, (ii) the local Peltier effect, and (iii) the inevitable Joule heating. In a steady state situation, those can be expressed as \citep{Gross,BookGrosso,Ashcroft:1976ud,HarmanHonig} 
\begin{equation}
	\dot{Q} = \boldsymbol{\nabla}(\kappa \boldsymbol{\nabla} T)  - T \Jvec \cdot \boldsymbol{\nabla} S + \Jvec (\sigmac^{-1} \Jvec), 
\end{equation}
where $- T \Jvec \cdot \boldsymbol{\nabla} S =-K_c \Jvec \boldsymbol{\nabla}T$. The Thomson coefficient $K_c$ is  related to the Seebeck and Peltier coefficient, via the first Thomson (or Kelvin) relation \citep{Thomson1,Thomson2,Gross,BookGrosso}
\begin{equation}
\label{eq:Thomson}
	K_c = \frac{d\Pi}{dT}-S = T\frac{dS}{dT}.
\end{equation}
Like the Peltier effect, the Thomson effect can be distinguished from (pure) Joule heating by current reversal. While the Joule heating scales quadratic with the charge current, the Thomson effect scales linear in the charge current and therefore can result in heating or cooling of the material.

Note that the Thomson effect allows one to determine the Seebeck or Peltier coefficient of a material without requiring a reference material (via Eq.\,(\ref{eq:Thomson})).
Furthermore, when modeling a system considering the temperature dependencies of the Peltier and Seebeck effects, the Thomson effect is included automatically and has not to be accounted for separately.

\begin{figure}[h]
	\includegraphics[width=\textfigureThomson]{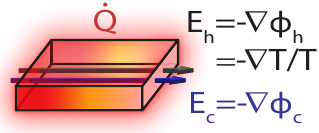} 
    \vspace{\captionhalfclose}
    \caption{\label{fig:Thomson} Schematic drawing of the Thomson effect assuming a positive Thomson coefficient.} 
\end{figure}

\subsection{Spin-related versions of the Seebeck, Peltier, and Thomson effects}\label{subsec:Spin_Seebeck_Peltier2}
In complete analogy with thermo-electric coupling, which leads to the Seebeck, Peltier and Thomson effect due to the fact that charge and heat transport are coupled, thermo-spin coupling leads to the spin counterparts of these phenomena due to the coupling of  spin and heat transport. 
 These are represented by the off-diagonal elements $L_{23}$ and $L_{32}$ of Eq.\,({\ref{eq:GeneralCollinearTransportEquation}}). Note that due to thermo-spin coupling an applied gradient of the spin-chemical potential leads to a "heat-polarized" spin current, while an applied temperature gradient results in "spin-polarized" heat current. 

Before we start to individually discuss the spin-related versions of the Seebeck, Peltier and Thomson effect, we note that in electrical conducting magnetic materials the charge- and spin-related Seebeck effect are difficult to disentangle, since charge carriers are transporting, in addition to heat, always charge and spin at the same time. That is, there is always a thermo-spin-electric coupling. Therefore, for magnetically ordered electrical conductors the nomenclature spin-dependent Seebeck effect (SdSE) has been introduced.\citep{BauerCaloritronics,Sinova,SSEUchida,SSEUchida2} The SdSE results from the fact that the charge carriers in a conducting magnetic metal have both charge and spin. Therefore, the usual Seebeck effect (thermo-electric coupling) always appears simultaneously with the spin-dependent Seebeck effect (thermo-spin coupling). That is, the SdSE represents a mixture of the charge- and spin-based Seebeck effects. In a two-spin-channel model, it can be  understood in terms of the joint action of the conventional Seebeck effect present for spin-up and spin-down charge carriers in a conducting magnetically ordered metal \citep{BauerCaloritronics,SdSE_Slachter,SdSE_Dejene}. In order to observe the pure SSE, the charge-related contribution has to be eliminated. This can be achieved in magnetic insulators, where charge transport is absent and  magnons -- the quantized excitations of a magnetically ordered system -- can carry both angular momentum (i.e. spin) and heat. That is, the magnon-based Seebeck effect is independent of the charge transport, resulting only from the coupling of spin and heat transport (thermo-spin coupling) \cite{TheorySSE,TheorySSE-erratum,SchreierSSE,ReviewSSE}. Since the carriers responsible for the appearance of the SSE in magnetic insulators are quantized spin waves (magnons), it  is sometimes also named as `spin-wave' or `magnonic' Seebeck effect.\citep{BauerCaloritronics} Strictly speaking, in magnetic metals, both electrons and magnons thus can contribute to a magnetic-property dependent Seebeck response - SdSE and SSE superimpose.

In analogy to the SdSE and the SSE, also the spin-dependent Peltier effect (SdPE) as well as the spin-dependent Thomson effect can be distinguished from the pure spin Peltier effect (SPE) and the pure spin Thomson effect, respectively. Whereas the former  are a mixture of the charge-related (thermo-electric coupling) and spin-related (thermo-spin coupling) Peltier and Thomson effects, the latter are purely spin-related effects (thermo-spin coupling).

Another issue with the nomenclature used in literature stems from the quantity considered as a response to the thermal drive. When the spin-related Seebeck effect is introduced in complete analogy to the charge-related Seebeck effect, it would describe the proportionality between the ''spin field'' $\Efield_s = - \boldsymbol{\nabla} \pots$ and the applied temperature gradient with the proportionality constant given by the spin Seebeck coefficient $S_s$:
\begin{equation}
    \Efield_s = -\boldsymbol{\nabla} \pots = S_s \boldsymbol{\nabla} T
\end{equation}
A general formulation of the purely spin-related  Seebeck effect according to Eqs.\,(\ref{eq:GeneralCollinearTransportEquation}) and (\ref{eq:GeneralCollinearTransportEquation2}) takes the form
\begin{equation}
\Jvecspin = -L_{23} \boldsymbol{\nabla} \poth.
\end{equation}
In a similar manner, the equivalent definition of the spin-related Peltier effect   to the charge-related  Peltier effect reads as 
\begin{equation}
\Jvecheat = \Pi_s \Jvecspin,
\end{equation}
where $\Pi_s$ is the spin Peltier coefficient. Both definitions are formally independent of the transport of charge.  

However, the spin-related effects have first been studied in metallic systems, where the charge- and spin-related Seebeck, Peltier and Thomson effects appear simultaneously. Moreover, charge-based quantities (electrical voltages) are typically measured in the experiments, making the situation even more complex. In the following subsections we discuss the situations in metallic systems, where always charge- and spin-related effects appear simultaneously and the respective effects are named "spin-dependent". Purely spin-related effects can be studied in magnetic insulators (absence of charge transport), where magnetic excitations of the spin systems can carry both spin and heat (thermo-spin coupling) and result in purely spin-related Seebeck, Peltier and Thomson effects.

\subsubsection{Spin-dependent Seebeck effect}\label{subsec:SpindependentSeebeckEffect}
\textit{The spin-dependent Seebeck effect (SdSE) describes the generation of a combined charge and spin current generated by a temperature gradient in a magnetic material.} 

This effect was first described by Ken-Ichi Uchida et al. in 2008 in a metallic ferromagnet \citep{Uchida1}, and soon after also in semiconductors \cite{JaworskiSSE, JaworskiSSE2}. The experimental reports triggered 
a multitude of experiments and discussions regarding  the magnitude of the effect as well as its precise microscopic origin \citep{TransverseSSE,SSEUchida}. The SdSE was also reported in semiconductors \citep{JaworksiSemiCond}.

In magnetically ordered electrical conductors, the spin-dependent Seebeck effect, can be understood by taking into account  that the applied temperature gradient causes both a charge and spin current, since the charge carriers in electrical conductors transport charge and spin at the same time. In a two-spin-channel model this can be described by the combined action of the charge-related Seebeck effect for the spin-up and spin-down charge carriers \citep{BauerCaloritronics,SdSE_Slachter,SdSE_Dejene}.
To discuss the spin-dependent Seebeck effect, we take the gradient of the spin chemical potential into account.
Then, by using Eqs.\, (\ref{eq:GeneralCollinearTransportEquation2}), (\ref{eq:spinpol6}), and (\ref{eq:Seebeck-twofluid2}), the spin current density in an electrical conductor is 
\begin{eqnarray}
\Jvecspintilde &=& \Jvec^\uparrow-\Jvec^\downarrow \nonumber\\
&=& 
-\left(\sigmac^\uparrow S^\uparrow-\sigmac^\downarrow S^\downarrow\right)\boldsymbol{\nabla}T - \left(\sigmac^\uparrow+\sigmac^\downarrow\right) \boldsymbol{\nabla} \muspin / 2 \charge
\end{eqnarray}
assuming a vanishing electrical field $\Efieldc = - \boldsymbol{\nabla} \mucharge / \charge$ as the boundary condition. Hence, in analogy to the Seebeck effect (see Sec.\,\ref{sec:Seebeck}), one expects for open spin-current circuit conditions ($\Jvecspintilde=0$) the build-up of a spin chemical potential gradient
\begin{equation}
    - \boldsymbol{\nabla} \muspin / (2 \charge) = \frac{\sigmac^\uparrow S^\uparrow-\sigmac^\downarrow S^\downarrow}{\sigmac^\uparrow+\sigmac^\downarrow}\boldsymbol{\nabla}T. \label{eq:32}
\end{equation}
Using $S$ according to Eq.\,(\ref{eq:Seebeck-twofluid3}) and $P^\prime$ according to Eq.\,(\ref{eq:SS_04}) together with $S^{\uparrow,\downarrow} \propto (\sigmac^{\uparrow,\downarrow})^\prime / \sigmac^{\uparrow,\downarrow}$ (Mott's relation), this leads to
\begin{equation}
- \boldsymbol{\nabla} \muspin / (2 \charge) = P^\prime S \; \boldsymbol{\nabla}T
\label{eq:SSE_02}
\end{equation}
and 
\begin{equation}
\Jvecspintilde = - \sigmac P^\prime S \boldsymbol{\nabla} T.
\label{eq:SSE_02b}
\end{equation}

\vspace{\figuretop} 
\begin{figure}[h]
	\includegraphics[width=\textfigures]{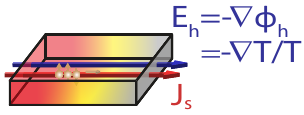} 
    \vspace{\captionhalfclose}
    \caption{\label{fig:SSE} Schematic drawing of the spin-dependent Seebeck effect (present in an electrically conducting magnetic material). In this schematic electrons with finite spin polarization are the carriers of the spin current. Thus, the temperature gradient generates a charge and spin current simultaneously, leading to a SdSE.  Here it is assumed that $P'S<0$. 
    }
\end{figure}

\subsubsection{Spin-dependent Peltier effect}\label{subsec:spinpeltier}
\textit{The spin-dependent Peltier effect describes the fact that the combined charge and spin current in an electrically conducting magnetic material is associated with a finite heat current.  From an experimental point of view, this effect is often described as heating or cooling of a junction of two dissimilar materials due to the presence of the combined charge/spin current.}

First experiments on the SdPE were performed in 2006 by  Gravier \textit{et al.}, indicating the existence of spin-dependent Peltier coefficients.\citep{SdPE_Gravier} In 2012,  Flipse \textit{et al.} were the first to directly measure cooling and heating by spin currents using the SdPE.\citep{SdPE_Flipse}

\begin{figure}[h]
	\includegraphics[width=\textfigures]{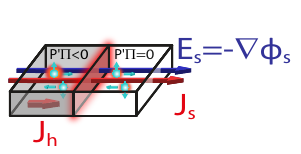} 
    \vspace{\captionhalfclose}
    \caption{\label{fig:SPE} Schematic drawing of the spin-dependent Peltier effect in a magnetic material (left with $P' \neq 0 $), connected to a non-magnetic material (right with $P' = 0 $), where only the charge-related Peltier effect occurs. This configuration is typical for experimental investigation of the SdPE.\citep{SdPE_Flipse} 
    In detail, if we assume  $P' \Pi <0 $ for the magnetic material, the indicated spin current direction results in a heating of the interface.  }
\end{figure}

In a magnetic conductor the flow of spin-polarized charge carriers is always associated with a spin current, leading to the simultaneous appearance of the charge- and spin-related Peltier effect. According to Eq.\,(\ref{eq:GeneralCollinearTransportEquation2}), in a two-spin-channel model the heat flow associated with the flow of spin-up and spin-down charge carriers can be expressed as 
\begin{equation}
\Jvecheat = \Jvecheat^{\uparrow} + \Jvecheat^{\downarrow} = - \sigmac P^\prime S T \boldsymbol{\nabla}\muspin /2\charge \; .\label{eq:SPE_06}
\end{equation}
Using (\ref{eq:32}) and (\ref{eq:SSE_02}), this can be expressed as

\begin{eqnarray}
\Jvecheat & = & - \frac{ \sigmac^{\uparrow} S^{\uparrow} -\sigmac^{\downarrow} S^{\downarrow} }{\sigmac^{\uparrow} S^{\uparrow} + \sigmac^{\downarrow} S^{\downarrow}} \; (\sigmac^{\uparrow} + \sigmac^{\downarrow}) \; ST \boldsymbol{\nabla}\muspin /2\charge \; \; .\label{eq:SPE_07}
\end{eqnarray}
Using further $\Jvecspintilde = \Jvec^\uparrow - \Jvec^\downarrow = -(\sigmac^\uparrow + \sigmac^\downarrow) \boldsymbol{\nabla}\muspin /2\charge$, 
leads to
\begin{equation}
\Jvecheat = \; P^\prime ST \Jvecspintilde = P^\prime \Pi \; \Jvecspintilde    \;
\label{eq:SPE_02}\end{equation}
We see that in analogy to the charge-related Peltier effect, a spin current is associated with a heat current with the proportionality constant given by the spin dependent Peltier coefficient $P^\prime \Pi$.  Again, inducing a heating or cooling effect -- the typical application of a generalized Peltier effect --  requires an (effective) interface separating two areas with different $P' \Pi$, which is realized  in experiments  using a heterostructure of two materials, as is shown in Fig.~\ref{fig:SPE}.

\vspace{\figuretop} 

\subsubsection{Spin-dependent Thomson effect}
\textit{The spin-dependent Thomson effect describes the generation (removal) of heat resulting in a heating (cooling) effect  in an electrically conducting magnetic material when a combined charge and spin current is sent through the material in the presence of a temperature gradient.  This is the analogue of the charge Thomson effect.}

Bakker \citep{FrankThesis} proposed the spin-dependent Thomson effect to be present in metallic conductors due to a finite temperature dependence of the spin-dependent Peltier ($P' \Pi$) or Seebeck ($P' S$) coefficients. 

Using the arguments presented in Sec.\,\ref{sec:thomson}, i.e. considering the Mott two-spin channel model \citep{Mott1936,OHandley} and non-vanishing spin-dependent Seebeck and Peltier coefficients, as well as the boundary condition of a vanishing charge current flow $\Jvec=0$, the heating or cooling rate per unit volume can be expressed as 
\begin{equation}
	\dot{Q} = \boldsymbol{\nabla}(\kappa \boldsymbol{\nabla} T)  - T \Jvecspintilde \cdot \boldsymbol{\nabla} \left( P^\prime S \right), 
\end{equation}
where the last term can be associated with the spin-dependent Thomson effect. This term can be rewritten in the form 
\begin{equation}
     - T \frac{\partial P^\prime S}{\partial T} \Jvecspintilde \boldsymbol{\nabla}T = - K_\mathrm{s} \Jvecspintilde \boldsymbol{\nabla}T, 
\end{equation}
with the spin-dependent Thomson coefficient $K_\mathrm{s}=T \frac{\partial P^\prime S}{\partial T}$. 

Taking the viewpoint of two independent spin channels present in the metal, the spin channel specific Thomson effect can be described by
\begin{equation}
	\dot{Q}^{\uparrow,\downarrow} = - K^{\uparrow,\downarrow} \Jvec^{\uparrow,\downarrow} \cdot \nabla T,
\end{equation}
where the total heating (or cooling) caused by the spin-dependent Thomson effect is calculated by the sum of the contributions of both spin channels.
The spin-dependent Thomson effect can be schematically drawn in the same way as the Thomson effect (see Fig.\,\ref{fig:Thomson}), replacing $\Efieldc$ by $\Efield_\mathrm{s}$. Note that whether the $\Jvec$ and $\Jvecspin$ are parallel or antiparallel depends on $K_\mathrm{s}$.


\section{Transverse transport phenomena}\label{sec:TransNew}

This section is dedicated to the transverse transport phenomena, which have already been introduced briefly in Sec.\,\ref{sec:Intro} and Sec.\,\ref{sec:Theory}. 
In analogy to the treatment of the collinear transport phenomena above,  we again focus on the linear response in the form  of charge, heat and spin currents. We focus on electrically conducting materials where heat, charge and spin are carried by the charge carriers at the same time.\footnote{Note that transverse transport phenomena can also be observed for magnetic insulators, where charge motion is absent. In this case, heat and spin are carried by solid state excitations, such as phonons and  magnons.} All effects discussed in this section are captured by the second term of the right hand side of  Eq.\,(\ref{eq:GeneralGeneralTransportEquation}), which reads in its expanded form
\begin{widetext}
\begin{equation}
\left(\begin{array}{c}
\Jvec \\ \Jvecheat \\ \Jvecspin
\end{array}\right)
=
\left(\begin{array}{ccc}
L_{11} & L_{12} & L_{13} \\
L_{21} & L_{22} & L_{23} \\
L_{31} & L_{32} & L_{33}
\end{array}\right)
\left(\begin{array}{c}
- \boldsymbol{\nabla} \potc \\ - \boldsymbol{\nabla} \poth \\ - \boldsymbol{\nabla} \pots
\end{array}\right)
+
\left(\begin{array}{ccc}
L^\perp_{11} & L^\perp_{12} & L^\perp_{13} \\
L^\perp_{21} & L^\perp_{22} & L^\perp_{23} \\
L^\perp_{31} & L^\perp_{32} & L^\perp_{33}
\end{array}\right)
\left(\begin{array}{c}
- \boldsymbol{\nabla} \potc \times \Bhat\\ 
- \boldsymbol{\nabla} \poth \times \Bhat\\ 
- \boldsymbol{\nabla} \pots \times \Bhat
\end{array}\right)
\; .\label{eq:GeneralTransportCoefficients_transverse}
\end{equation}
\end{widetext}
Here, 
we only consider effects up to first order in the generalized magnetic field $\Bvec$. Evidently, the cross products describe a perpendicular motion of all response quantities with respect to the gradient of the charge, heat, or spin chemical potential, and the orientation of the generalized magnetic field $\Bhat=\Bvec/|\Bvec|$, which is in strong contrast to the effects addressed in Sec.\,\ref{sec:Col}. 
While various microscopic effects are at the origin of the transverse response, we can identify and classify two sub-categories, namely a \textbf{\textit{charge-sensitive}} and a \textbf{\textit{spin-sensitive}} deflection. While the \textit{charge-sensitive} deflection direction solely depends on sign the   of the charge, the \textit{spin-sensitive} deflection direction is linked to the spin orientation, such as spin-up and spin-down.

A well-known example is the transverse \textit{charge-sensitive deflection} by the Lorentz force  $\mathbf{F}_\mathrm{L} = \charge \left( \mathbf{v}\times  \Bvec \right)$ acting on charge carriers moving with the velocity $\mathbf{v}$. Here, the role of the generalized magnetic flux density $\Bvec$ is taken by the applied static magnetic field $\mu_0 \Hvec$, with $\mu_0$ the magnetic permeability. For the situation where  charge carriers move  with an average drift velocity $\mathbf{v}$ in a direction  with a  finite perpendicular component to the applied magnetic field, the Lorentz force causes an average deflection orthogonal to both $\Bvec$ and $\mathbf{v}$. In a solid state environment,  the average Lorentz force and hence transverse deflection is determined by the average longitudinal drift velocity of the charge carriers, caused either by a gradient of the electrical potential (i.e. an electric field) or temperature.  Thus, this simple picture already allows one to rationalize the (charge) Hall effect (see Sec.\,\ref{sec:AHE}) and the thermal Hall effect (see Sec.\,\ref{sec:Righi-Leduc}), as well as the Nernst effect (see Sec.\,\ref{sec:Nernst}) and the Ettingshausen effect (see Sec.\,\ref{sec:Ett}). 

As this simple model provides  an intuitive understanding of the origin of the discussed transverse transport phenomena, we will depict this mechanism in all figures. However, we emphasize that the microscopic mechanisms leading to non-collinear transport behavior are typically of much more complex origin.

In the example discussed above, the generalized magnetic flux density $\Bvec$ is assumed to arise solely from the externally applied static magnetic field $\mu_0 \Hvec$. This simplified scenario enables a further classification of effects: transverse transport phenomena associated with $\mu_0 \Hvec$ are designated with the prefix  \textit{\textbf{ordinary}}.

\textit{Spin-sensitive deflection} occurs in solids with broken time-reversal symmetry. A prime example for such a system is a ferromagnetic conductor with a finite magnetization $\Mvec$ and spin-orbit coupling. Here, spin-sensitive deflection is best known in the context of the anomalous Hall effect (AHE, see Sec.\,\ref{sec:AHE}).\citep{AHEreview,Handbook,OHandley,AHEreview2, Maekawa:2017gn}  In contrast to charge-sensitive deflection, spin-sensitive deflection cannot be rationalized in terms of a simple classical picture. Theoretical studies distinguish between intrinsic and extrinsic mechanisms. 

For metallic ferromagnets with a moderate conductivity, the intrinsic spin-deflection mechanism has been identified as dominant \cite{Onda.2008, NagaosaRMP.2010} and can be described in terms of a Berry phase\citep{BerryOriginal} related effect. The Hall response thus is governed by a quantum-mechanical property of the perfect crystal, which is also reflected by the bandstructure of the material, and thus is not related to any extrinsic scattering process. The broken time-reversal symmetry of the sample in combination with finite spin-orbit coupling modifies the charge carrier transport and allows to link it to the spin-sensitive deflection. This coupling can be expressed as a finite Berry curvature within the Brillouin zone, which in turn can be viewed as an emergent magnetic flux density (for details see Refs.\,\citep{AHEreview,Gross}). This viewpoint has the advantage, that the transverse charge carrier motion can be rationalized in the frame of a Lorentz force, even if the role of the external magnetic field is taken by this emergent magnetic flux density, reflecting the intrinsic material properties. Hence, this property can also be expressed in form of a generalized magnetic flux density $\Bvec$. 

Extrinsic spin-deflection effects such as skew scattering and side jump originate from disorder. They tend to dominate in highly conductive ferromagnets.\citep{AHEreview,Handbook,OHandley,AHEreview2} These scattering processes deflect the charge carriers depending on their spin-orientation and as a result cause charge carrier motion along the transverse direction. As the
deflection is spin-dependent the resulting transverse current
is a spin current. However, due to the imbalance of the spin-up and spin-down charge carrier density for materials with  $\Mvec \ne 0$, the transverse current is also a charge current. 

Effectively, both the intrinsic and extrinsic mechanisms result in transverse motion of the charge carriers with opposite sign for the two spin species. 
In the following, all transverse 
transport phenomena related to spin-sensitive deflection in materials with finite magnetization $\Mvec$ ($\Bvec$ then is represented by $\mu_0 \Mvec$) are labeled with the prefix {\it\bfseries anomalous}.

The concepts developed for explaining the anomalous Hall effect can be naturally transferred to materials with $\Mvec=0$. We later name these materials with the term `non-magnetic', which indicates the absence of a magnetically ordered phase under the conditions of interest. The vanishing magnetization $\Mvec=0$ requires the introduction of an additional effective field  taking the role of $\Mvec$. This field depends on the spin direction $\sigmapol$. Again, we can absorb this field in a generalized effective field $\Bvec$ which takes then the orientation of $\sigmapol$ and the strength of the spin-selective deflection encoded in its amplitude. As $\sigmapol$ is intimately connected to the relativistic spin-orbit interaction, we will label all effects associated with $\sigmapol$ as \textit{\textbf{spin-orbitronic}} \footnote{It is noteworthy that the coupled transport phenomena described in (\ref{eq:GeneralTransportCoefficients_transverse}) can be further extended to the transport of orbital momentum, which we do not cover here. We refer the reader to Refs.\,\citep{Jo:2024, Wang:2025, Go:2018, Go:2020, Choi.2023, Ding:2020} for further details. }.

In summary, we can relate the transverse deflection of charge, heat and spin independently from the type of the gradient in the respective potential, to a generalized magnetic flux density $\Bvec$. This $\Bvec$ can either be represented by the applied external magnetic field $\mu_0 \Hvec$, related to the internal magnetization direction $\mu_0 \Mvec$, linked to the spin quantization direction (spin polarization) $\sigmapol$ or, in a more complex situation, a combination of those. Here, the transverse transport phenomena based on $\Bvec \propto \mu_0 \Hvec$ describe the \textit{\textbf{ordinary effects}}, phenomena where $\Bvec$ is represented by $\mu_0 \Mvec$ are the so-called \textit{\textbf{anomalous effects}}, and the case where $\Bvec$ is associated with $\sigmapol$ sets the basis for effects labeled with the prefix \textit{\textbf{spin-orbitronic}}.

Table\,\ref{tab:trans} presents a schematic overview of all expected transverse transport phenomena in metallic conductors reflecting the transport matrix $\mathbf{L}^\perp$ in Eq.\,(\ref{eq:GeneralTransportCoefficients_transverse}). According to the applied driving force, we can distinguish between \textbf{\textit{galvanomagnetic}} (caused by a gradient in the electrochemical potential), \textbf{\textit{thermomagnetic}} (caused by a temperature gradient) and \textbf{\textit{spinmagnetic effects}} (caused by a gradient in the spinchemical potential). These three subgroups are represented by the three columns of Table\,\ref{tab:trans}. Linked to the response channel (charge, heat, or spin), we can identify  three subgroups which form the three rows in Table\,\ref{tab:trans}. We assign these groups the prefixes 'charge', 'thermo' and 'spin' for further classification, resulting e.g. in the spin-galvanomagnetic, spin-thermomagnetic, or thermo-spinmagnetic effect. However, historically,  the names of famous scientists have been used to denote the responses in the charge and heat channel. The charge- and thermo-galvanomagnetic effects presented in the left column of Table\,\ref{tab:trans} are called Hall and Ettingshausen effect, respectively. The thermo- and charge-thermomagnetic effects (see middle column of Table\,\ref{tab:trans}) are called Righi-Leduc and Nernst effect, respectively. Of course, all these effects come with an ordinary, anomalous and spin-orbitronic version. 
We see that due to the additional spin degree of freedom the originally two (charge- and heat-related) galvanomagnetic effects (Hall and Ettingshausen effect) have to be completed by a third spin-related galvanomagnetic effect, which in literature is -- in an illogical way -- usually called spin Hall effect. In the same way, we have to add a third spin-related effect to the former two (charge- and heat-related) thermomagnetic effects (Nernst- and Righi-Leduc effect). In literature, this spin-thermomagnetic effect is called the spin Nernst effect. It is natural to name the transverse spin-related effect caused by a longitudinal gradient of the electrochemical potential \textbf{\textit{spin-galvanomagnetic effect}} (SGME) (ordinary, anomalous and orbitronics spin-galvanomagnetic effect) and that caused  by a longitudinal temperature gradient \textbf{\textit{spin-thermomagnetic effect}} (STME) (ordinary, anomalous and orbitronic spin Nernst effect).  
All galvanomagnetic effects are found in the left column of Table\,\ref{tab:trans}. All thermomagnetic effects are found in the middle column of Table\,\ref{tab:trans}. 

In order to continue with the logic of this nomenclature we denote the transverse effects caused by a longitudinal gradient of the spinchemical potential as \textbf{\textit{spinmagnetic effects}}. We suggest to name the transverse electric response \textbf{\textit{electro-spinmagnetic effect}} (ESME) or \textbf{\textit{charge-spinmagnetic effect}} (in literature commonly called inverse spin Hall effect) and the transverse thermal response \textbf{\textit{thermo-spinmagnetic effect}} (TSME) (in literature called spin Ettingshausen effect). The spin-related spinmagnetic effect is the pure spin Hall effect. All spinmagnetic effects are found in the right column of Table\,\ref{tab:trans}. 

According to our classification of the transverse effects into ordinary, anomalous, and spin-orbitronic, in combination with Eq.\,(\ref{eq:GeneralTransportCoefficients_transverse}), we expect a total $3\times9=27$ transverse transport phenomena in this section. While some of these have been reported, others have not been studied to the best of our knowlede, although their presence is expected according to simple symmetry considerations. We note that the unambiguous experimental verification of all effects might be challenging, as some of the effects will appear in combination with others  and might even share the same symmetry. For clarity, Table\,\ref{tab:trans} labels phenomena which have been discussed in literature, i.e. have been experimentally observed, in solid black. In contrast, names set in gray have not yet been reported in literature, to the best of our knowledge. 

In the following, we will discuss only effects being linear in the magnetic field and exclude those being quadratic in the magnetic field such as the ordinary magnetoresistance or the Maggi-Righi-Leduc effect (thermal analogue of the ordinary magnetoresistance).

\subsection{The charge, the thermal and the pure spin Hall effect}
\label{sec:Hall_Effect}
In analogy to the collinear transport effects presented in Sec.\,\ref{sec:Col}, we first discuss effects  for which the longitudinal drive and the transverse response are related to the same transport entities (e.g. longitudinal temperature gradient drives transverse heat current). The related transverse transport phenomena are named Hall effects, which  are represented by the diagonal matrix elements $L^\perp_{ii}$ in Eq.\,(\ref{eq:GeneralTransportCoefficients_transverse}). They quantify the transverse response in form of charge, heat or spin current density driven by a longitudinal gradient of the corresponding potential.
Using the definitions of Eq.\,(\ref{eq:GeneralCollinearTransportEquation2}), the corresponding charge, heat and spin current densities are 
\begin{widetext}
\begin{eqnarray}
\Jvec & = & \Jveclong + \Jvectrans = \sigmac \left( -\boldsymbol{\nabla} \mucharge/q \right) + {\sigmac}_\mathrm{H} \left( -\boldsymbol{\nabla} \mucharge/q \times \Bhat \right)
    \label{eq:Hall_conductivities_2a}\\
\Jvecheat & = & \Jhveclong + \Jhvectrans = \kappa T \left( -\boldsymbol{\nabla}T/T \right) + \kappa_\mathrm{H} \,T \left( -\boldsymbol{\nabla}T/T \times \Bhat \right) \;
    \label{eq:Hall_conductivities_2b}\\
\Jvecspintilde & = & \Jsvectildelong + \Jsvectildetrans = \sigmac \left( -\boldsymbol{\nabla} \mu_\mathrm{s}/2q \right) + {\sigmac}_\mathrm{H}  \left( -\boldsymbol{\nabla} \mu_\mathrm{s}/2q \times \Bhat \right),
    \label{eq:Hall_conductivities_2c}
\end{eqnarray}
\end{widetext}
where we have separated the stimulated current densities into a longitudinal $\mathbf{J}^\mathrm{long}$ and transverse $\mathbf{J}^\mathrm{trans}$ component, and have introduced the charge and thermal Hall conductivities ${\sigmac}_\mathrm{H}=L^\perp_{11}$, ${\sigmac}_\mathrm{H} \hbar/(2\charge) = L^\perp_{33}$ and $\kappa_\mathrm{H}=L^\perp_{22}/T$. 
It is important to note that  Eqs.\,(\ref{eq:Hall_conductivities_2a})-(\ref{eq:Hall_conductivities_2c}) are valid only under specific boundary conditions. For example, to obtain the charge current density $\Jvec$ given in Eq.\,(\ref{eq:Hall_conductivities_2a}), it is required that the longitudinal thermal gradient and the spin-chemical potential vanish. Equivalent boundary conditions are in place for the heat $\Jvecheat$ and spin $\Jvecspin$ current cases. If other boundary conditions apply, Eq.\,(\ref{eq:GeneralTransportCoefficients_transverse}) needs to be considered as a whole.

From Eq.\,(\ref{eq:GeneralTransportCoefficients_transverse}) and Eqs.\,(\ref{eq:Hall_conductivities_2a})-(\ref{eq:Hall_conductivities_2c}) we expect the generation of current densities with directions along and perpendicular to the applied gradient of the respective potentials. This requires the expansion of the scalar transport coefficients used in Sec.\,\ref{sec:Col} to a  $3\times 3$ conductivity tensor for each transport quantity. Here, the diagonal elements reflect the collinear conductivities and the Hall conductivities appear as off-diagonal elements. We further use the convention that the generalized magnetic field $\Bvec$ is oriented along the $z$-axis and that the applied potential gradient is directed along the $x$-axis (cf. Fig.\,\ref{fig:Hall}). Then the conductivity tensor of an isotropic material can be written as (we do this exemplary only for the charge conductivity)\citep{Gross,BookHall,OHandley,BookGrosso,BookDavies}
\begin{equation}
\boldsymbol{\sigmavec} =
\left(\begin{array}{ccc}
\sigmac & {\sigmac}_\mathrm{H}      & 0\\
-{\sigmac}_\mathrm{H}     & \sigmac & 0\\
0 & 0 & \sigmac
\end{array}\right) \; .
\label{eq:Hall}
\end{equation}
Here, $\sigmac  = \sigma_{xx}=\sigma_{yy}=\sigma_{zz}$ is the (longitudinal) conductivity of the isotropic material and ${\sigmac}_\mathrm{H}=\sigma_{xy}=-\sigma_{yx}$ the (transverse) Hall conductivity. The respective electrical, thermal and spin resistivities $\rhoc$, $\rho_\mathrm{h}$ and $\rho_\mathrm{s}$ are obtained by inverting the corresponding conductivity tensor.

\newcommand{\Halignn}{\hspace{0.192em}}
\newcommand{\Salignn}{\hspace{0.368em}}

\begin{table*}
\textsf{
    \begin{tabularx}{\textwidth}{ r?Y|Y|SY } 
    	\diagbox[width=\firstcolumn]{\color{response}\textbf{Response}}{\color{drive}\textbf{Drive} \hspace{2pt}} &  \textcolor{drive}{\textbf{Electrical}} & \textcolor{drive}{\textbf{Thermal}} & \textcolor{drive}{\textbf{Spin}} \\
   		\midrule[1.5pt]
   		\multirow{4}{*}[\centerfirstt]{\textcolor{response}{\textbf{Electrical}} \hspace{10pt}}
        & ${\color{response}\Jvectrans} \propto {\color{drive}\mingradmuc} \times \Bvec$ & ${\color{response}\Jvectrans} \propto {\color{drive}\mingradT} \times \Bvec$ & ${\color{response}\Jvectrans} \propto {\color{drive}\mingradmus} \times \Bvec$ \\
        & \textleft{\textbf{(Charge) Hall effect} \\ $\Hvec$: \hspace{0.192em}(Ordinary) Hall effect \\ $\Mvec$: Anomalous Hall effect \\ {\color{grey-text}$\sigmapol$: \Salignn spin-orbitronic Hall effect}} & 
        \textleft{\textbf{Nernst effect} \\ $\Hvec$: \Halignn (Ordinary) Nernst effect \\ $\Mvec$: Anomalous Nernst effect \\ {\color{grey-text}$\sigmapol$: \Salignn spin-orbitronic Nernst effect}} & 
        \textleft{{\color{grey-text}\textbf{Electro-spinmagnetic effect}} \\ {\color{grey-text}$\Hvec$: \Halignn Ordinary ESME} \\ {\color{grey-text}$\Mvec$: Anomalous ESME} \\ $\sigmapol$: {\color{grey-text}\Salignn spin-orbitronic ESME} \\ \textcolor{white}{$\sigmapol$:} \Salignn Lit: Inverse spin Hall effect} \hspace{25pt}{\color{white}.} \\ 
        \hline
    	\multirow{5}{*}[\centerfirstt]{\textcolor{response}{\textbf{Thermal}} \hspace{10pt}} 
        & ${\color{response}\Jhvectrans} \propto {\color{drive}\mingradmuc} \times \Bvec$ & ${\color{response}\Jhvectrans} \propto {\color{drive}\mingradT} \times \Bvec$ & ${\color{response}\Jhvectrans} \propto {\color{drive}\mingradmus} \times \Bvec$ \\
        & \textleft{\textbf{Ettingshausen effect} \\ $\Hvec$: \Halignn \hspace{-.75em} (Ordinary) Ettingshausen effect \\ $\Mvec$: \Halignn\hspace{-.75em} Anomalous Ettingshausen effect \\ {\color{grey-text}$\sigmapol$: \Salignn \hspace{-.75em}\, spin-orbitronic} \\ {\color{grey-text}\hspace{.5cm} \Halignn \hspace{-.75em} Ettingshausen effect}} & 
        \textleft{\textbf{Thermal Hall/Righi-Leduc effect} \\ $\Hvec$: \Halignn (Ordinary) Thermal Hall effect \\ {\color{white}$\Hvec$:} \Halignn (Ordinary) Righi-Leduc effect \\ $\Mvec$: Anomalous thermal Hall effect \\ {\color{white}$\Mvec$:} Anomalous Righi-Leduc effect \\ {\color{grey-text}$\sigmapol$: \Salignn spin-orbitronic }\\ {\color{white}$\sigmapol$:} \Salignn {\color{grey-text}thermal Hall effect}} & 
        \textleft{{\color{grey-text}\textbf{Thermo-spinmagnetic effect}} \\ {\color{grey-text}$\Hvec$: \Halignn Ordinary TSME} \\ {\color{grey-text}$\Mvec$: Anomalous TSME} \\ $\sigmapol$: \Salignn {\color{grey-text}\textit{spin-orbitronic TSME}} \\ \textcolor{white}{$\sigmapol$:} \Salignn Lit: Spin Ettingshausen effect} \hspace{16pt}{\color{white}.} \\[20pt]  
        \hline
    	\multirow{4}{*}[\centerfirstt]{\textcolor{response}{\textbf{Spin}} \hspace{10pt}}
        & ${\color{response}\Jsvectrans} \propto {\color{drive}\mingradmuc} \times \Bvec$ & ${\color{response}\Jsvectrans} \propto {\color{drive}\mingradT} \times \Bvec$ & ${\color{response}\Jsvectrans} \propto {\color{drive}\mingradmus} \times \Bvec$ \\
        & \textleft{{\color{grey-text}\textbf{Spin-galvanomagnetic effect}} \\ {\color{grey-text}$\Hvec$: \Halignn Ordinary SGME} \\ {\color{grey-text} $\Mvec$: Anomalous SGME} \\ $\sigmapol$: \Salignn {\color{grey-text}spin-orbitronic SGME} \\ \textcolor{white}{$\sigmapol$:} \Salignn Literature: Spin Hall effect} & 
        \textleft{{\color{grey-text}\textbf{Spin-thermomagnetic effect}} \\ {\color{grey-text}$\Hvec$: \Halignn Ordinary STME} \\ {\color{grey-text}$\Mvec$: Anomalous STME}\\ $\sigmapol$: {\color{grey-text}\Salignn spin-orbitronic STME} \\ \textcolor{white}{$\sigmapol$:} \Salignn Literature: Spin Nernst effect} & 
        \textleft{\textbf{Pure spin Hall effect} \\ $\Hvec$: \Halignn \textit{(Ordinary) Pure spin Hall effect}  \\ $\Mvec$: \textit{Anomalous pure spin Hall effect} \\  {\color{white}$\Mvec$:} Magnon Hall effect \\ {\color{grey-text}$\sigmapol$: \Salignn spin-orbitronic } \\ {\color{white}$\sigmapol$:} \Salignn  {\color{grey-text}pure spin-Hall effect}} \hspace{25pt}{\color{white}.} \\ 
    \end{tabularx} 
    }    
    \caption{\label{tab:trans}
    Schematic overview of the linear transverse transport phenomena, i.e. the generalized transport response is orthogonal and proportional to the applied longitudinal gradient of the respective potential ($\gradmuc$, $\gradT$, and $\gradmus$) in the presence of a perpendicularly aligned generalized magnetic field $\Bvec$, as discussed in Sec.\,\ref{sec:TransNew}. All effects caused by an applied gradient of the electrochemical potential (left column) are denoted as galvanomagnetic effects, those caused by an applied temperature gradient (middle column) are named thermomagnetic effects, and those caused by an applied gradient of the spinchemical potential (right column) are denoted as spinmagnetic effects. Each of these three types of effect has a charge-, heat- and spin-related version, entering in the three horizontal rows.  To obtain a given transverse effect, only the component of $\Bvec$ quoted should be considered, e.g. $\Bvec=\mu_0\Hvec$ for the (ordinary) Hall effect. All effects named with black roman lettering  have been already reported in literature, while those labelled in italics  have only been found as side-notes. Grey-colored entries have not been reported to the best of our knowledge, but from symmetry reasons they are entered in this table. Whether they can exist or not is unclear. For more details we refer to the text. 
    }
\end{table*}

\subsubsection{(Charge) Hall effect}\label{sec:AHE}
\textit{The (charge) Hall effect describes the generation of a transverse charge current by a londitudinal gradient of the electrochemical potential in the presence of an external magnetic field (ordinary Hall effect) or finite magnetization (anomalous Hall effect), which is perpendicular to both the applied gradient and the current response.}

The  ordinary Hall effect (OHE) was discovered in 1879 by Edwin Herbert Hall.\citep{Hall} Soon after, in 1881, Hall reported a `rotational coefficient' for nickel and cobalt,\citep{HallAHE,AHE2} which was the first observation of, what nowadays is referred to as, the anomalous Hall effect (AHE). The Hall effect is typically measured as Hall electric field using an open circuit boundary condition for the transverse direction. For the anomalous Hall effect, the role of the external magnetic field is taken by the magnetization of the material or a similar quantity breaking the time reversal symmetry and the charge carrier `deflection' is attributed to the spin-orbit interaction.   \citep{IntroductionSpintronics,OHandley,BookGrosso,Gross}.  Note that the understanding of the AHE is still evolving, in particular with more and more studies reporting the presence of AHE in antiferromagnets (altermagnets) with vanishing M. \citep{SmeikalPRX:2022,SmeikalPRX:2022b}

\begin{figure}[h]
    \includegraphics[width=\Axes]{Axes.pdf}
	\includegraphics[width=\textfiguresPerp]{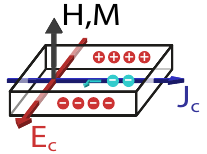}
    \vspace{\captionhalfclose}
    \caption{\label{fig:Hall} Schematic representation of the ordinary Hall effect ($\Hvec$) and anomalous Hall effect ($\Mvec$), assuming open circuit conditions for the generated charge current and short-circuit conditions for the heat current (not shown) along the $y$-direction. Here, the Hall coefficient is assumed to be negative, which corresponds to having negatively charged charge carriers (electrons).}
\end{figure}

Figure\,\ref{fig:Hall} shows the semi-classical picture of the ordinary Hall effect, where the effect is rationalized as a deflection of the charge carriers due to the Lorentz force from their average drift direction along the applied gradient of the average potential, i.e. the electric field. The charges, which drift with the average velocity $\mathbf{v}$, experience the average Lorentz force $\mathbf{F}_\mathrm{L} = \charge (\mathbf{v} \times \Bvec)$, which  results in an average charge current transverse to the drift velocity and the magnetic flux density.\citep{IntroductionSpintronics,OHandley,BookGrosso,Gross} 

For the ordinary Hall effect, where $\Bvec\approx\mu_0\Hvec$, the electrical field perpendicular to the applied longitudinal potential gradient (electric field ${\Efieldc}^\mathrm{trans}$) can be derived from Eq.\,(\ref{eq:Hall_conductivities_2a}) to 
\begin{equation}
{\Efieldc}_\mathrm{,OHE}^\mathrm{trans} = - \frac{{\sigmac}_\mathrm{H} (\mu_0 \Hvec)}{\sigmac } \; \left[ -\boldsymbol{\nabla} \mucharge / \charge \times \Hhat \right].
\label{eq:Ehall_2}
\end{equation}
Here, we have introduced the unit vector $\Hhat=\Hvec/|\Hvec|$ along the direction of the applied magnetic field, and indicate that ${\sigmac}_\mathrm{H}$ is magnetic field dependent. In addition, we assume vanishing longitudinal as well as transverse temperature gradients, and that the charge carrier scattering time is much shorter than the inverse cyclotron frequency (ordinary Hall effect). \cite{Gross2nd,BookGrosso,BookDavies}
Equation (\ref{eq:Ehall_2}) also introduces the so-called Hall angle $\theta_\mathrm{OHE}={\sigmac}_\mathrm{H} (\mu_0 \Hvec)/\sigmac$.  

For the boundary conditions above, and identifiying the longitudinal   charge current density as ${\Jvec}^\mathrm{long} = \sigmac \left( -\boldsymbol{\nabla} \mucharge/\charge \right)$, we can  rewrite  Eq.\,(\ref{eq:Ehall_2}) as
\begin{eqnarray}
{\Efieldc}_\mathrm{,OHE}^\mathrm{trans} & = & \frac{{\sigmac}_\mathrm{H} (\mu_0 \Hvec) }{\sigmac^2  } \; \left[ \Hhat \times {\Jvec}^\mathrm{long} \right] = \rho_\mathrm{OHE} \; \left[ \Hhat \times {\Jvec}^\mathrm{long} \right]
\nonumber \\
&=& R_\mathrm{OHE} \; \left[ \mu_0\Hvec \times {\Jvec}^\mathrm{long} \right],
\label{eq:Ehall}
\end{eqnarray}
where we have introduced the ordinary Hall coefficient $R_\mathrm{OHE}=\rho_\mathrm{OHE}/(\mu_0 H)$. Note that $\rho_\mathrm{OHE}$ is obtained 
by inverting $\sigmavec$ (Eq.\,(\ref{eq:Hall})) and that this expression considers only linear terms in $\mu_0 H$. 

In a single band picture the Hall coefficient $R_\mathrm{OHE}= \frac{1}{n \charge}$ is determined by the charge carrier density $n$ and the charge carrier type.\citep{BookGrosso,Gross,Kittel,Hurd,Campbell,Ashcroft:1976ud} It is negative for electrons ($\charge=-e$). 

In analogy to Eq.\,(\ref{eq:Ehall}), the anomalous Hall effect and the corresponding anomalous Hall coefficient $R_\textrm{AHE}$ are defined via \citep{OHandley,ReviewSMR,AHEreview}
\begin{equation}
  {\Efieldc}_\mathrm{, AHE}^\mathrm{trans} =  \rho_\mathrm{AHE} \; \left[ \Mhat \times {\Jvec}^\mathrm{long} \right] = R_\textrm{AHE} \, \left[ \mu_0 \Mvec \times{\Jvec}^\mathrm{long} \right],
\end{equation}
where the unit vector $\Mhat$ represents the direction of the magnetization  (or a similar quantity breaking time reversal symmetry). Here, the intrinsic magnetization of the material $\mu_0 \Mvec$ is represented in form of the generalized magnetic field $\Bvec$. Even though the OHE and AHE can be described by very similar equations,  
the microscopic origin of the AHE is non-trivial (see also in the introduction to Sec.\,\ref{sec:TransNew}).\citep{AHEreview,AHEreview2,Nakatsuji.2015,Nayak:2016}  

Considering the geometry depicted in Fig.\,\ref{fig:Hall} with $\Mvec \neq 0$, the transverse anomalous Hall field is directed along the $y$-axis. If, in addition, a magnetic field $\mu_0\Hvec$ is applied collinear with $\Mvec$, the anomalous and ordinary Hall effect can be expressed in the combined Hall resistivity \citep{OHandley,Handbook,IntroductionSpintronics,SmitAHE,AHEreview2}
\begin{equation}
	\rho_\mathrm{H}=R_\mathrm{OHE} \mu_0 H + R_\textrm{AHE} \, \mu_0 M.
\end{equation}

We note that in a magnetic material the driven charge current is always accompanied by a finite spin current as the charge carriers transport charge and spin simultaneously. This results in the simultaneous occurrence of the AHE and the spin Hall effect and we could coin this phenomenon as "spin-dependent" AHE (see Sec.\,\ref{sec:SHE}).\citep{SHEreview,Hirsch} In the same way, as charge carriers also can transport heat,  the occurrence of an (anomalous) Hall effect implies the presence of the (anomalous) Ettingshausen effect (discussed in Sec.\,\ref{sec:Ett}). However, the Hall effect is usually measured under the boundary condition (thermal short circuit) that the transverse temperature gradient vanishes (see above). For the boundary condition $J_\mathrm{h}^\mathrm{trans}=0$, we obtain the so-called adiabatic Hall effect which indeed contains a correction term due to the presence of the  Ettingshausen effect. 

Finally, the spin-orbitronic charge Hall effect is expected to disappear since in a non-magnetic metal the spin-selective deflection of the longitudinal flow of spin-up and spin down charge carriers
result in a vanishing transverse charge current.

\subsubsection{Righi-Leduc or Thermal Hall effect} \label{sec:Righi-Leduc}
\textit{The Righi-Leduc or thermal Hall effect describes the generation of a transverse heat current by a londitudinal temperature gradient in the presence of an external magnetic field (ordinary thermal Hall effect) or finite magnetization (anomalous thermal Hall effect), which is perpendicular to both the applied gradient and the current response. }

The Righi-Leduc/thermal Hall effect describes the thermal analogue of the ordinary and anomalous charge Hall effect. Augusto Righi \citep{Righi} and Anatole Leduc \citep{Leduc} discovered the effect independently in 1887. Initially, Righi and Leduc investigated the effect only in bismuth.\citep{Righi,Leduc} However, it was later observed that an `anomalous' Righi-Leduc effect is present in magnetic materials even in the absence of an applied magnetic field.\citep{Zahn,SmithLeduc} 

We note that a heat current in a solid is not exclusively carried by charge carriers but also by various other entities such as phonons or magnons. \citep{MurakamiReview,BookZiman,Sanders:1977ha} However, for clean metals, the electronic heat conductivity typically dominates the heat transport and thus we will focus on this contribution here.\footnote{This is of course different, e.g.~in paramagnetic insulators, where the heat transport is mostly due to phonons. Even in those systems a thermal Hall effect was observed, the so-called  phonon Hall effect.\citep{PhononHall,PhononHall2,PhononHall3} Similarly, for the heat transport by neutral particles in polyatomic gases, a thermal Hall effect was found and named the Senftleben-Beenakker effect.\citep{Senftleben:1930,Beenakker:1962} Moreover, in magnetic insulators heat can also be  transported by  magnons.\citep{magnonHall,magnonHall2,Katsura,Onose} Also in this case, a magnon-mediated Righi-Leduc effect is obtained. Unfortunately, in literature this effect is sometimes confusingly called 'magnon Hall effect'.\citep{magnonHall,magnonHall2}} 

\begin{figure}[h]
	\includegraphics[width=\textfiguresPerp]{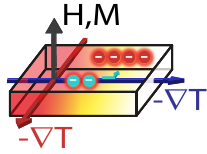} 
    \vspace{\captionhalfclose}
    \caption{\label{fig:RighiLeduc} Schematic representation of the Righi-Leduc effect ($\Hvec$) and anomalous Righi-Leduc effect ($\Mvec$), also known as the thermal Hall effect and anomalous thermal Hall effect, respectively, in open circuit conditions for the transverse heat current. Here, the (anomalous) Righi-Leduc coefficient is assumed to be positive.}
\end{figure}

The ordinary Righi-Leduc effect in metallic systems can be understood intuitively:  An applied longitudinal temperature gradient drives a longitudinal heat current carried by charge carriers. Those are deflected in a direction orthogonal to the applied magnetic field and the heat current direction as a result of the Lorentz force. This transverse flow of charge carriers is accompanied by a transverse heat flow (see Fig.~\ref{fig:RighiLeduc}). 

In experiments it is easier to measure temperature differences than heat currents. Therefore, open boundary conditions ($J_\mathrm{h}^\mathrm{trans} =0$) are used resulting in the build up of a transverse temperature gradient $\boldsymbol{\nabla}T^\mathrm{trans}$. For these boundary conditions, Eq.\,(\ref{eq:Hall_conductivities_2b}) reads

\begin{equation}
\Jvecheat^\mathrm{trans} = 0 = -\kappa \boldsymbol{\nabla} T^\mathrm{trans} + \kappa_\mathrm{H} (\mu_0 \Hvec) \, T \left( -\boldsymbol{\nabla}T^\mathrm{long}/T \times \mathbf{\hat{h}} \right),
\label{eq:ORL01}
\end{equation}
with the magnetic field dependent thermal Hall coefficient $\kappa_\mathrm{H}$ and $\mathbf{\hat{h}}=\Hvec / |\Hvec|$. Resolving for $\boldsymbol{\nabla}T^\mathrm{trans}$ yields
\begin{eqnarray}
\label{eq:Righi}
	 \boldsymbol{\nabla} T^\mathrm{trans} &=& \frac{\kappa_\mathrm{H} (\mu_0 H)}{\kappa} \left[ \Hhat  \times \mathbf{\nabla} T^\mathrm{long} \right] \nonumber\\
	 &=&  R_\mathrm{ORL} \left[ \mu_0\Hvec \times \mathbf{\nabla} T^\mathrm{long} \right],
\end{eqnarray}
where $R_\mathrm{ORL}=\kappa_\mathrm{H}(\mu_0 H)/(\kappa \mu_0 H)$ is the material dependent ordinary Righi-Leduc coefficient, which scales inversely proportional with the thermal conductivity of the material.\citep{Campbell,Gross} This coefficient can be seen as the thermal analogue of the charge Hall coefficient $R_\mathrm{OHE}$.

The anomalous Righi-Leduc effect is observed in solid-state systems with a finite magnetization, such as ferromagnetic metals, and it can be treated completely analogous to its ordinary counterpart by replacing $\mu_0 \Hvec$ with $\mu_0 \Mvec$, leading to \cite{Zahn,SmithLeduc,AnomalousThermal,Li:2017}
\begin{equation}
 	\boldsymbol{\nabla} T^\mathrm{trans} = R_\mathrm{ARL} \left[ \mu_0\Mvec \times \mathbf{\nabla} T^\mathrm{long} \right],
 \end{equation}
where $R_\mathrm{ARL}$ is the anomalous Righi-Leduc coefficient. 

We note that the thermal current is always accompanied by a finite charge current as the charge carriers transport charge and heat simultaneously. This results in the simultaneous occurrence of the (anomalous) thermal Hall effect and the (anomalous) Nernst effect (which is discussed in Sec.\,\ref{sec:Nernst}). 

Finally, we also could expect a spin-orbitronic Righi-Leduc effect in non-magnetic metals. The spin-sensitive transverse deflection of the charge carriers flowing in the longitudinal direction driven by the applied temperature gradient results in a pure transverse spin current. This pure spin current is not accompanied by any transverse flow of charge carriers, however, a finite transverse heat current can be generated in case one spin species carries more heat than the other.

\subsubsection{Pure spin-Hall effect}\label{sec:purespinHalleffect}
\textit{The pure spin-Hall effect describes the generation of a transverse spin current by a longitudinal gradient of the spin-chemical potential in the presence of an external magnetic field (ordinary pure spin Hall effect) or finite magnetization (anomalous pure spin Hall effect), which is perpendicular to both the applied gradient and the current response. }

The pure spin Hall effect is the spin analogue to the charge and thermal Hall effect (see Table\,\ref{tab:trans}), where the response current and the applied gradient of the potential is associated with the same entity. In this spirit, one would therefore expect that the  effect described above should be named the spin Hall effect. Unfortunately, in literature the name spin Hall effect is already used for the spin-galvanomagnetic effect discussed below
(cf. Sec.\,\ref{sec:SHE}), which describes the transverse 
spin current response generated by a longitudinal electric field. Therefore, to discern between the two effects we refer to the effect discussed in this section as the pure spin Hall effect. We  distinguish between the ordinary, the anomalous, and the spin-orbitronic version of the effect as in the previous versions and also  its magnonic version.

\begin{figure}[h]
	\includegraphics[width=\textfiguresPerp]{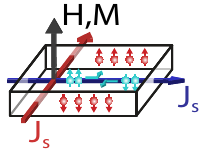} 
    \vspace{\captionhalfclose}
    \caption{\label{fig:RealSHE} Schematic representation of the pure spin Hall effect. Here a negative pure spin Hall coefficient is assumed in analogy to the (Lorentz force-based) Hall effect/Hall coefficient. 
    }
\end{figure}

The ordinary and anomalous pure spin Hall effect  can be understood intuitively as follows: The applied gradient in the spin accumulation, $-\boldsymbol{\nabla}\mu_\mathrm{s}/2q$, drives a longitudinal spin current density ${\Jvecspintilde}^\mathrm{long}={\Jvecspintilde}^\mathrm{\uparrow,long}-{\Jvecspintilde}^\mathrm{\downarrow,long}$ mediated by charge carriers, which are deflected in transverse direction due to the Lorentz force.\footnote{The spin up / down charge carrier density originating form a gradient in the  chemical potentials of the respective chemical potential gradient ${\Jvecspintilde}^{\uparrow,\downarrow}= \sigmac (-\boldsymbol{\nabla}\mu^{\uparrow,\downarrow}/2\charge)=\sigmac (-\boldsymbol{\nabla}(\frac{2 \mucharge\pm\muspin}{4 \charge})$} The generated transverse flow of charge carriers, in turn, is associated with a transverse spin current density ${\Jvecspintilde}^\mathrm{trans}$ (cf. Fig.~\ref{fig:RealSHE}): 
\begin{eqnarray}
{\Jvecspintilde}^\mathrm{trans}&=& \left({\Jvec}^\uparrow- {\Jvec}^\downarrow\right)^\mathrm{trans} \nonumber\\ 
&=& {\sigmac}_\mathrm{H}\left[-\boldsymbol{\nabla}\mucharge^{\mathrm{\uparrow,long}}/\charge \times \Hhat \right]-{\sigmac}_\mathrm{H}\left[-\boldsymbol{\nabla}\mucharge^{\mathrm{\downarrow,long}}/\charge \times \Hhat \right]\nonumber\\
&=& {\sigmac}_\mathrm{H} \left[-\boldsymbol{\nabla}\muspin^{\mathrm{long}}/\charge \times \Hhat \right] \label{eq:purespinHallEq1}
\end{eqnarray}
We see that the Hall coefficient ${\sigmac}_\mathrm{H}$ determines the charge and the  spin current response. This is expected since they result from the deflection of the same carriers (transporting spin and charge) due to the Lorentz force. We point out that for the case of ${\Jvecspintilde}^{\mathrm{\uparrow,long}}=-{\Jvecspintilde}^{\mathrm{\downarrow,long}}$ the transverse charge current density $\Jvec^{\mathrm{trans}}$ vanishes, while ${\Jvecspintilde}^{\mathrm{long}}$ assumes the value of $2{\Jvecspintilde}^{\mathrm{\uparrow,long}}$. Note that Eq.\,(\ref{eq:purespinHallEq1}), exclusively considers the contribution of a gradient in the spin chemical potential to the spin current density. We therefore expect no term mixing charge and spin current densities. For the ordinary pure spin Hall effect, ${\sigmac}_\mathrm{H}$ is the charge Hall conductivity underlining the shared deflection mechanisms in both cases. 
In the same way, the anomalous pure spin Hall effect is obtained in metals with $\Mvec\neq 0$. Here, ${\sigmac}_\mathrm{H}$ assumes the expression expected for charge carriers in a magnetic system. Note that for the anomalous case charge and spin current densities coexist, due to the imbalance of the spin dependent conductivites expressed by $P$ (see Eq.\,\ref{eq:SS_04}). However, Eq.\,(\ref{eq:purespinHallEq1}) does not include this aspect as the equation was derived by restricting the discussion to pure spin transport.

We do not expect any spin-orbitronic pure spin Hall effect in non-magnetic metals. Since spin-up and spin-down charge carriers have the same density, a longitudinal spin current could be established only by the counterflow of spin-up and spin-down charge carriers. Spin-selective transverse deflection would then generate a pure transverse charge current without any accompanied spin current. This is just the inverse spin Hall effect discussed in Sec.\,(\ref{sec:iSHE}). 

\paragraph{Note on the magnon Hall effect}
In materials with finite $\Mvec$ the transport of spin or angular momentum is not exclusive for charge transport. Also the quantized excitations of the spin system (magnons) can transport angular momentum and therefore can  give rise to a Hall effect in the sense of the pure spin Hall effect. In this case, the effect can  be referred to as the magnon Hall effect, as both the drive and the response entities are carried by magnons.\citep{Jin2023}

In 2010, first reports related to the  `magnon Hall effect' were published by Hosho Katsura \textit{et al.}\citep{Katsura} (theory) and Yoshinori Onose  \textit{et al.}\citep{Onose} (theory and experiment). However, Onose states that the magnon Hall effect is analogues to the anomalous Righi-Leduc/thermal Hall effect, as a transverse temperature gradient is created in a magnetic material and in later publications, the `magnon Hall effect' is usually referred to as the thermal Hall effect of magnons.\citep{Katsura,Ideue,Matsumoto,MurakamiReview} 
Therefore, the 'magnon Hall effect' in the viewpoint of these authors is not a pure spin Hall effect. 

One should note that the naming of the `magnon Hall effect' in literature is inconsistent, as magnons are not exclusively transported via temperature gradients, but can also be driven by diffusion or a gradient in the spin chemical potential.\citep{MeijsBachelor} Therefore it would be more to the point to name the `magnon Hall effect' driven by a thermal gradient the `magnon thermal Hall effect'. This would have the benefit that it leaves the name `magnon Hall effect' to describe the generation of a transverse, magnon-mediated spin current due to an applied potential  gradient driving a longitudinal magnon current  (see Table.\,\ref{tab:trans}). This nomenclature would also be the consequent continuation of the names `(charge) Hall effect' and `thermal Hall effect', which describe pure charge and pure heat related transport.

For consistent denotation of other magnon-related effects we suggest to use the established nomenclature given in Table\,\ref{tab:trans} and add `magnon' to the name in order to underline that the transport phenomenon is related to magnons.

\subsection{Spin-galvanomagnetic effect}\label{sec:SHEandiSHE}

In Sec.\,\ref{sec:Hall_Effect}, the transverse response currents caused by longitudinal gradients of their corresponding potential (i.e. the diagonal elements $L^\perp_{ii}$ from Eq.\,(\ref{eq:GeneralTransportCoefficients_transverse})) were discussed. We will next focus on the effects related to the off-diagonal elements $L^\perp_{13}=-L^\perp_{31}$, which relate a charge current (or a gradient in the electrochemical potential) to the generation of a transverse spin current (and vice versa) when a generalized magnetic field $\Bvec$ is applied.

The scenario where the role of the generalized magnetic field is taken by the spin polarization ($\sigmapol$) is well known in literature as the (inverse) spin Hall effect.\citep{SHEreview} However, as discussed in Sec.\,\ref{sec:purespinHalleffect}, this naming is misleading in the sense that the effect converts between two entities and does not describe the deflection of a pure spin current. Nevertheless, to not confuse the reader, we will adopt this slightly illogical denotation in the following subsections. However, as a general name for the transverse transport phenomena connecting the entities charge and spin we introduce the name `spin-galvanomagnetic effect', as a continuation of the historical name `galvanomagnetic effects' represented by the first column of Table\,\ref{tab:trans}.

In analogy to Eqs.~(\ref{eq:Hall_conductivities_2a})-(\ref{eq:Hall_conductivities_2c}), the equations focusing on the coupled charge and spin current transport can be deduced from Eq.\,(\ref{eq:GeneralTransportCoefficients_transverse}) and expressed as

\begin{widetext}
\begin{eqnarray}
    \Jvec &=&   L_{11} \left( -\boldsymbol{\nabla} \mucharge/\charge \right) 
    + L_{13} \left( -\boldsymbol{\nabla} \muspin/2\charge \right) 
    +L^\perp_{11} \left( -\boldsymbol{\nabla} \mucharge/\charge \times \Bhat \right)  
    +  L^\perp_{13} \left( -\boldsymbol{\nabla} \muspin/2\charge \times \Bhat \right)
     \; \label{eq:spin_Hall_a}\\
      \Jvecspin &=& L_{33} \left( -\boldsymbol{\nabla} \muspin/2\charge \right) + L_{31} \left( -\boldsymbol{\nabla} \mucharge/\charge \right)  +  L^\perp_{33} \left( -\boldsymbol{\nabla} \muspin/2\charge \times \Bhat \right) +L^\perp_{31} \left( -\boldsymbol{\nabla} \mucharge/\charge \times \Bhat \right)\; ,
    \label{eq:spin_Hall_b}
\end{eqnarray}
\end{widetext}
where we focus on the impact of the gradients of the charge and spin chemical potentials, assuming the absence of thermal gradients. For open boundary conditions, the current densities above translate into spin and charge accumulations, where the latter are frequently used for spin-to-charge current conversion applications. 

\subsubsection{Spin-galvanomagnetic effect (Spin Hall effect)}\label{sec:SHE}
\textit{The spin-galvanomagnetic effect describes the generation of a transverse spin current by an applied longitudinal gradient in the electrochemical potential in the presence of a generalized magnetic field which is perpendicular to both the applied gradient and the current response.} 


The conceptual idea of a current induced spin accumulation was developed in 1971 by Michel Dyakonov and Vladimir Perel \citep{DyakonovPerel1,DyakonovPerel2} based on the idea of asymmetric Mott scattering.\citep{Mott:1929} The effect was explained by spin-orbit effects in the scattering of charge carriers. Many years later, in 1999, Jorge Hirsch separately developed the same idea and proposed that a longitudinal charge current in a paramagnetic metal generates a transverse spin imbalance.\citep{Hirsch} Within this concept one can distinguish between intrinsic and extrinsic mechanisms.\citep{Hirsch,Zhang:2000,Sinova:2004,Murakami:2003,Shchelushkin:2005,SHEreview, Maekawa:2017gn,BookTakahashi} Later this effect was experimentally demonstrated in semiconductors \citep{SHE1,SHE2} and  metals.\citep{SHE-metals2, SHEmetal,SHE-metals}  Nowadays, there are hundreds of experimental and theoretical reports referring to this effect as the spin Hall effect (SHE): A detailed review of the SHE is given in Ref.\, \citep{SHEreview}.

As for the previously discussed transverse effects, we can distinguish three scenarios depending on the origin of the generalized magnetic field. In Sec.\,\ref{par:AHBE} we will discuss the `ordinary' and `anomalous' spin-galvanomagnetic effect, followed by Sec.\,\ref{par:SHE}, where we focus on the `spin-orbitronic' spin-galvanomagnetic effect, which in literature is also known as the spin-Hall effect. 

\paragraph{(Anomalous) spin-galvanomagnetic effect} \label{par:AHBE}
The ordinary spin-galvanomagnetic effect describes the generation of a transverse spin current by an applied longitudinal gradient in the electrochemical potential (i.e. electric field) in the presence of a perpendicularly oriented magnetic field $\mu_0 \Hvec$. In contrast to Eq.\,(\ref{eq:purespinHallEq1}), where the transverse spin current density originates from a longitudinal gradient in the spin chemical potential, we now consider a longitudinal gradient in the electrochemical potential as the origin. That is, the resulting transport phenomenon is related to spin-electric coupling. The longitudinal spin current density caused by a longitudinal electric field $\Efieldc = \mingradmuc /q$ can  therefore be written as $\Jvecspintilde=\sigmac P (-\boldsymbol{\nabla}\mucharge/\charge$). The transverse spin current in turn is then given by \begin{eqnarray}
{\Jvecspintilde}^\mathrm{trans} = - \theta_\mathrm{OHE} \sigmac P  \left[  \Hhat  \times \boldsymbol{\nabla}\mucharge/\charge \right].\label{eq:spin_galvanic_effect_01}
\end{eqnarray}
Note that Eq.\,(\ref{eq:spin_galvanic_effect_01}) shares the Hall angle with the ordinary charge Hall effect, as the charge carrier deflection resulting in the transverse spin current originates from the Lorentz force. The presence of $P$ in Eq.\,(\ref{eq:spin_galvanic_effect_01}) indicates that a finite difference in the conductivity of the two spin channels is required to observe the ordinary spin-galvanomagnetic effect. Therefore, for a non-magnetic material with a vanishing net spin polarization of the charge carriers we expect this effect to vanish. However, in materials with $\Mvec \neq 0$, the ordinary spin-galvanomagnetic effect is naturally present and can be understood in terms of the finite spin polarization of the charge carriers at the Fermi energy being deflected due to the ordinary Hall effect.  

Moreover, even for $\mathbf{H}=0$ for magnetically ordered materials with finite $\Mvec$, we can rewrite Eq.\,(\ref{eq:spin_galvanic_effect_01}) to describe the anomalous spin-galvanomagnetic effect:
\begin{eqnarray}
\Jvecspintilde^\mathrm{trans} &=& - \theta_\mathrm{ASGME}\,\sigmac P \; \left[ \Mhat \times \boldsymbol{\nabla}\mucharge/\charge \right] \; .
\label{eq:spin_galvanic_effect_05}
\end{eqnarray}
Here, $\Mhat$ is the unit vector parallel to $\Mvec$ and the angle $\theta_\mathrm{ASGME}$ depends on the details of the involved microscopic mechanisms.
More details about possible detection schemes can be found in Ref.\,\citep{Wang:2021}. Experiments demonstrated that the anomalous spin Hall effect can be used to study  magnon transport effects in magnetic insulators as well as angular momentum transport non-conducting materials   \citep{Das:2017bc, schlitz:2024}

\begin{figure}[h]
	\includegraphics[width=\textfiguresPerp]{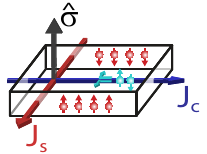} 
    \vspace{\captionhalfclose}
    \caption{\label{fig:SHE} Schematic representation of the spin-Hall effect (or spin-orbitronic spin-galvanomagnetic effect). Here shown for a positive spin-Hall angle. The accumulation of spin polarized electrons at either side of the material indicate open boundary conditions in the transverse direction.}
\end{figure}

\paragraph{Spin-orbitronic spin-galvanomagnetic effect (spin Hall effect)} \label{par:SHE}
The probably most interesting case is the spin-orbitronic spin-galvanomagnetic effect in non-magnetic metals, which in literature is well known as the spin Hall effect (SHE) and will be referred to as such in the rest of this section. Like the AHE, the SHE hinges on spin-orbit interactions and can have both intrinsic and extrinsic origins.\citep{SHEreview} In a simple picture, electrons are deflected depending on their spin configuration, whereby the direction of the deflection is opposite for spin-up and spin-down electrons. This results in a pure spin current ($\Jvec^{\uparrow,\mathrm{trans}} = - \Jvec^{\downarrow,\mathrm{trans}}$) and ultimately, assuming open boundary conditions in the transverse direction, in an accumulation of spin-up and spin-down electrons on either side of the material (see Fig.\,\ref{fig:SHE}).\citep{bookHans,SHEreview,SHE1,SHE2}  The generated spin-orbitronic transverse spin current density is given by
\begin{eqnarray}
\Jvecspintilde^\mathrm{trans} & = &  - \theta_\mathrm{SHE} \left( \sigmac^\uparrow +  \sigmac^\downarrow \right) \; \left(\boldsymbol{\hat{\sigma}} \times \boldsymbol{\nabla}\mucharge/q \right)
\nonumber \\
& = & - \theta_\mathrm{SHE}\,  \sigmac \; \left(\boldsymbol{\hat{\sigma}} \times \boldsymbol{\nabla}\mucharge/q \right)\; ,
   \label{eq:spin_galvanic_effect_06}
\end{eqnarray}
where $\boldsymbol{\hat{\sigma}}$ is the unit vector parallel to the spin direction and the angle $\theta_\mathrm{SHE}$ is the spin Hall angle, describing the strength of the spin-dependent transverse deflection. As mentioned before, we want to emphasize that $\Jvecspintilde$ is defined in units of the charge current density and that $\Jvecspintilde=\frac{2\charge}{\hbar}{\Jvecspin}$ is used to convert $\Jvecspintilde$ to an angular momentum current density $\Jvecspin$. Converting $\Jvecspintilde$ to $\Jvecspin$ and expressing the gradient of the electrochemical potential by the  charge current density ($\Jvec=-\sigmac \boldsymbol{\nabla}\mucharge/q$) Eq.\,(\ref{eq:spin_galvanic_effect_06}) yields 
\begin{equation}
    \Jvecspin^\mathrm{trans} = \frac{\hbar}{2\charge} \theta_\mathrm{SHE} \left( \sigmapol \times \Jvec \right).
\end{equation}

The magnitude and sign of $\theta_\mathrm{SHE}$ depend on the involved microscopic mechanisms. Since spin-orbitronic effects are based on spin-orbit-interactions, which scale proportional to the atomic number to the power of 4 ($Z^4$), large $\theta_\textrm{SHE}$ of the order of $0.01 - 0.3$ are found for heavy metals like platinum, tantalum and tungsten or Bi$_2$Se$_3$.\citep{SHE-metals,SHE-metals2,bookHans, Leiva:2021, Lu:2022} Interestingly, $\theta_\textrm{SHE}$ is positive for platinum, whereas it is negative for tantalum and tungsten.\citep{bookHans} Moreover, the SHE also can be present in magnetically ordered materials.\citep{Das:2017bc, Wimmer:2019ho}

\subsubsection{Electro-spinmagnetic effect (inverse spin Hall effect)}\label{sec:iSHE}
\textit{The electro-spinmagnetic effect describes the appearance of a transverse charge current by an applied longitudinal  gradient in the spin-chemical potential in the presence of a generalized magnetic field which is perpendicular to both the applied gradient and the current response.} 

First experimental proof of the spin-orbitronic electro-spinmagnetic effect (in literature well known as the inverse spin Hall effect (ISHE)) was given in 1984 by A. Bakun and coworkers, \citep{ISHEfirst} who realized an experiment proposed  in 1983 by Nikita Averkiev and coworkers \citep{ISHEfirst2} by using circularly polarized light to excite a pure spin current in a direct-band semiconductor. The ISHE turned out to be a powerful tool to electrically detect pure spin currents, which not only can be optically excited \citep{OpticalISHE}, but also for example generated by spin pumping \citep{Urban:2001hr,Heinrich:2003en,Tserkovnyak:2002bm,Tserkovnyak:2002ju,Tserkovnyak:2005fr,bookHans,SaitohAPL2006,ISHEspinpumping,ISHEspinpumping2,CzeschkaPRL,WoltersdorfBack,MosendzPRL,CastelVlietstra,CastelVlietstra2,Maekawa:2017gn,Mizukami:2002hh} or the spin(-dependent) Seebeck effect \citep{SSEUchida,Uchida1,Weiler-Thermal-Landscape,Gepraegs-SSE,Vlietstra-LSSE,SchreierSSE}
(see Sec.\,\ref{subsec:SpindependentSeebeckEffect}).

The electro-spinmagnetic effect can be viewed as the reciprocal effect of the spin-galvanomagnetic effect, as is shown in Table\,\ref{tab:trans},  originating also from spin-orbit interactions \citep{SHEreview,bookHans,Hirsch,SchreierThesis,SchreierSign,VlietstraThesis}. 
In analogy to the spin-galvanomagnetic effect, the electro-spinmagnetic effect appears when a generalized magnetic field is present perpendicular to the applied gradient in the spin chemical potential. 

Again, we can distinguish three scenarios depending on the origin of the generalized magnetic field. In Sec.\,\ref{par:IHBE} we will discuss the `ordinary' and `anomalous' electro-spinmagnetic effect, followed by the `spin-orbitronic' electro-spinmagnetic effect, which in literature is also known as the inverse spin-Hall effect, in Sec.\,\ref{par:ISHE}. 

\paragraph{(Anomalous) electro-spinmagnetic effect.} \label{par:IHBE}
The ordinary electro-spinmagnetic effect describes the generation of a transverse charge current by an applied longitudinal gradient in the spin chemical potential (i.e. a spin current flow) in the presence of a perpendicularly oriented magnetic field $\mu_0 \Hvec$.
Here, the longitudinal charge current density due to the applied longitudinal spin-chemical potential gradient is given by
\begin{eqnarray}
\Jvec^\mathrm{long} & = & \Jvec^{\uparrow,\mathrm{long}} + \Jvec^{\downarrow,\mathrm{long}} = \sigmac P \; \left(-\boldsymbol{\nabla}\muspin/2\charge\right).
    \label{eq:spin_galvanic_effect_07}
\end{eqnarray}
The corresponding transverse deflected charge current is 
\begin{eqnarray}
\Jvec^\mathrm{trans} & = &  - \theta_\mathrm{OHE} \sigmac P \; \left(\Hhat \times \boldsymbol{\nabla}\muspin/2\charge \right) \; ,
   \label{eq:spin_galvanic_effect_08}
\end{eqnarray}
where $\Hhat$ is the unit vector parallel to $\Hvec$. Similar to the ordinary spin-galvanomagnetic effect, the spin deflection is governed by the ordinary charge Hall angle $\theta_\mathrm{OHE}$ and a material with finite $P$ is required to observe the effect. 

Correspondingly, in magnetically ordered materials with finite $\Mvec$, we expect an inverse anomalous spin-galvanomagnetic effect, 
\begin{eqnarray}
\Jvec^\mathrm{trans} &=&  - \theta_\mathrm{ASGME} P \sigmac \left(\Mhat \times \boldsymbol{\nabla}\muspin/2\charge \right) \; ,
\label{eq:spin_galvanic_effect_09}
\end{eqnarray}
where $\Mhat$ is the unit vector parallel to $\Mvec$ and the angle $\theta_\mathrm{ASGME}$ depends on the details of the involved microscopic mechanism. 

\begin{figure}[h]
	\includegraphics[width=\textfiguresPerp]{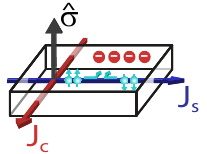} 
    \vspace{\captionhalfclose}
    \caption{\label{fig:ISHE} Schematic representation of the electro-spinmagnetic effect usually named inverse spin Hall effect. Here drawn for a positive spin-Hall angle.} 
\end{figure}

\paragraph{Spin-orbitronic electro-spinmagnetic effect (inverse spin Hall effect).} \label{par:ISHE}
The spin-orbitronic electro-spinmagnetic effect in non-magnetic metals, is in literature well known as the inverse spin Hall effect (ISHE) and will be referred to as such in the rest of this subsection. A pure ISHE is obtained in the situation where $\Jvec^{\uparrow,\mathrm{long}} = - \Jvec^{\downarrow,\mathrm{long}}$, resulting in a vanishing longitudinal charge current and pure spin current, as is schematically shown in Fig.\,\ref{fig:ISHE}. In this case the generated transverse charge current is given by 
\begin{eqnarray}
\Jvec^\mathrm{trans} & = &  - \theta_\mathrm{SHE} \,\sigmac \; \left(\boldsymbol{\hat{\sigma}} \times \boldsymbol{\nabla}\muspin/2\charge \right)\; .
   \label{eq:spin_galvanic_effect_10}
\end{eqnarray}
Here, $-\sigmac \boldsymbol{\nabla}\muspin/2\charge$ is often identified with $\Jvecspintilde$, which is defined in units of the charge current density. Using this definition, Eq.\,(\ref{eq:spin_galvanic_effect_10}) reads 
\begin{equation}
	\Jvec^\mathrm{trans}= \theta_\textrm{SHE}\, \left( \sigmapol \times \Jvecspintilde \right).
\end{equation}
Alternatively, if the spin current is written as angular momentum flow density, Eq.\,(\ref{eq:spin_galvanic_effect_10}) takes the form 
\begin{equation}
	\Jvec^\mathrm{trans}= \frac{2\charge}{\hbar}\theta_\textrm{SHE}\, \left( \sigmapol \times \Jvecspin \right).
\end{equation}

The combination of the SHE and ISHE in a single device, gives rise to more intricate phenomena, such as the in 2013 discovered spin Hall magnetoresistance (SMR),\citep{TheorySMR,BauerSMR,VlietstraSMR,AlthammerSMR,HahnSMR,ReviewSMR} and allows to study magnon transport in magnetic insulators,\citep{LudoMMR,GoennenweinMMR,LudoMMRtheory,Lebrun:2018kt,Wimmer:2020hh,Kamra:2020dm} and even control magnon flow \cite{Cornelissen:2018bh, Wimmer:2019gr}.

Finally, we note that the simple sketches presented in Figs. \ref{fig:ISHE} and \ref{fig:SpinEtt} suggest a simultaneous occurrence of the ISHE and the spin-Ettingshausen effect (see also Sec.\,\ref{sec:thermospinmagneticeffect}) (assuming a finite spin-Hall and spin-Nernst angle as well as a finite Seebeck coefficient).

\subsection{Nernst and Ettingshausen effect}\label{sec:NernstEtt}

So far, we have discussed the transverse transport phenomena based on the same transport entities (Sec.\,\ref{sec:Hall_Effect}), as well as the coupling between the entities charge and spin (Sec.\,\ref{sec:SHEandiSHE}).
This section is dedicated to the coupling between the entities charge and heat, which gives rise to transverse transport phenomena represented by the off-diagonal matrix-elements $L^\perp_{12}(\Bvec)=L^\perp_{21}(-\Bvec)$ in Eq.\,(\ref{eq:GeneralTransportCoefficients_transverse}) and can be found in Table~\ref{tab:trans} as the Nernst and Ettingshausen effect.

In analogy to Eqs.~(\ref{eq:spin_Hall_a}) and (\ref{eq:spin_Hall_b}), the equations focusing on the coupled charge and heat current transport can be expressed as
\begin{widetext}
\begin{eqnarray}
    \Jvec & = & L_{11} \left( -\boldsymbol{\nabla} \mucharge/q \right) + L_{12}\left( -\boldsymbol{\nabla}T/T \right) + L^\perp_{11} \left( -\boldsymbol{\nabla} \mucharge/q \times \Bhat \right) + L^\perp_{12} \left( -\boldsymbol{\nabla}T/T \times \Bhat  \right) \;
    \label{eq:Nernst_Ettingshausen_a}\\
    \mathbf{J}_\mathrm{h} & = & L_{22}\left( -\boldsymbol{\nabla}T/T \right) + L_{21} \left( -\boldsymbol{\nabla} \mucharge/q \right) + L^\perp_{22} \left( -\boldsymbol{\nabla}T/T \times \Bhat \right) +  L^\perp_{21} \; \left( -\boldsymbol{\nabla} \mucharge/q \times \Bhat \right) \; ,
    \label{eq:Nernst_Ettingshausen_b}
\end{eqnarray}
\end{widetext}
assuming no gradients in the spin chemical potential are present. 

Historically, transport coefficients have been introduced with an experimental perspective,  as setups typically impose the charge current density and temperature gradient via external circuits, while the resulting electric field and heat current density are measured. This leads to \citep{Gross}
\begin{widetext}
\begin{eqnarray}
    \Efieldc & = & \sigmac^{-1} \Jvec + S \boldsymbol{\nabla}T - \frac{{\sigmac}_\mathrm{H} }{\sigmac^2}  \left( \Jvec \times \Bhat \right)  - N_\mathrm{c} \left|\Bvec\right|\left( \boldsymbol{\nabla}T \times \Bhat \right) \;
    \label{eq:Nernst_Ettingshausen_a1}\\
    \Jvecheat & = & \Pi \Jvec - \kappa \boldsymbol{\nabla}T - \kappa P_\mathrm{c}  \left|\Bvec\right|\left( \Jvec \times \Bhat \right) - \kappa_\mathrm{H} \left( \boldsymbol{\nabla}T \times \Bhat \right) \; ,
    \label{eq:Nernst_Ettingshausen_b1}
\end{eqnarray}
\end{widetext}
where $P_\mathrm{c}$ and $N_\mathrm{c}$ represent 
the Ettingshausen and Nernst coefficient, respectively. The relation between these coefficients and the transport coefficients $L_{ij}$ and $L^\perp_{ij}$ can be derived by elementary algebra. Higher order effects in $\Bvec$ are neglected.

\subsubsection{(Anomalous) Nernst Effect}
\label{sec:Nernst}
\textit{The (anomalous) Nernst effect describes the appearance of a transverse charge current  (or an electrical field for open circuit conditions) due to an applied longitudinal temperature gradient in the presence of an external magnetic field (ordinary Nernst effect) or internal magnetization (anomalous Nernst effect) which is perpendicular to both the applied longitudinal gradient and the transverse response.}

In the same period when Righi and Leduc discovered the thermal Hall effect (1887), 
Albert von Ettingshausen and his student Walther Nernst discovered the Nernst effect, which sometimes also is called Nernst-Ettingshausen effect.\citep{Nernst,Nernst2} Nernst and Ettingshausen also studied the Nernst effect in magnetic materials, revealing the effect now known as the anomalous Nernst effect (ANE).\citep{Nernst,Nernst2,Li:2017}

\begin{figure}[h]
	\includegraphics[width=\textfiguresPerp]{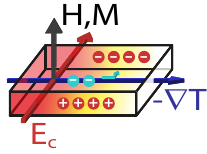} 
    \vspace{\captionhalfclose}
    \caption{\label{fig:Nernst} Schematic representation of the Nernst effect ($\Hvec$) and anomalous Nernst effect ($\Mvec$), in open circuit conditions, assuming a positive (anomalous) Nernst coefficient.}
\end{figure}

Historically, according to the usually used experimental configuration, the ordinary Nernst coefficient $N_\mathrm{ONE}$ has been introduced as the proportionality constant between the measured transverse electric field ${\Efieldc}_\mathrm{,ONE}$ and the applied longitudinal temperature gradient (see Fig.\,\ref{fig:Nernst}). Assuming that no net charge current flows in the longitudinal and transverse direction (electric open circuit, ${\Jc}_{,x}={\Jc}_{,y}=0$) and short-circuited transverse thermal boundary conditions (no transverse temperature gradient), the total electric field, according to Eq.\,(\ref{eq:Nernst_Ettingshausen_a1}) can be expressed as \citep{Gross}
\begin{equation}
{\Efieldc}_{\mathrm{,ONE}} = S \boldsymbol{\nabla}T + N_\mathrm{ONE} \; \left[ \mu_0 \Hvec \times \boldsymbol{\nabla} T \right] \; . 
\label{eq:Nernst_Ettingshausen_03}
\end{equation}
With $-\boldsymbol{\nabla}T$ oriented along the $x$-direction and $\mu_0 \Hvec$ along the $z$-direction, the resulting transverse Nernst field along the $y$-direction, under the boundary conditions $\partial T/\partial y = 0$ (thermal short-circuited), can be deduced from Eq.\,(\ref{eq:Nernst_Ettingshausen_03}), resulting in\citep{Gross,Seeger}
\begin{equation}
{\Ec}_{\mathrm{,ONE},y} = N_\mathrm{ONE} \, \mu_0 H \, \frac{\partial T}{\partial x} \:.
  \label{eq:Nernst_Ettingshausen_04}
\end{equation}
Note that $N_\mathrm{ONE}$ can be regarded as a measure of the transverse heat or entropy flow associated with a longitudinal charge particle flow.\citep{Behnia2,Gross,Seeger} Since in the temperature gradient positive and negative charge carriers flow in the same direction but are deflected by the Lorentz force in opposite direction, the sign of the Nernst field is independent of the charge carrier type (i.e. whether positive or negative charge carriers are transported).\cite{BehniaBook,Gross,Seeger}

Like the anomalous Hall effect (AHE), the anomalous Nernst effect (ANE) is governed by the internal magnetization of the material rather than an externally applied magnetic field.\cite{Gross} Therefore, it is not surprising that both effects share a common microscopic origin. In detail, their origin can be explained by the presence of spin-orbit interactions causing (extrinsic) skew-scattering and side-jump events, as well as by the (intrinsic) Berry phase.\citep{Schmid,AHEreview,ANE_AFM,ANEHeremans,ANE3,ANE4}

Completely analogous to the ordinary Nernst effect, for the anomalous Nernst effect we obtain \citep{ANEeq,Weiler-Thermal-Landscape,SakurabaANE, Reichlova:2019, Ikhlas:2017}
\begin{equation}
{\Efieldc}_{\mathrm{,ANE}} = S \boldsymbol{\nabla}T + N_\mathrm{ANE}\, \left[ \mu_0 \Mvec \times \boldsymbol{\nabla} T \right] \; ,
\label{eq:Nernst_Ettingshausen_05}
\end{equation}
leading to
\begin{equation}
{\Ec}_{\mathrm{,ANE},y} = N_\mathrm{ANE} \, \mu_0 M \, \frac{\partial T}{\partial x} \:
\label{eq:Nernst_Ettingshausen_06}
\end{equation}
under the same boundary conditions.

Note that although the net charge current flow vanishes for the chosen boundary conditions, a heat current still can be observed, as charge carriers with higher energy or temperature move on average in the opposite direction compared to charge carriers with lower energy. Therefore, in addition to the presence of ${\Ec}_{\mathrm{,ANE},y}$, heat current or temperature gradient can arise depending on the chosen boundary conditions. Thus, the (anomalous) Nernst effect occurs simultaneously with the (anomalous) Righi-Leduc/thermal Hall effect (Sec.\,\ref{sec:Righi-Leduc}).

Finally, we do not expect a pure spin-orbitronic Nernst effect in non-magnetic metals. The spin-selective transverse deflection of the charge carriers flowing in the longitudinal direction driven by the applied temperature gradient results in a pure transverse spin current. Since this pure spin current is not accompanied by any transverse charge flow, no electric Nernst field is expected. 

\subsubsection{(Anomalous) Ettingshausen effect}\label{sec:Ett}
\textit{The (anomalous) Ettingshausen effect describes the appearance of a transverse heat current  (a temperature gradient for open circuit conditions) due to a longitudinal gradient in the electrochemical potential (a charge current) in the presence of an external magnetic field (ordinary Ettingshausen effect) or internal magnetization (anomalous Ettingshausen effect) which is perpendicular to both the applied longitudinal gradient and the transverse response. }

In 1886, Walther Nernst and Albert von Ettingshausen also discovered the Ettingshausen effect,\citep{Ettingshausen} the inverse of the Nernst effect\citep{Campbell} (See Table\,\ref{tab:trans}). As for the other thermomagnetic and galvanomagnetic effects, this effect was also studied in magnetic materials,\citep{Ettingshausen} where the `anomalous' behavior was observed, which could not be explained with the same theory as for non-magnetic materials.\citep{Butler}

\begin{figure}[h]
 	\includegraphics[width=\textfiguresPerp]{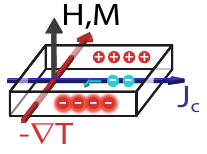} 
     \vspace{\captionhalfclose}
     \caption{\label{fig:Ettingshausen} Schematic representation of the Ettingshausen ($\Hvec$) and anomalous Ettingshausen effect ($\Mvec$), in open circuit conditions, assuming a positive (anomalous) Ettingshausen coefficient. }
 \end{figure}

In analogy to the Peltier coefficient, $\Jvecheat=\Pi \Jvec$, the ordinary Ettingshausen coefficient $P_\mathrm{OEE}$ can be introduced as the proportionality constant between the transverse heat current and the applied longitudinal electrical current. According to Eq.\,(\ref{eq:Nernst_Ettingshausen_b1}), assuming that there is no longitudinal temperature gradient (thermal short-circuited),  the total heat current can be expressed as \cite{Gross}
\begin{equation}
\Jvecheat = \Pi \Jvec + \kappa P_\mathrm{OEE} \; \left[ \mu_0 \Hvec \times \Jvec \right] \; .
\label{eq:Nernst_Ettingshausen_07}
\end{equation}
Figure\,\ref{fig:Ettingshausen} shows $\Jvec$ oriented along the $x$-direction and $\mu_0 \Hvec$ along the $z$-direction. Then, the transverse temperature gradient along the $y$-direction is obtained with $J_{h,y}=\kappa (\partial T/\partial y)$ under the transverse boundary conditions ${\Jc}_{,y}=0$ (electric open circuit) as \citep{Gross,Seeger}
\begin{equation}
\left(\frac{\partial T}{\partial y} \right)_\mathrm{OEE} = P_\mathrm{OEE} \, \mu_0 H \, {\Jc}_{,x}  \; .
\label{eq:Nernst_Ettingshausen_08}
\end{equation}
For simple metals such as Cu, Ag or Au, $P_\mathrm{OEE} \sim 10^{-16}\tfrac{\mathrm{K}\cdot \mathrm{m}}{\mathrm{T}\cdot \mathrm{A}}$ is small and therefore the Ettingshausen effect is difficult to measure. It is particularly large for compensated metals such as bismuth.\cite{Behnia2} There, positive and negative charge carriers flow in opposite directions and cancel their contribution to the longitudinal charge current. They are, however, (charge-sensitively) deflected in the same transverse direction by the Lorentz force, leading to a large transverse heat current.

For the anomalous Ettingshausen effect (AEE) we obtain analogously
\begin{equation}
\Jvecheat = \Pi \Jvec + \kappa P_\mathrm{AEE}  \; \left[ \mu_0\Mvec \times \Jvec \right] \; .
\label{eq:Nernst_Ettingshausen_09}
\end{equation}
leading to
\begin{equation}
\left(\frac{\partial T}{\partial y} \right)_\mathrm{AEE} = P_\mathrm{AEE} \, \mu_0M \, {\Jc}_{,x}  \; ,
\label{eq:Nernst_Ettingshausen_10}
\end{equation}
under the same boundary conditions. Experiments have shown the occurrence of the anomalous Ettingshausen effect in FePt thin films.\citep{AEE_Uchida}

Finally, the spin-orbitronic Ettingshausen effect is expected to disappear since in a non-magnetic metal the spin-selective deflection of the longitudinal flow of spin-up and spin-down charge carriers with the same density results in a pure transverse spin current (SHE, see Sec.\,\ref{sec:SHE}). Since this pure spin current is not accompanied by any net transverse flow of charge carriers, no charge carrier assisted heat current is expected. As for the Righi-Leduc effect, a finite transverse heat current could only be observed, if the two spin species carry different amounts of heat.

Finalizing this subsection we note that due to Onsager's reciprocity relation ($L^\perp_{ij}(\Bvec) = L^\perp_{ji}(-\Bvec)$)\citep{Onsager,Onsager2} the Nernst and Ettingshausen coefficients are related to each other via the Bridgman relation\citep{Bridgman,Behnia2,BehniaBook}
\begin{equation}
P_\mathrm{OEE,AEE} =  N_\mathrm{OEE,AEE} \, T / \kappa \; ,
\label{eq:Nernst_Ettingshausen_11}
\end{equation}
in the same way as the Peltier and Seebeck coefficient are related via the Thomson (Kelvin) relation Eq.\,(\ref{eq:thomsonrelation}).

%
\subsection{Spin-related counterparts of the Nernst and Ettingshausen effect}

The fact that moving charge carriers also transport two other entities, namely spin and heat, results in spin-thermo-electric coupling. In the presence of a generalized magnetic field, the coupling of spin and heat transport results in transverse spin-thermomagnetic effects. These spin-thermomagnetic effects are the spin-related counterparts of the charge-related Nernst- and Ettingshausen effect. All thermomagnetic effects are listed in the middle column of Table\,\ref{tab:trans}. The spin-thermomagnetic effects are represented by the off-diagonal matrix-elements $L^\perp_{23}(\Bvec)=L^\perp_{32}(-\Bvec)$ in Eq.\,(\ref{eq:GeneralTransportCoefficients_transverse}) and are known as the spin-Nernst and the spin-Ettingshausen effect. 

As for the collinear spin-Seebeck and spin-Peltier effect, it is appropriate to add the prefix `dependent' to the name of the transverse effects, whenever the quantity spin is transported by charge carriers and, hence, spin- and charge-related effects cannot be separated. That is, the addition of `dependent' in the name of the effect (e.g. spin-dependent Nernst effect) implies a charge-carrier related approach (as we use in this article). In contrast, leaving this prefix out (e.g. spin Nernst effect) means that the effect originates from pure spin transport without associated charge transport. This is for example the case for spin transport by magnons in magnetic insulators.\citep{BauerCaloritronics} Unfortunately, in literature, this naming is not consistently adopted.

In analogy to Eqs.~(\ref{eq:Nernst_Ettingshausen_a}) and (\ref{eq:Nernst_Ettingshausen_b}) the relation between the spin and heat current can be obtained from Eq.\,(\ref{eq:GeneralTransportCoefficients_transverse}) and expressed as

\begin{widetext}
\begin{eqnarray}
    \Jvecspin &=& L_{33} \left( -\boldsymbol{\nabla} \muspin/2q \right) +  L_{32} \left( -\boldsymbol{\nabla}T/T \right)  +  L^\perp_{33} \left( -\boldsymbol{\nabla} \muspin/2q \times \Bhat \right)+ L^\perp_{32} \left( -\boldsymbol{\nabla}T/T \times \Bhat \right)\;
    \label{eq:spin_Nernst_Ettingshausen_a}\\
    \Jvecheat &=& L_{22} \left( -\boldsymbol{\nabla}T/T \right)  
    + L_{23} \left( -\boldsymbol{\nabla} \muspin/2q \right) 
    + L^\perp_{22} \left( -\boldsymbol{\nabla}T/T \times \Bhat \right)
    +  L^\perp_{23} \left( -\boldsymbol{\nabla} \muspin/2q \times \Bhat \right) \; ,
    \label{eq:spin_Nernst_Ettingshausen_b}
\end{eqnarray}
\end{widetext}
where we focus on the perturbation caused by an applied temperature gradient and the gradient in the spin chemical potential, while we assume a vanishing electrochemical potential. In analogy to Eqs.~(\ref{eq:Nernst_Ettingshausen_a1}) and (\ref{eq:Nernst_Ettingshausen_b1}), the spin field $\mathbf{\tilde{E}}_\mathrm{s}$ and the heat current density $\mathbf{J}_\mathrm{h}$ can be expressed as 
\begin{widetext}
\begin{eqnarray}
    \mathbf{\tilde{E}}_\mathrm{s} & = & \sigmac^{-1} \mathbf{\tilde{J}}_\mathrm{s} - \frac{{\sigmac}_\mathrm{H} }{\sigmac^2}  \left( \mathbf{\tilde{J}}_\mathrm{s} \times \Bhat \right) + S_\mathrm{s} \boldsymbol{\nabla}T - N_\mathrm{s} \left( \boldsymbol{\nabla}T \times \Bvec  \right) \; .
    \label{eq:spin_Nernst_Ettingshausen_a1}\\
    \mathbf{J}_\mathrm{h} & = & \Pi_\mathrm{s} \mathbf{\tilde{J}}_\mathrm{s} - \kappa \boldsymbol{\nabla}T - \kappa P_\mathrm{s}  \left( \mathbf{\tilde{J}}_\mathrm{s} \times \Bvec  \right) - \kappa_H \left( \boldsymbol{\nabla}T \times \Bhat  \right) \; .
    \label{eq:spin_Nernst_Ettingshausen_b1}
\end{eqnarray}
\end{widetext}
where $S_\mathrm{s}=\sigmac P' S$ and $\Pi_\mathrm{s}=P' S T$ are the coefficients for the spin Seebeck and the spin Peltier effect, respectively.

\subsubsection{Spin-thermomagnetic effect (spin-dependent Nernst effect)}
\textit{The spin-thermomagnetic effect describes the appearance of a transverse spin current component (a transverse spin accumulation in open circuit conditions) due to an applied longitudinal temperature gradient in the presence of a generalized magnetic field which is perpendicular to both the applied longitudinal gradient and the transverse response.} 

As for the previously discussed transverse effects, we can again distinguish three scenarios, depending on the origin of the generalized magnetic field. In Sec.\,\ref{par:STME} we will discuss the `ordinary' and `anomalous' spin-thermomagnetic effect, followed by the `spin-orbitronic' spin-thermomagnetic effect, which in literature is also known as the spin-dependent Nernst effect, in Sec.\,\ref{par:SNE}.

\paragraph{(Anomalous) spin-thermomagnetic effect.}\label{par:STME}
The ordinary spin-thermomagnetic effect describes the generation of a transverse spin current by an applied longitudinal temperature gradient in the presence of a perpendicularly oriented external magnetic field $\mu_0 \Hvec$. This effect is the spin analogue of the charge-thermomagnetic effect (Nernst effect). We intentionally do not name it spin Nernst effect here as this name has been specifically used in literature for the spin-orbitronic spin Nernst effect (see below).   

A longitudinal temperature gradient causes a collinear spin current density. For cases where the spin is transported by charge carriers, the equivalent electrical current density can be expressed as  (cf. Eq.~\ref{eq:GeneralCollinearTransportEquation2}))
\begin{eqnarray}
    \Jvecspintilde^\mathrm{long}  &=&  \Jvec^{\uparrow,\mathrm{long}} -  \Jvec^{\downarrow,\mathrm{long}} \nonumber\\
     &=& \left( \sigmac^\uparrow S^\uparrow -  \sigmac^\downarrow S^\downarrow \right) \; \left( -\boldsymbol{\nabla}T \right).
    \label{eq:spin_Nernst_Ettingshausen_03}
\end{eqnarray}
The transverse deflection of the charge carriers results in a transverse spin current density, which we again express as an equivalent charge current density
\begin{eqnarray} 
\Jvecspintilde^\mathrm{trans}  &=& \Jvec^{\uparrow,\mathrm{trans}} -  \Jvec^{\downarrow,\mathrm{trans}} \nonumber\\
&=& - \theta_\mathrm{OHE} \left( \sigmac^\uparrow S^\uparrow -  \sigmac^\downarrow S^\downarrow \right) \left(\Hhat \times \boldsymbol{\nabla}T\right),
   \label{eq:spin_Nernst_Ettingshausen_04}
\end{eqnarray}
where $\Hhat$ is the unit vector parallel to $\Hvec$. In the same way as Eq.\,(\ref{eq:32}) this can be rewritten into
\begin{equation}
\Jvecspintilde^\mathrm{trans} = - \theta_\mathrm{OHE} P^\prime S \sigmac \; (\Hhat \times \boldsymbol{\nabla}T),
\label{eq:spin_Nernst_Ettingshausen_05}
\end{equation}
with $\sigmac=\sigmac^\uparrow + \sigmac^\downarrow$, $S=[\sigmac^\uparrow S^\uparrow +\sigmac^\downarrow S^\downarrow ]/[\sigmac^\uparrow + \sigmac^\downarrow]$, and $P^\prime$ according to Eq.\,(\ref{eq:SS_04}). We see that the same Hall angle enters into the expressions for the transverse spin current as for the  transverse charge current in the ordinary Hall effect. This is obvious as the same transverse (charge-sensitive) deflection  the Lorentz force causes both effects. We also point out that transverse spin and charge currents appear at the same time when applying a longitudinal temperature gradient. Therefore, in the same way as for the collinear Seebeck effect, charge- and spin-related effects cannot be separated in experiments measuring transverse electric fields.

In magnetically ordered materials with finite $\Mvec$ we obtain the anomalous spin-thermomagnetic effect (ASTME) to
\begin{equation}
\mathbf{\tilde{J}}_\mathrm{s}^\mathrm{trans} =  - \theta_\mathrm{ASTME} P^\prime S \sigmac \; (\Mhat \times \boldsymbol{\nabla}T)  \; ,
\label{eq:spin_Nernst_Ettingshausen_06}
\end{equation}
where $\Mhat$ is the unit vector parallel to $\Mvec$ and the angle $\theta_\mathrm{ASTME}$ depends on $\Mvec$ and the details of the underlying intrinsic and/or extrinsic deflection mechanisms.

\paragraph{Spin-orbitronic spin-thermomagnetic effect (spin(-dependent) Nernst effect).}\label{par:SNE}
The spin-orbitronic spin-thermomagnetic effect is in literature know as the spin-dependent Nernst effect and spin Nernst effect (SNE) and was first theoretically described in 2008.\citep{SNEtheory1,SNEtheory2} A few years later, in 2017, the existence of the effect was experimentally confirmed.\citep{NernstWMI,SNE2, Bose:2018, Bose:2019} As discussed before, we here will include the prefix `dependent' to emphasize the fact that we focus in the following  on systems where the spin is transported by charge carriers and therefore spin- and charge-related thermomagnetic effects can no longer be separated. The prefix 'dependent' is omitted when the spin transport is not associated by charge transport. This is for example the case in magnetically ordered insulators, where spin is transported by magnons.\citep{BauerCaloritronics}

\begin{figure}[h]
	\includegraphics[width=\textfiguresPerp]{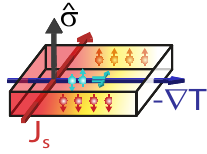} 
    \vspace{\captionhalfclose}
    \caption{\label{fig:SNE} Schematic representation of the spin-orbitronic spin-thermomagnetic (spin-dependent Nernst) effect. Here drawn assuming a negative Seebeck coefficient (as in Fig.\,\ref{fig:Seebeck}) and a positive spin Nernst angle.} 
\end{figure}

A schematic representation of the spin-dependent Nernst effect (SdNE) is shown in Fig.\,\ref{fig:SNE}: Spin-up and spin-down charge carriers of equal density are moving along the applied temperature gradient and are (spin-selectively) deflected in opposite transverse directions. As a result, a pure spin current in the transverse direction ($\Jvec^{\uparrow,\mathrm{trans}} = - \Jvec^{\downarrow,\mathrm{trans}}$) is generated. The resulting transverse spin current density is given by
\begin{eqnarray}
\Jvecspintilde^\mathrm{trans} & = &  - \theta_\mathrm{SNE} \left( \sigmac^\uparrow S^\uparrow +  \sigmac^\downarrow S^\downarrow \right) \; \left(\boldsymbol{\hat{\sigma}} \times \boldsymbol{\nabla}T\right) \; ,
   \label{eq:spin_Nernst_Ettingshausen_07}
\end{eqnarray}
where $\boldsymbol{\hat{\sigma}}$ is the unit vector in the direction of the spin polarization. The magnitude and sign of the spin Nernst angle $\theta_\mathrm{SNE}$ depend on the underlying microscopic, spin-dependent deflection mechanisms, which enter via the parameter `$a$' in the effective spin-orbitronic field $a\boldsymbol{\hat{\sigma}}$. With $S=[\sigmac^\uparrow S^\uparrow +\sigmac^\downarrow S^\downarrow ]/[\sigmac^\uparrow + \sigmac^\downarrow]$,  Eq.\,(\ref{eq:spin_Nernst_Ettingshausen_07}) can be rewritten into
\begin{eqnarray}
\Jvecspintilde^\mathrm{trans} &=&  - \theta_\mathrm{SNE} S \sigmac (\boldsymbol{\hat{\sigma}} \times \boldsymbol{\nabla}T) \; .
\label{eq:spin_Nernst_Ettingshausen_08}
\end{eqnarray}
Since the underlying microscopic mechanisms are based on spin-orbit interactions which scale proportional to $Z^4$, the angle $\theta_\mathrm{SNE}$ is expected to be large for heavy metals with large atomic number $Z$. Experimentally it was found that in Pt and W the spin Nernst angle $\theta_\mathrm{SNE}$ is as large as about $0.2$ and has the opposite sign compared to the spin Hall angle $\theta_\mathrm{SHE}$.\citep{NernstWMI,SNE-Kim} Intuitively, this can be understood by the fact that in spin-dependent Nernst experiments electrons are propagating parallel to $\mathbf{E}_\mathrm{h}=-\boldsymbol{\nabla}T$. In contrast, in experiments on the spin Hall effect (cf. subsection~\ref{par:SHE}) electrons are moving anti-parallel to $\Efieldc=-\boldsymbol{\nabla} \mucharge/q$. If the microsopic mechanisms for the two cases deflect the spin-up and spin-down  electrons in the same direction, the resulting transverse spin currents are expected to have the opposite signs corresponding to opposite signs of $\theta_\mathrm{SNE}$ and $\theta_\mathrm{SHE}$. Moreover, since positive and negative charge carriers move in the same direction in a temperature gradient and the spin-dependent transverse deflection is the same, the sign of $\theta_\mathrm{SNE}$ is expected to be independent of the charge carrier type. This is analogous to the charge Nernst coefficient, which also has a sign independent of the carrier type.

\subsubsection{Thermo-spinmagnetic effect (spin-dependent Ettingshausen effect)}\label{sec:thermospinmagneticeffect}

\textit{The thermo-spinmagnetic effect can be viewed as the reciprocal of the spin-thermomagnetic effect. It describes the generation of a transverse heat current flow (a temperature gradient in open circuit conditions) by an applied longitudinal gradient  in the spin chemical potential (a spin current) in the presence of a generalized magnetic field which is perpendicular to both the applied longitudinal gradient and the transverse response.} 

Analogue to the spin(-dependent) Nernst effect, the spin-orbitronic thermo-spinmagnetic  effect can be denoted as  the spin(-dependent) Ettingshausen effect. The potential presence of this effect has been mentioned in two side notes in Refs.\,\citep{Gross,LudoMMRtheory}.

Again, we split the discussion of the thermo-spinmagnetic effect depending on the origin of the generalized magnetic field. In Sec.\,\ref{par:ISTME}, we discuss the `ordinary' and `anomalous' thermo-spinmagnetic effect, followed by the `spin-orbitronic' thermo-spinmagnetic effect or spin(-dependent) Ettingshausen effect in Sec.\,\ref{par:SEE}.

\paragraph{(Anomalous) thermo-spinmagnetic effect.}\label{par:ISTME}
Similar to the previous sections we rationalize the thermo-spinmagnetic effect by considering the longitudinal charge current density generated by the applied gradient in the spin-chemical potential  which  is given by
\begin{equation}
\Jvec^\mathrm{long}  =  \sigmac P \; \left( -\boldsymbol{\nabla} \muspin/2q \right).
    \label{eq:spin_Nernst_Ettingshausen_09}
\end{equation}
The transverse deflection of the charge carriers results in the transverse charge current density
\begin{equation}
\Jvec^\mathrm{trans}  =  - \theta_\mathrm{OHE} \sigmac P \; \left( \Hhat \times \boldsymbol{\nabla} \muspin/2q \right) ,
\label{eq:spin_Nernst_Ettingshausen_10}
\end{equation}
which, according to $\Jvecheat = \Pi \Jvec$ and $\Pi=ST$, is associated with the transverse heat current density via 
\begin{equation}
\mathbf{J}_\mathrm{h}^\mathrm{trans}  = - \theta_\mathrm{OHE} \sigmac S T P \;  \left(\Hhat \times \boldsymbol{\nabla} \mu_\mathrm{s}/2q \right),
   \label{eq:spin_Nernst_Ettingshausen_11}
\end{equation}
where again $\Hhat$ is the unit vector parallel to $\Hvec$.

In magnetically ordered materials with finite $\Mvec$ we obtain the anomalous thermo-spinmagnetic (ATSME)effect to
\begin{equation}
\mathbf{J}_\mathrm{h}^\mathrm{trans}  = - \theta_\mathrm{ATSME} \sigmac S T P \; \left(\Mhat \times\boldsymbol{\nabla} \mu_\mathrm{s}/2q \right).
   \label{eq:spin_Nernst_Ettingshausen_12}
\end{equation}
where again $\Mhat$ is the unit vector parallel to $\Mvec$ and the angle $\theta_\mathrm{ATSME}$ depends on the details of the underlying microscopic mechanisms.

\begin{figure}[h]
	\includegraphics[width=\textfiguresPerp]{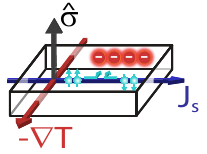} 
    \vspace{\captionhalfclose}
    \caption{\label{fig:SpinEtt} Schematic representation of the spin-orbitronic thermo-spinmagnetic effect (spin-dependent Ettingshausen effect), in open circuit conditions. Here drawn assuming a negative Seebeck coefficient and a positive spin Nernst angle (as in Fig.\,\ref{fig:SNE}). 
    }
\end{figure}

\paragraph{Spin-orbitronic thermo-spinmagnetic effect (spin(-dependent) Ettingshausen effect).}\label{par:SEE}
Analogue to the Nernst and Ettingshausen effect, the spin-orbitronic thermo-spinmagnetic effect can be viewed as the reciprocal effect of the spin-orbitronic spin-thermomagnetic effect, or spin(-dependent) Nernst effect. Therefore, we suggest to name this effect as spin(-dependent) Ettingshausen effect (SDEE) (see Table\,\ref{tab:trans}). 
A schematic representation of this effect is shown in Fig.\,\ref{fig:SpinEtt}.

To observe the spin-dependent Ettingshausen effect in a non-magnetic metal, a pure spin current (associated with no net charge current) has to be generated in the longitudinal direction, requiring $\Jvec^{\uparrow,\mathrm{long}} = -  \Jvec^{\downarrow,\mathrm{long}}$. Spin-up and spin-down charge carriers, which flow in the same direction are spin-selectively deflected in opposite transverse directions. Therefore, the opposite transverse flow of charge carriers with spin-up and spin-down character results in a pure transverse spin current. As a result, we obtain a pure transverse charge current $\Jvec^\mathrm{trans} = \theta_\mathrm{SEE} ( \Jvec^{\uparrow,\mathrm{long}} + \Jvec^{\downarrow,\mathrm{long}})$, which translates into a transverse heat current
\begin{equation}
\Jvecheat^\mathrm{trans}  = - \theta_\mathrm{SEE}  \sigmac S T \; \left(\boldsymbol{\hat{\sigma}} \times\boldsymbol{\nabla} \muspin/2q \right) \; ,
   \label{eq:spin_Nernst_Ettingshausen_13}
\end{equation}
where the magnitude and sign of $\theta_\mathrm{SEE}$ depend on the involved microscopic mechanisms.

For the experimental observation of the pure spin-dependent Ettingshausen effect, a pure spin current (associated with no charge current) has to be generated in longitudinal direction and the resulting transverse thermal gradient associated with the charge current has to be measured. To the best of our knowledge, this has not yet been realized experimentally.

\section{
Planar galvanomagnetic and thermomagnetic effects in magnetically ordered materials} 
\label{sec:Planar}

In section \ref{sec:TransNew}, we discussed transport phenomena, where the generalized magnetic field, and hence the magnetization $\Mvec$, was oriented perpendicular to the plane spanned by the driving force (applied potential gradients) and the response currents. In contrast, in this section we focus on the case of magnetically ordered materials with their magnetization $\Mvec$ oriented within the plane spanned by the driving force and the response currents. A schematic overview of the effects described in this section is given in Table\,\ref{tab:planar}. Note that the  planar effects discussed in this section are even in magnetization, which means that they do not change sign with inversion of $\Mvec$. This renders them qualitatively different from the effects described in section \ref{sec:TransNew}, which all are odd in $\Mvec$. The reason is that the former are of second order in $\Mvec$, whereas the latter are of first order in $\Mvec$. Although we restricted our discussion so far only to first order effects, we include the second order effects here as they have the potential to contribute to the transport phenomena highlighted in the previous sections. Each of the phenomena mentioned in Table\,\ref{tab:planar} is briefly discussed in the following subsections.

\newcommand{\LongSpace}{\hspace{-.015em}} 

\begin{table*}
\textsf{
    \begin{tabularx}{\textwidth}{ r ? Y | SZ } 
    	\diagbox[width=\firstcolumn]{\color{response} \textbf{Response}}{\color{drive}\textbf{Drive} \hspace{2pt}} &  \textcolor{drive}{\textbf{Electrical}} & \textcolor{drive}{\textbf{Thermal}} \\
   		\midrule[1.5pt]
   		\multirow{3}{*}[\centerfirstt]{\textcolor{response}{\textbf{Electrical}} \hspace{10pt}}
		& ${\color{response}\Efield_\mathrm{mod}} \propto \left(\hat{\mathbf{m}}\cdot {\color{drive} \Jvec} \right)\hat{\mathbf{m}}$  & ${\color{response}\Efield_\mathrm{mod}} \propto \left(\hat{\mathbf{m}}\cdot {\color{drive} \nabla T} \right)\hat{\mathbf{m}}$ \\
        & \textleft{  {\textcolor{response}{  $\textrm{E}_\textrm{long}$}} \LongSpace : Anisotropic magnetoresistance \\ {\textcolor{response}{ $\textrm{E}_\textrm{trans}$}}: Planar Hall effect} & \textleft{{\textcolor{response}{ $\textrm{E}_\textrm{long}$}} \LongSpace : Anisotropic magneto-thermopower \\ \textcolor{white}{ $\textrm{E}_\textrm{long}$ \LongSpace :} (Anisotropic magneto-Seebeck effect) \\ {\textcolor{response}{ $\textrm{E}_\textrm{trans}$}}: Planar Nernst effect} \hspace{30pt}{\color{white}.} \\ 
        \hline
    	\multirow{3}{*}[\centerfirstt]{\textcolor{response}{\textbf{Thermal}} \hspace{10pt}}
		& ${\color{response}\mingradT_\mathrm{mod}} \propto \left(\hat{\mathbf{m}}\cdot {\color{drive} \Jvec} \right)\hat{\mathbf{m}}$ & ${\color{response}\mingradT_\mathrm{mod}} \propto \left(\hat{\mathbf{m}}\cdot {\color{drive} \mingradT} \right)\hat{\mathbf{m}}$ \\
         & \textleft{{\textcolor{response}{ $-\nabla T_\textrm{long}$}} \LongSpace : Anisotropic magneto-Peltier effect \\  {\textcolor{response}{ $-\nabla T_\textrm{trans}$}}: Planar Ettingshausen effect} & \hspace{3pt} 
         \textleft{{\textcolor{response}{ $-\nabla T_\textrm{long}$}} \LongSpace : Anisotropic magneto-thermoresistance \\ {\textcolor{response}{ $-\nabla T_\textrm{trans}$}}: Planar Righi-Leduc/Planar thermal Hall effect}  \\
\end{tabularx} 
    }    
    \caption{\label{tab:planar}
	Overview of the planar galvanomagnetic (left column) and thermomagnetic (right column) effects in magnetically ordered materials, as discussed in Sec.\,\ref{sec:Planar}. The modulation of the response ($\Efield_\mathrm{mod}$ and $\mingradT_\mathrm{mod}$) due to these effects is determined by the angle $\alpha$ between the magnetization direction and the (electrical or thermal) drive. For each drive/response combination, two effects are defined: $X_\textrm{long}$ describes the effect where the response is parallel to the drive and $X_\textrm{trans}$ describes the effect where the response is perpendicular to the drive (where $X=\Efield$ or $\mingradT$, e.g. $E_\textrm{long}\propto\Efield \cdot \Jvec$ and $E_\textrm{trans}\propto\Efield \times \Jvec$).
	}
\end{table*}

\subsection{Anisotropic magnetoresistance (AMR) and planar Hall effect (PHE)}
\textit{The anisotropic magnetoresistance and planar Hall effect describe the dependence of the longitudinal (AMR) and transverse resistance (PHE) of a magnetic material on the angle between its magnetization and the applied current direction.}

\begin{figure}[h]
	\includegraphics[height=\planarFigs]{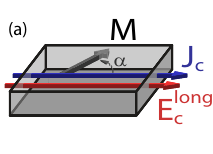}
    \includegraphics[height=\planarFigs]{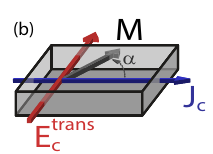}
    \vspace{\captionhalfclose}
    \caption{\label{fig:AMR} Schematic representation of (a) the anisotropic magnetoresistance (AMR) and (b) the planar Hall effect (PHE), in open circuit conditions, assuming $\rho_{\perp} < \rho_{\parallel}$. 
    }
\end{figure}

In 1857, William Thomson observed that the electric resistivity of nickel varies when measured along or perpendicular to the magnetization direction, and thus discovered the anisotropic magnetoresistance (AMR).\citep{Thomson} More generally, the AMR can be described as a change in the longitudinal resistance of a magnetic material (where `longitudinal' means `detected along the direction of the applied electrochemical potential difference'), as a function of the angle $\alpha$ between the applied potential gradient and the magnetization direction, as is schematically shown in Fig.\,\ref{fig:AMR}(a). 

In 1954, Colman Goldberg and R. Davis reported a new galvanomagnetic effect, which they called the planar Hall effect (PHE).\citep{PHE} This effect appears as a transverse voltage, which is measured in transverse direction relative to the applied longitudinal gradient in the electrochemical potential (cf. Fig.\,\ref{fig:AMR}(b)). Note that the historical denotation 'planar Hall effect' unfortunately is misleading and is not in agreement with the general nomenclature defined in this article. Whereas the (anomalous) Hall effect in a magnetically ordered material is of first order in the magnetization and therefore changes sign on the inversion of $\Mvec$, the PHE is of second order in $\Mvec$ and therefore does not change sign on inversion of $\Mvec$. It shares the same microscopic origin as the anisotropic magnetoresistance (AMR), namely spin-orbit interactions in the magnetic material.\citep{PHEmagnetometer,OHandley} 

To include the AMR and PHE in the conductivity tensor discussed in context of the Hall effect in Sec\,\ref{sec:Hall_Effect}, the conductivity parallel and perpendicular to $\Mvec$ needs to be distinguished. Traditionally, AMR and PHE consider the corresponding resistivities\footnote{The conductivity and resistivity tensor are linked via Ohm's law (Eq.~(\ref{eq:ohmslaw})).}:  $\rho_\perp$, the resistivity for a perpendicular configuration of the  magnetization and current ($\Mvec \perp \Jvec$) and $\rho_\parallel$ for the collinear orientation of $\Mvec$ and $\Jvec$ ($\Mvec \parallel \Jvec$). Typically, $\rho_{\perp} < \rho_{\parallel}$ in  3d-ferromagnets.\citep{OHandley,Smit, Gross} 
Combining Ohm's law (Eq.~(\ref{eq:ohmslaw})) with the resistivity tensor $\rhovec$ yields \citep{OHandley,Schmid} 
\begin{equation}
\label{eq:AMR}
		\Efieldc = \rho_{\parallel} \mathbf{J}_{\parallel} + \rho_{\perp} \mathbf{J}_{\perp} + \rho_\textrm{AHE} \,\Mhat \times \Jvec,
\end{equation}
where $\Mhat$ denotes the unit vector of the magnetization $\Mvec$ and for $\rho_\textrm{AHE}$ we refer to Sec.\,\ref{sec:AHE}. We can further decompose the current density $\Jvec$ into the components parallel to the magnetization direction $\mathbf{J}_{\parallel}=(\Mhat \cdot \Jvec)\Mhat$ and  perpendicular to the magnetization direction $\mathbf{J}_{\perp}=\Jvec-\mathbf{J}_{\parallel}$. Then, Eq.\,(\ref{eq:AMR}) takes the form \cite{OHandley,Schmid}
\begin{equation}
\label{eq:AMR2}
	\Efieldc = (\Mhat \cdot \Jvec)\Mhat[\rho_{\parallel}-\rho_{\perp}]+\rho_{\perp} \Jvec + \rho_\textrm{AHE} \,\Mhat \times \Jvec.
\end{equation}
While the first two terms on the right-hand-side describe both the AMR and PHE, the third term covers the (anomalous) Hall effect. Note that Eqs. (\ref{eq:AMR}) and (\ref{eq:AMR2}) are independent of the  specific sample geometry (i.e. the directions of the given vectors are free to choose).

For the typical measurement configuration, a thin-film Hall bar with an applied electrochemical potential gradient along the $x$-direction, the response originating from the AMR and PHE is characterized in the form of resistivities along ($\rho_\textrm{long}$) and perpendicular ($\rho_\textrm{trans}$) to the current density ($\Jvec$), respectively:  $\rho_\textrm{long}=\Efield \cdot \Jvec/ \lvert \Jvec\rvert^2$ (AMR) and $\rho_\textrm{trans}=\Efield \times \Jvec / \lvert \Jvec \rvert^2$ (PHE). Expressing the resistivities in terms of the angle $\alpha$ between $\Jvec$ and $\Mvec$ (cf. Fig.\,\ref{fig:AMR})  results in the longitudinal resistivity, associated with the AMR,\citep{OHandley,AMReq,ReviewSMR}
\begin{equation} \label{eq:rholong}
 	\rho_\textrm{long}(\alpha)=\rho_{\perp}+(\rho_{\parallel}-\rho_{\perp})\cos^2(\alpha),
\end{equation}
and the transverse resistivity, associated with the PHE, \citep{PHEmagnetometer,ReviewSMR}
\begin{equation}\label{eq:rhotrans}
	\rho_\textrm{trans}(\alpha)=(\rho_{\parallel}-\rho_{\perp})\cos(\alpha)\sin(\alpha).
\end{equation}
Note that (\ref{eq:rholong}) and (\ref{eq:rhotrans}) have been derived for isotropic materials. Of course,  in crystalline materials, additional AMR and PHE terms can arise from the (lower) crystal symmetry. \cite{Limmer:2006ju, Rushforth:2007, Ranieri:2008}  We also note that  the PHE, as well as the AMR, not only exist in conducting magnetic materials. Moreover,  Liu and coworkers reported the presence of a magnon PHE in the electrically insulating magnetic material yttrium iron garnet.\citep{Liu:2017}

\subsection{Anisotropic magneto-thermopower (AMTP) and planar Nernst effect (PNE)}
\textit{The anisotropic magneto-thermopower and planar Nernst effect describe the generation of an electric current (electric field in open circuit conditions) along (AMTP) or perpendicular (PNE) to an applied temperature gradient, which depends on the relative orientation of the magnetization and the direction of the applied temperature gradient. }

The planar Nernst effect (PNE) can be viewed as the planar analog of the Nernst effect 
and is sometimes also referred to as transverse (magneto-)thermopower. 
Its existence was first observed in 1966 by Vu Dinh Ky.\citep{PNE,PNE2} 


Similar to the AMR and PHE, the AMTP and PNE can be derived from Eq.~(\ref{eq:Seebeck-std}), taking into account the individual components of the Seebeck tensor $\mathbf{S}$.\citep{Schmid} In analogy to Eqs.\,(\ref{eq:rholong}) and (\ref{eq:rhotrans}), the electric fields generated by the AMTP\citep{Pu2006,Kuschel2017,Schmid} and the PNE\citep{PNE2,ANEeq,Pu2006,Kuschel2017,Schmid,PNEexp} can be expressed as
\begin{align}
	E_c^\mathrm{long} =  \nabla T [S_{\perp} + (S_{\parallel}-S_{\perp}) \cos^2(\alpha)] \enskip& \textrm{(AMTP)} \\
    E_c^\mathrm{trans} = \nabla T (S_{\parallel}-S_{\perp}) \cos(\alpha) \sin(\alpha) \enskip& \textrm{(PNE)},
\end{align}
where $S_{\parallel}$ and $S_{\perp}$ are the Seebeck coefficients of an isotropic material in the direction parallel and perpendicular to the applied temperature gradient $\nabla T$, respectively. Analogue to before, $\alpha$ is the angle between the temperature gradient and the magnetization direction. As a longitudinal electric field is generated by a longitudinal temperature gradient as for the Seebeck effect, this effect also can be named anisotropic magneto-Seebeck effect.

Unfortunately, in literature the nomenclature used for  the AMTP is not consistent, e.g. the following names are used for one and the same effect: longitudinal magnetothermopower,\citep{Pu2006} longitudinal thermopower,\citep{Pu2006} anisotropic thermopower,\citep{Pu2006} anisotropic magnetothermo(electric)power (abbreviated as AMTP)\citep{Kuschel2017} and (AMTEP)\citep{ANEeq,Schmid}), and longitudinal AMTP\citep{Kuschel2017}. 
Following the denotation used for the AMR we propose to consistently refer to this effect in future as anisotropic magneto-thermopower, abbreviated as AMTP. Noteably, AMTP in singly crystalline samples show a more complex behavior reflecting the underlying  crystalline symmetry, which is captured by additional terms \cite{Ritzinger:2021}.

\begin{figure}[h]
	\includegraphics[height=\planarFigs]{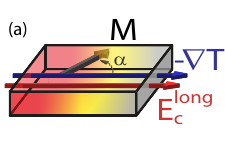}
    \includegraphics[height=\planarFigs]{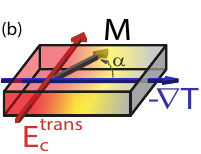}
    \vspace{\captionhalfclose}
    \caption{\label{fig:AMRP} Schematic representation of (a) the anisotropic magneto-thermopower (AMTP) and (b) the planar Nernst effect (PNE), in open circuit conditions, assuming $S_\parallel<S_\perp<0$. 
    }
\end{figure}

\subsection{Other planar effects}\label{sec:Other}
So far we have discussed the (longitudinal and transverse) planar electrical effects, appearing when we apply a longitudinal electrochemical potential or temperature gradient. The resulting effects are listed in the upper row of Table\,\ref{tab:planar}. In analogy, we also expect  corresponding (longitudinal and transverse) planar thermal effects, appearing when we apply a longitudinal electrochemical potential or temperature gradient. The resulting effects are listed in the lower row of Table\,\ref{tab:planar}. 

The longitudinal thermal analogue of the longitudinal AMR is the 
anisotropic magneto-thermoresistance (AMTR).\citep{AMTR} This effect describes the longitudinal thermal response generated by an applied longitudinal temperature gradient in the same way as the AMR describes the longitudinal electrical response generated by an applied longitudinal gradient of the electrochemical potential.  The AMTR was already measured in 1850 by G. Maggi.\citep{Maggi,Campbell} 
They observed a change in the thermal resistance of a (magnetic) material when varying the relative orientation between the applied thermal gradient and the magnetization.\citep{Campbell} \footnote{Interestingly, a similar effect as the AMTR was observed in non-magnetic materials.\cite{MRLinWolfram,MRLinTungsten,MRLsemicond,Campbell} This effect, a change in thermal conductivity of a material when a transverse magnetic field is applied, is called after its discoverers as the Maggi-Righi-Leduc effect\citep{Maggi,Campbell,Righi,Leduc}, and describes the thermal analogue of the ordinary magnetoresistance (and thus could also be named as magneto-thermoresistance). }

\begin{figure}[h]
	\includegraphics[height=\planarFigs]{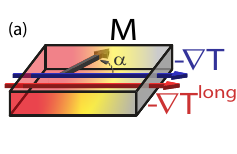}
    \includegraphics[height=\planarFigs]{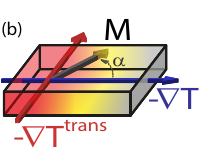}
    \vspace{\captionhalfclose}
    \caption{\label{fig:planarRLeduc} Schematic representation of (a) the anisotropic magneto-thermoresistance (AMTR) and (b) the planar Righi-Leduc/planar thermal Hall effect, in open circuit conditions.}
\end{figure}

The thermal analogue of the  (electrical) PHE is the (transverse) planar Righi-Leduc effect or planar thermal Hall effect. This effect describes the transverse thermal response generated by an applied longitudinal temperature gradient in the same way as the PHE describes the transverse electrical response generated by an applied longitudinal gradient of the electrochemical potential. (see Fig.~\ref{fig:planarRLeduc}b). It is found that the measured transverse temperature gradient depends on the angle between the applied longitudinal temperature gradient and the magnetization.  The presence of this effect has been  theoretically predicted by Jean-Eric Wegrowe and coworkers in 2014\citep{Wegrowe} and experimentally observed by Benjamin Madon and coworkers in 2016.\citep{Madon}

\begin{figure}[h]
	\includegraphics[height=\planarFigs]{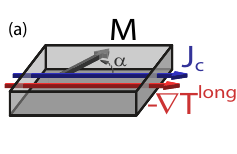}
    \includegraphics[height=\planarFigs]{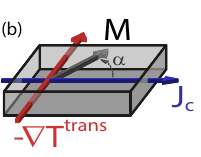}
    \vspace{\captionhalfclose}
    \caption{\label{fig:planarEtt} Schematic representation of (a) the anisotropic magneto-Peltier effect (AMP) and (b) the planar Ettingshausen effect, in open circuit conditions.}
\end{figure}

The two remaining planar effects describe the longitudinal and transverse thermal response caused by an applied gradient in the electrochemical potential (cf. Table\,\ref{tab:planar} and Fig.\,\ref{fig:planarEtt}). In analogy to the anisotropic magneto-thermopower or anisotropic magneto-Seebeck effect  (longitudinal gradient in the electrochemical potential generated by a longitudinal temperature gradient) we denote the corresponding effect as anisotropic magneto-Peltier effect (AMPE) (longitudinal temperature gradient generated by a longitudinal gradient in the electrochemical potential). In the same way, we use the name planar Ettingshausen effect (PEE) (transverse temperature gradient generated by a longitudinal gradient in the electrochemical potential) in analogy to the planar Nernst effect  (transverse gradient in the electrochemical potential generated by a longitudinal temperature gradient). Regarding the planar Ettingshausen effect, the authors are not aware of any published work besides the mentioning of the effect in form of some side-notes in Refs.\,\citep{Gross,LudoMMRtheory,AMP_Uchida}  An experimental proof for the anisotropic magneto-Peltier effect has been given by Uchida and coworkers.\citep{AMP_Uchida}

Finally we want to note that all spin-related effects are missing in Table\,\ref{tab:planar}. To the best of our knowledge, only the magnon planar Hall effect \citep{Liu:2017} has been reported so-far, warranting more efforts in this direction. 

\section{Summary}
In summary, we have presented a comprehensive discussion of the 
transport phenomena, resulting from the coupled transport of charge, heat, and spin in metallic conductors, which are described by the generalized transport matrix for charge, heat, and spin. In our discussion, we also point to some inconsistencies in the nomenclature used for the identification of  the various transport phenomena, which originates from the fact that the historical denotation of the effects  did not account for the full scheme of all effects known today. We structured this work  into (i) collinear, (ii) transverse, and (iii) planar transport phenomena. Within these categories, we introduced and discussed:

(i) Electrical conductivity, thermal conductivity, spin conductivity, spin-polarized transport, Seebeck effect, Peltier effect, Thomson effect, spin(-dependent) Seebeck effect, spin(-dependent) Peltier effect, and spin(-dependent) Thomson effect.

(ii) Ordinary, anomalous, and spin-orbitronic form of the Hall effect, thermal Hall/Righi-Leduc effect, pure spin Hall effect, spin-galvanomagnetic effect and the reciprocal electro-spinmagnetic effect (including the spin Hall effect and the inverse spin Hall effect), Nernst effect, Ettingshausen effect, spin-thermomagnetic effect and the reciprocal thermo-spin magnetic effect (including the spin Nernst effect and the spin Ettingshausen effect).

(iii) Anisotropic magnetoresistance and planar Hall effect, anisotropic magneto-thermopower and planar Nernst effect, anisotropic magneto-thermoresistance and planar thermal Hall effect, and anisotropic magneto-Peltier effect and planar Ettingshausen effect.

Additionally, the appendix visualizes the direct link between the generalized transport matrix and the above described phenomena and emphasizes the chosen categorization.

\section*{Acknowledgements}
We acknowledge financial support by the Deutsche Forschungsgemeinschaft (DFG, German Research Foundation) via DFG Priority program 1538 ‘Spin-Caloric Transport’ (Projects GO 944/4 and GR 1132/18),  the Transregio “Constrained Quantum Matter” (TRR 360, Project-ID 492547816), the Research Unit ChiPS ( Project-ID 541503763)  the German Excellence Initiative via the ”Nanosystems Initiative Munich (NIM)”, and Germany’s Excellence Strategy EXC-2111-390814868. NV acknowledges the Laura-Bassi stipend of the Technical University of Munich. 
\section*{Appendix A}
\label{appendix}
\begin{figure*}[]
		\includegraphics[width=\textwidth]{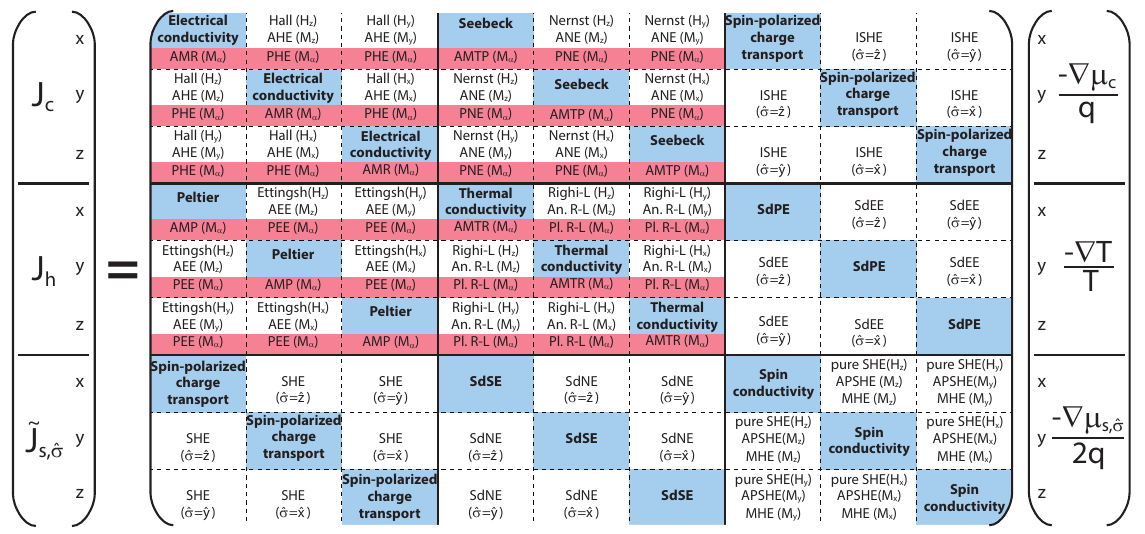}
		\caption{\label{fig:BIG}
        Overview of all transport phenomena described in this paper (except the (spin-)Thomson effect), based on the general transport matrix given in Eq.\,(\ref{eq:GeneralGeneralTransportEquation}). This figure shows how all transport phenomena are included by expanding the scalar matrix elements given in Eq.\,(\ref{eq:GeneralGeneralTransportEquation}) to tensor quantities. Collinear transport phenomena are marked in blue and set in bold letter font, transverse/cross-product transport phenomena are marked in white and planar effects are marked in red. 
        The variable displayed in brackets behind the name of each phenomenon denotes the additionally required external magnetic field ($\Hvec$), internal magnetic order ($\Mvec$), or spin polarization ($\sigmapol$) with their corresponding direction ($x$, $y$, $z$, or `$\alpha$' representing an angle between the generalized drive and the direction of the magnetization). The $x$, $y$, and $z$ denoted along with the drive ($-\nabla \mucharge$, $-\nabla T$, and $-\nabla \muspin{_{,\bm{\hat{\sigma}}}}$) and response ($\Jvec$, $\Jvecheat$, and ${\Jvecspintilde}{_{,\bm{\hat{\sigma}}}}$) terms, define their spatial (flow) direction.     
        Besides the abbreviations used in the main text, this table employs due to space limitations the following abbreviations: Ettingshausen effect (Ettingsh), planar Ettingshausen effect (PEE), Righi-Leduc/Thermal Hall effect (Righi-L), anomalous Righi-Leduc/thermal Hall effect (An. R-L), planar Righi-Leduc/thermal Hall effect (Pl. R-L),  pure spin Hall effect (pure SHE), anomalous pure spin Hall effect (APSHE), and magnon Hall effect (MHE). 
}
\end{figure*}

All effects described in this paper can be deduced from the general transport equation given by Eq.\,(\ref{eq:GeneralGeneralTransportEquation}) or Eq.\,(\ref{eq:GeneralTransportCoefficients_transverse}). Nevertheless, expanding the scalar transport coefficients  to tensor quantities in order to include all (anomalous) thermoelectric, thermomagnetic, galvanomagnetic, and related spin-dependent effects is beyond the scope of this article. However, to give some guidance to the reader, we have compiled the matrix depicted in Fig.\,\ref{fig:BIG}. Here, it can be seen that each initially scalar matrix-element in Eq.\,(\ref{eq:GeneralTransportCoefficients_transverse}) actually transforms to tensors. These include the scalar, the transverse and planar effects described in this article. For example,  $\sigmac$ of Eq.\,(\ref{eq:GeneralGeneralTransportEquation}) is expanded to a tensor including next to Ohm's law also the (anomalous) Hall effect, the planar Hall effect, and the AMR. 

In Fig.\,\ref{fig:BIG}, the collinear transport phenomena, which are described in Sec.\,\ref{sec:Col}, are marked in blue and bold font face. The transverse transport phenomena, which are described in Sec.\,\ref{sec:TransNew} are shown as white entries. The red marked phenomena are the planar effects, which are discussed in Sec.\,\ref{sec:Planar}. 

The reader should note that the spin current itself is a tensor quantity. Taking into account all possible terms for ${\Jvecspin}_{,\hat{\sigma}}$ (${\Jvecspin}_{,\hat{x}}$, ${\Jvecspin}_{,\hat{y}}$, ${\Jvecspin}_{,\hat{z}}$), each with flow directions $x$,$y$ and $z$) the matrix would grow to a $5\times 5$ matrix. In that case spin-flip processes and non-cross-product spin-processes would also be included.  
By this overview, we want to offer the reader a helping hand in recognizing, disentangling, and understanding the basic origin of the different charge, heat, and spin related transport phenomena.

\sloppy
\normalem 

\bibliography{References_Effects}

\end{document}